# Hypertableau Reasoning for Description Logics

**Boris Motik**　　　　　　　　　　　　　　　　　　BORIS.MOTIK@COMLAB.OX.AC.UK
**Rob Shearer**　　　　　　　　　　　　　　　　　　ROB.SHEARER@COMLAB.OX.AC.UK
**Ian Horrocks**　　　　　　　　　　　　　　　　　　IAN.HORROCKS@COMLAB.OX.AC.UK
*Computing Laboratory, University of Oxford*
*Wolfson Building*
*Parks Road*
*Oxford OX1 3QD*
*United Kingdom*

## Abstract

We present a novel reasoning calculus for the description logic $\mathcal{SHOIQ}^+$—a knowledge representation formalism with applications in areas such as the Semantic Web. Unnecessary nondeterminism and the construction of large models are two primary sources of inefficiency in the tableau-based reasoning calculi used in state-of-the-art reasoners. In order to reduce nondeterminism, we base our calculus on hypertableau and hyperresolution calculi, which we extend with a blocking condition to ensure termination. In order to reduce the size of the constructed models, we introduce *anywhere pairwise blocking*. We also present an improved nominal introduction rule that ensures termination in the presence of nominals, inverse roles, and number restrictions—a combination of DL constructs that has proven notoriously difficult to handle. Our implementation shows significant performance improvements over state-of-the-art reasoners on several well-known ontologies.

## 1. Introduction

Description Logics (DLs) (Baader, Calvanese, McGuinness, Nardi, & Patel-Schneider, 2007) are a family of knowledge representation formalisms with well-understood formal properties. DLs have been applied to numerous problems in computer science such as information integration and metadata management. Recent interest in DLs has been spurred by their application in the Semantic Web: the DL $\mathcal{SHOIQ}$ provides the logical underpinning for the Web Ontology Language (OWL) (Patel-Schneider, Hayes, & Horrocks, 2004), and the DL $\mathcal{SROIQ}$ (Kutz, Horrocks, & Sattler, 2006) is used in OWL 2—an extension of OWL currently being standardized by the World Wide Web Consortium.

A central component of most DL applications is an efficient and scalable reasoner. Modern reasoners, such as Pellet (Parsia & Sirin, 2004), FaCT++ (Tsarkov & Horrocks, 2006), and RACER (Haarslev & Möller, 2001), are typically based on tableau calculi (Baader & Nutt, 2007), which demonstrate the (un)satisfiability of a knowledge base $\mathcal{K}$ via a constructive search for an abstraction of a model of $\mathcal{K}$. Numerous optimizations have been developed in an effort to reduce the size of the search space (Horrocks, 2007). Despite major advances in tableau reasoning algorithms, however, ontologies are still encountered in practice that cannot be handled by existing DL reasoners. Two main sources of complexity in tableau calculi have been identified in the literature (Donini, 2007).

This first source of complexity is known as *or-branching*: given a disjunctive assertion

$(C \sqcup D)(s)$, a tableau algorithm nondeterministically guesses that either $C(s)$ or $D(s)$ holds. To show the unsatisfiability of $\mathcal{K}$, *every* possible guess must lead to a contradiction: if assuming that $C(s)$ holds leads to a contradiction, the algorithm must backtrack and assume that $D(s)$ holds, which can give rise to exponential behavior. General concept inclusions (GCIs)—implications of the form $C \sqsubseteq D$—are the main source of disjunctions: to ensure that $C \sqsubseteq D$ holds, a tableau algorithm adds a disjunction $(\neg C \sqcup D)(s)$ to each individual $s$ in the model. Various *absorption* optimizations (Horrocks, 1998; Tsarkov & Horrocks, 2004; Hudek & Weddell, 2006; Horrocks, 2007) have been developed to reduce the nondeterminism in tableau calculi.

The second source of complexity in tableau calculi is known as *and-branching*: the expansion of a model due to existential quantifiers can generate very large models. Apart from memory consumption problems, and-branching can increase or-branching by increasing the number of individuals to which GCIs are applied.

In this paper, we present a reasoning calculus that addresses both sources of complexity. We focus on the DL $\mathcal{SHOIQ}^+$, which is obtained by extending $\mathcal{SHOIQ}$ with local reflexivity and disjoint, reflexive, irreflexive, symmetric, and asymmetric roles. $\mathcal{SROIQ}$ further extends $\mathcal{SHOIQ}^+$ with *generalized role inclusions* of the form $R_1 \circ \ldots \circ R_n \sqsubseteq R$. Generalized role inclusions can be encoded using standard GCIs as proposed by Demri and de Nivelle (2005); thus, by adding a suitable preprocessing phase, the results from this paper should allow us to handle $\mathcal{SROIQ}$ (and hence OWL 2) as well.

Our algorithm can be viewed as a hybrid of resolution and tableau, and is related to the hypertableau (Baumgartner, Furbach, & Niemelä, 1996) and hyperresolution (Robinson, 1965) calculi. It first preprocesses a $\mathcal{SHOIQ}^+$ knowledge base into a set of *DL-clauses*—universally quantified implications containing DL concepts and roles as predicates. The main derivation rule for DL-clauses is hyperresolution: an atom from the consequent of a DL-clause is derived *only if* all atoms from the DL-clause antecedent can be matched to already derived consequences. Hyperresolution is very effective at restricting or-branching. Consider, for example, the following example:

$$(1) \qquad R(x, y_1) \wedge S(x, y_2) \rightarrow A(x) \vee B(y_1) \vee C(y_2)$$

This DL-clause derives a disjunction only if it is applied to assertions of the form $R(a, b)$ and $S(c, d)$ where $a = c$. The presence of variables in (1) allows us to simultaneously work with individuals $a$, $b$, $c$ and $d$, and to check whether $a = c$. In contrast, derivation rules in tableau algorithms consider at most pairs of individuals; consequently, no absorption technique we are aware of can localize nondeterminism only to the individuals that satisfy the mentioned constraints. As we discuss in detail in Section 3.3.1, our calculus generalizes all known absorption variants. Furthermore, in contrast to absorption techniques, our algorithm is guaranteed to exhibit no nondeterminism on Horn knowledge bases (Hustadt, Motik, & Sattler, 2005) such as GALEN, NCI, and SNOMED CT (see Section 7). Finally, our calculus provides a uniform proof-theoretic framework that can handle several useful extensions of commonly used DLs (see Section 4.1.3).

Hyperresolution decides many fragments of first-order logic (e.g., Fermüller, Leitsch, Hustadt, & Tammet, 2001; Fermüller, Tammet, Zamov, & Leitsch, 1993), as well as description and modal logics (e.g., Georgieva, Hustadt, & Schmidt, 2003; Hustadt & Schmidt, 1999). Unlike most of these fragments, $\mathcal{SHOIQ}^+$ allows for cyclic GCIs of the form $C \sqsubseteq \exists R.C$, on which hyperresolution can generate infinite paths of successors. To ensure termination, we use the pairwise blocking technique (Horrocks, Sattler, & Tobies, 2000b) to detect cyclic computations. Due to hyper-inferences, the soundness and correctness proofs by Horrocks et al. (2000b) do not carry over immediately to our calculus; in fact, certain simpler blocking conditions applicable to weaker DLs cannot be straightforwardly transferred to our setting. To limit and-branching, we extend the blocking condition by Horrocks et al. to *anywhere pairwise blocking*: an individual can be blocked by another individual that is not necessarily its ancestor, which can reduce the sizes of the constructed models. Anywhere blocking has already been used with single blocking (Buchheit, Donini, & Schaerf, 1993; Baader, Buchheit, & Hollunder, 1996; Donini & Massacci, 2000; Donini, Lenzerini, Nardi, & Schaerf, 1998); however, to the best of our knowledge, it has been neither used with the more sophisticated pairwise blocking nor tested in practice.

Ensuring termination of a tableau decision procedure for DLs with nominals, inverse roles, and number restrictions has proven notoriously difficult. This problem was finally solved by Horrocks and Sattler (2007) by extending the tableau calculus with a *nominal introduction* rule. In certain situations, this rule guesses and introduces new nominals, and is thus a potential source of inefficiency in practice. In this paper, we present a variant of this rule that is simpler and more efficient.

We have implemented our calculus in a new reasoner called HermiT.[1] Even with a rather naïve implementation, the deterministic treatment of GCIs significantly reduces classification times for several real-world ontologies. Furthermore, pairwise anywhere blocking seems to be very effective in limiting model sizes and it allows HermiT to classify several ontologies that, to the best of our knowledge, no other reasoner can handle.

## 2. Preliminaries

We now define the syntax and the semantics of the description logic $\mathcal{SHOIQ}^+$. A *signature* is a triple $\Sigma = (N_R, N_C, N_I)$ consisting of mutually disjoint sets of *atomic roles* $N_R$, *atomic concepts* $N_C$, and *individuals* $N_I$. The set of *roles* is then $N_R \cup \{R^- \mid R \in N_R\}$. The function $\mathsf{inv}(\cdot)$ is defined on the set of roles as follows, where $R$ is an atomic role: $\mathsf{inv}(R) = R^-$ and $\mathsf{inv}(R^-) = R$. An *RBox* $\mathcal{R}$ is a finite set of axioms of the form $R_1 \sqsubseteq R_2$ (*role inclusion*), $\mathsf{Dis}(S_1, S_2)$ (*role disjointness*), $\mathsf{Ref}(R)$ (*reflexivity*), $\mathsf{Irr}(S)$ (*irreflexivity*), $\mathsf{Sym}(R)$ (*symmetry*), $\mathsf{Asy}(S)$ (*asymmetry*), and $\mathsf{Tra}(R)$ (*transitivity*), where $R$, $R_1$, and $R_2$ are roles, and $S$, $S_1$, and $S_2$ are *simple* roles, as defined next. Let $\sqsubseteq^*_{\mathcal{R}}$ be the reflexive-transitive closure of the following relation: $\{(R_1, R_2) \mid R_1 \sqsubseteq R_2 \in \mathcal{R} \text{ or } \mathsf{inv}(R_1) \sqsubseteq \mathsf{inv}(R_2) \in \mathcal{R}\}$. A role $R$ is *transitive* in $\mathcal{R}$ if a role $R'$ exists such that $R' \sqsubseteq^*_{\mathcal{R}} R$, $R \sqsubseteq^*_{\mathcal{R}} R'$, and either $\mathsf{Tra}(R') \in \mathcal{R}$ or $\mathsf{Tra}(\mathsf{inv}(R')) \in \mathcal{R}$. A role $S$ is *simple* if no transitive role $R$ exists such that $R \sqsubseteq^*_{\mathcal{R}} S$. The set of *concepts* is the smallest set containing $\top$ (the *top concept*), $\bot$ (the *bottom concept*), $A$ (*atomic concept*), $\{a\}$ (*nominal*), $\neg C$ (*negation*), $C \sqcap D$ (*conjunction*), $C \sqcup D$ (*disjunction*), $\exists R.C$ (*existential restriction*), $\forall R.C$ (*universal restriction*), $\exists S.\mathsf{Self}$ (*local reflexivity*), $\geq n\, S.C$ (*at-least restriction*), and $\leq n\, S.C$ (*at-most restriction*), for $A$ an atomic concept, $a$ an individual, $C$ and $D$ concepts, $R$ a role, $S$ a simple role, and $n$ a nonnegative integer. A *TBox* $\mathcal{T}$ is a finite set of *general concept inclusions* (GCIs) $C \sqsubseteq D$ for $C$ and $D$ concepts. An *ABox* $\mathcal{A}$ is a finite set of assertions of the form $C(a)$ (*concept assertion*), $R(a, b)$

---

1. http://www.hermit-reasoner.com/

Table 1: Model-Theoretic Semantics of $\mathcal{SHOIQ}^+$

| Interpretation of Concepts and Roles | |
|---|---|
| $\top^I = \triangle^I$ | $\bot^I = \emptyset$ |
| $\{s\}^I = \{s^I\}$ | $(\neg C)^I = \triangle^I \setminus C^I$ |
| $(C \sqcap D)^I = C^I \cap D^I$ | $(C \sqcup D)^I = C^I \cup D^I$ |
| $(R^-)^I = \{\langle y, x \rangle \mid \langle x, y \rangle \in R^I\}$ | $(\exists S.\mathsf{Self})^I = \{x \mid \langle x, x \rangle \in S^I\}$ |
| $(\exists R.C)^I = \{x \mid \exists y : \langle x, y \rangle \in R^I \wedge y \in C^I\}$ | |
| $(\forall R.C)^I = \{x \mid \forall y : \langle x, y \rangle \in R^I \rightarrow y \in C^I\}$ | |
| $(\geq n\, S.C)^I = \{x \mid \sharp\{y \mid \langle x, y \rangle \in S^I \wedge y \in C^I\} \geq n\}$ | |
| $(\leq n\, S.C)^I = \{x \mid \sharp\{y \mid \langle x, y \rangle \in S^I \wedge y \in C^I\} \leq n\}$ | |

| Satisfaction of Axioms in an Interpretation | | | |
|---|---|---|---|
| $I \models C \sqsubseteq D$ | iff $C^I \subseteq D^I$ | $I \models R_1 \sqsubseteq R_2$ | iff $R_1^I \subseteq R_2^I$ |
| $I \models \mathsf{Ref}(R)$ | iff $\forall x \in \triangle^I : \langle x, x \rangle \in R^I$ | $I \models \mathsf{Irr}(S)$ | iff $\forall x \in \triangle^I : \langle x, x \rangle \notin S^I$ |
| $I \models \mathsf{Sym}(R)$ | iff $R^I \subseteq (\mathsf{inv}(R))^I$ | $I \models \mathsf{Asy}(S)$ | iff $S^I \cap (\mathsf{inv}(S))^I = \emptyset$ |
| $I \models \mathsf{Tra}(R)$ | iff $(R^I)^+ \subseteq R^I$ | $I \models \mathsf{Dis}(S_1, S_2)$ | iff $S_1^I \cap S_2^I = \emptyset$ |
| $I \models C(a)$ | iff $a^I \in C^I$ | $I \models R(a, b)$ | iff $\langle a^I, b^I \rangle \in R^I$ |
| $I \models a \approx b$ | iff $a^I = b^I$ | $I \models a \not\approx b$ | iff $a^I \neq b^I$ |

**Note:** $\sharp N$ is the number of elements in $N$, and $R^+$ is the transitive closure of $R$.

(*role assertion*), $a \approx b$ (*equality assertion*), and $a \not\approx b$ (*inequality assertion*), where $C$ is a concept, $R$ is a role, and $a$ and $b$ are individuals. A $\mathcal{SHOIQ}^+$ knowledge base $\mathcal{K}$ is a triple $(\mathcal{R}, \mathcal{T}, \mathcal{A})$. With $|\mathcal{K}|$ we denote the size of $\mathcal{K}$—that is, the number of symbols required to encode $\mathcal{K}$ on the input tape of a Turing machine (numbers can be coded in binary).

An *interpretation* for $\mathcal{K}$ is a tuple $I = (\triangle^I, \cdot^I)$, where $\triangle^I$ is a nonempty set, and $\cdot^I$ assigns an element $a^I \in \triangle^I$ to each individual $a$, a set $A^I \subseteq \triangle^I$ to each atomic concept $A$, and a relation $R^I \subseteq \triangle^I \times \triangle^I$ to each atomic role $R$. The function $\cdot^I$ is extended to concepts and roles as shown in the upper part of Table 1. $I$ is a *model* of $\mathcal{K}$, written $I \models \mathcal{K}$, if it satisfies all axioms of $\mathcal{K}$ as shown in the lower part of Table 1. The basic inference problem for $\mathcal{SHOIQ}^+$ is checking whether $\mathcal{K}$ is *satisfiable*—that is, checking whether a model of $\mathcal{K}$ exists. A concept $C$ *subsumes* a concept $D$, written $\mathcal{K} \models C \sqsubseteq D$, if $C^I \subseteq D^I$ for each model $I$ of $\mathcal{K}$. It is easy to see that $\mathcal{K} \models C \sqsubseteq D$ if and only if $\mathcal{K} \cup \{(C \sqcap \neg D)(a)\}$ is unsatisfiable, where $a$ is an individual that does not occur in $\mathcal{K}$ (Baader & Nutt, 2007).

The *negation-normal form* $\mathsf{nnf}(C)$ of a concept $C$ is the concept obtained from $C$ by using de Morgan's laws, the dualities between existential and universal restrictions, and the dualities between at-least and at-most restrictions to push negations inwards so that they occur only in front of atomic concepts, nominals, and local reflexivity concepts. The concept $\mathsf{nnf}(C)$ is logically equivalent to $C$, and it can be computed from $C$ in time linear in the size of $C$ (Baader & Nutt, 2007). We use $\dot{\neg}C$ to denote $\mathsf{nnf}(\neg C)$.

As mentioned in Section 1, extending $\mathcal{SHOIQ}^+$ with general role inclusions would yield $\mathcal{SROIQ}$ (Kutz et al., 2006)—the DL that underpins OWL 2. $\mathcal{ALCHOIQ}^+$ is obtained from $\mathcal{SHOIQ}^+$ by disallowing transitivity axioms. $\mathcal{SHIQ}^+$ is obtained from $\mathcal{SHOIQ}^+$ by disallowing nominals. $\mathcal{SHOQ}^+$ is obtained from $\mathcal{SHOIQ}^+$ by disallowing inverse roles. $\mathcal{SHOIQ}$ and $\mathcal{SHIQ}$ are obtained from $\mathcal{SHOIQ}^+$ and $\mathcal{SHIQ}^+$, respectively, by disallow-ing local reflexivity, role disjointness, reflexivity, irreflexivity, symmetry, and asymmetry axioms. Finally, $\mathcal{SHOI}$ is obtained from $\mathcal{SHOIQ}$ by disallowing at-least and at-most restrictions.

## 3. Motivation and Algorithm Overview

In this section, we present an overview of the main aspects of our algorithm. We explain in Section 3.1 the root causes of the scalability problems encountered in tableau algorithms, and in Section 3.2 we outline how we address them. Finally, in Section 3.3 we discuss the relationship between our algorithm and some related approaches.

### 3.1 Causes of Scalability Problems in Tableau Algorithms

To show that a knowledge base $\mathcal{K} = (\mathcal{R}, \mathcal{T}, \mathcal{A})$ is satisfiable, a tableau algorithm constructs a *derivation*—a sequence of ABoxes $\mathcal{A}_0, \mathcal{A}_1, \ldots, \mathcal{A}_n$ where $\mathcal{A}_0 = \mathcal{A}$ and each $\mathcal{A}_i$ is obtained from $\mathcal{A}_{i-1}$ by an application of one *derivation rule*.[2] The derivation rules make the information implicit in the axioms of $\mathcal{R}$ and $\mathcal{T}$ explicit, and thus evolve the ABox $\mathcal{A}$ towards a (representation of a) model of $\mathcal{K}$. The algorithm terminates either if no derivation rule is applicable to some $\mathcal{A}_n$, in which case $\mathcal{A}_n$ represents a model of $\mathcal{K}$, or if $\mathcal{A}_n$ contains an obvious contradiction, in which case the model construction has failed. The following derivation rules are commonly used in DL tableau calculi.

- $\sqcup$-rule: Given $(C_1 \sqcup C_2)(s)$, derive either $C_1(s)$ or $C_2(s)$.

- $\sqcap$-rule: Given $(C_1 \sqcap C_2)(s)$, derive $C_1(s)$ and $C_2(s)$.

- $\exists$-rule: Given $(\exists R.C)(s)$, derive $R(s,t)$ and $C(t)$ for $t$ a fresh individual.

- $\forall$-rule: Given $(\forall R.C)(s)$ and $R(s,t)$, derive $C(t)$.

- $\sqsubseteq$-rule: Given a GCI $C \sqsubseteq D$ and an individual $s$, derive $(\neg C \sqcup D)(s)$.

The $\sqcup$-rule is nondeterministic: if $(C_1 \sqcup C_2)(s)$ is true, then $C_1(s)$ or $C_2(s)$ or both are true. Therefore, tableau calculi make a nondeterministic guess and choose either $C_1$ or $C_2$; if one choice leads to a contradiction, the algorithm must backtrack and try the other choice. Thus, $\mathcal{K}$ is unsatisfiable only if all choices lead to a contradiction. We next discuss two sources of complexity inherent in the tableau derivation rules.

### 3.1.1 Or-Branching

Handing disjunctions through reasoning by case is often called *or-branching*. The $\sqsubseteq$-rule adds a disjunction for each GCI to each individual in an ABox and is thus a major source of or-branching and inefficiency (Horrocks, 2007). Consider, for example, the knowledge base $\mathcal{K}_1 = (\emptyset, \mathcal{T}_1, \mathcal{A}_1)$, with $\mathcal{T}_1$ and $\mathcal{A}_1$ specified as follows:

$$(2) \quad \begin{aligned} \mathcal{T}_1 &= \{\exists R.A \sqsubseteq A\} \\ \mathcal{A}_1 &= \{\neg A(a_0),\ R(a_0, b_1),\ R(b_1, a_1), \ldots,\ R(a_{n-1}, b_n),\ R(b_n, a_n),\ A(a_n)\} \end{aligned}$$

---

2. Some formalizations of tableau algorithms work on *completion graphs* (Horrocks & Sattler, 2007), which have a natural correspondence to ABoxes.

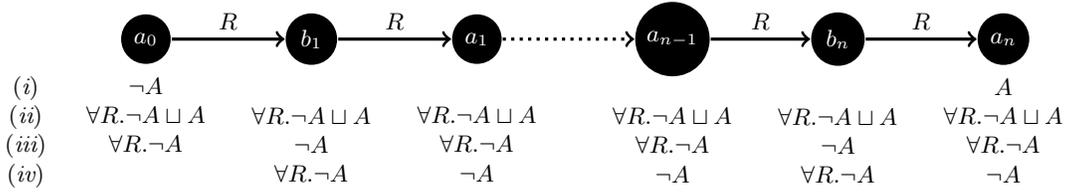

Figure 1: Or-Branching Example

The ABox $\mathcal{A}_1$ is graphically shown in Figure 1. The individuals occurring in the ABox are represented as black dots, an assertion of the form $A(a_0)$ is represented by placing $A$ next to the individual $a_0$, and an assertion of the form $R(a_0, b_1)$ is represented as an $R$-labeled arrow from $a_0$ to $b_1$. Initially, $\mathcal{A}_1$ contains only the concept assertions shown in line $(i)$.

To satisfy the GCI in $\mathcal{T}_1$, a tableau algorithm applies the $\sqsubseteq$-rule, thus adding the assertions shown in line $(ii)$ of Figure 1. Tableau algorithms are usually free to choose the order in which they process the assertions in an ABox; in fact, finding an order that exhibits good performance in practice requires advanced heuristics (Tsarkov & Horrocks, 2005b). Let us assume that the algorithm chooses to process the assertions on $a_i$ before those on $b_j$. Hence, by applying the derivation rules to all $a_i$, a tableau algorithm derives the assertions shown in line $(iii)$ of Figure 1; after that, by applying the derivation rules to all $b_i$, the algorithm derives the assertions shown in line $(iv)$ of Figure 1. The ABox now contains both $A(a_n)$ and $\neg A(a_n)$, which is a contradiction. Thus, the algorithm needs to backtrack its most recent choice, so it flips its guess on $b_{n-1}$ to $A(b_{n-1})$. This generates a contradiction on $b_{n-1}$, so the algorithm backtracks from all guesses for $b_i$, changes the guess on $a_n$ to $A(a_n)$, and repeats the work for all $b_i$. This also leads to a contradiction, so the algorithm must revise its guess for $a_{n-1}$; but then, two guesses are again possible for $a_n$. In general, after revising a guess for $a_i$, all possibilities for $a_j$, $i < j \leq n$, must be reexamined, which results in exponential behavior. None of the standard backtracking optimizations (Horrocks, 2007) are helpful: the problem arises because the order in which the individuals are processed makes the guesses on $a_i$ independent from the guesses on $a_j$ for $i \neq j$.

The GCI $\exists R.A \sqsubseteq A$, however, is not inherently nondeterministic: it is equivalent to the Horn clause $\forall x, y : [R(x, y) \land A(y) \rightarrow A(x)]$, which can be applied bottom-up to derive the assertions $A(b_n), A(a_{n-1}), \ldots, A(a_0)$ and eventually reveal a contradiction on $a_0$. These inferences are deterministic,[3] so we can conclude that $\mathcal{K}_1$ is unsatisfiable without any backtracking. This example suggests that the processing of GCIs in tableau algorithms can be "unnecessarily" nondeterministic. Hustadt et al. (2005) have identified a class of knowledge bases without "unnecessary" nondeterminism: knowledge bases expressed in the description logic Horn-$\mathcal{SHIQ}$ can always be translated into Horn clauses, suggesting that reasoning without any nondeterminism is possible in principle. Ideally, a practical DL reasoning procedure should exhibit no nondeterminism on Horn knowledge bases.

---

3. More precisely, each inference is deterministic, but the order in which the inferences are performed is don't-care nondeterministic.

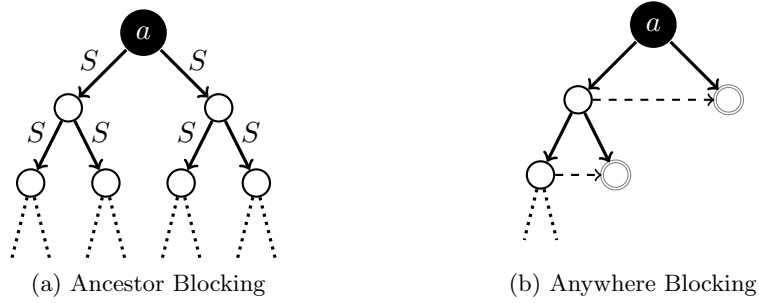

(a) Ancestor Blocking        (b) Anywhere Blocking

Figure 2: And-Branching Example

In the context of tableau calculi, various *absorption* optimizations (Horrocks, 2007) have been developed to control the nondeterminism arising in the application of GCIs. We discuss these optimizations in depth in Section 3.3.1.

### 3.1.2 And-Branching

The introduction of new individuals in the $\exists$-rule is often called *and-branching*, and it is another major source of inefficiency in tableau algorithms (Donini, 2007). Consider, for example, the (satisfiable) knowledge base $\mathcal{K}_2 = (\emptyset, \mathcal{T}_2, \mathcal{A}_2)$, with $\mathcal{T}_2$ and $\mathcal{A}_2$ specified as follows (where $n$ and $m$ are integers):

$$
\begin{aligned}
\mathcal{T}_2 \;=\; \{ \quad & A_1 \sqsubseteq \geq 2\,S.A_2, \; \ldots, \; A_{n-1} \sqsubseteq \geq 2\,S.A_n, \; A_n \sqsubseteq A_1, \\
& A_i \sqsubseteq (B_1 \sqcup C_1) \sqcap \ldots \sqcap (B_m \sqcup C_m) \text{ for } 1 \leq i \leq n \;\} \\
\text{(3)} \qquad \mathcal{A}_2 \;=\; \{ \quad & A_1(a) \;\}
\end{aligned}
$$

At-least restrictions are dealt with in tableau algorithms by the $\geq$-rule, which is quite similar to the $\exists$-rule: from $(\geq n\,R.C)(s)$, the $\geq$-rule derives $R(s, t_i)$ and $C(t_i)$ for $1 \leq i \leq n$, and $t_i \not\approx t_j$ for $1 \leq i < j \leq n$. Thus, the assertion $A_1(a)$ implies the existence of at least two individuals in $\mathcal{A}_2$, which imply the existence of at least two individuals in $\mathcal{A}_3$, and so on. Given $\mathcal{K}_2$, a tableau algorithm thus constructs a binary tree, shown in Figure 2a, in which each individual is labeled with some $A_i$ and an element of $\Pi = \{B_1, C_1\} \times \ldots \times \{B_m, C_m\}$. All individuals in the tree at depth $n$ are instances of $A_n$; because of the GCI $A_n \sqsubseteq A_1$, these individuals must be instances of $A_1$ as well, so we can repeat the whole construction and generate an even deeper tree. Clearly, a naïve application of the tableau rules does not terminate if the TBox contains existential quantifiers in cycles.

To ensure termination is such cases, tableau algorithms employ *blocking* (Baader & Nutt, 2007), which is based on an important observation about the shape of ABoxes that can be derived from some input ABox $\mathcal{A}$. The individuals in $\mathcal{A}$ are called *named* (shown as black circles), and they can be connected by role assertions in an arbitrary way. The individuals introduced by the $\exists$- and $\geq$-rules are called *blockable* (shown as white circles). For example, if $\exists R.C(a)$ is expanded into $R(a, s)$ and $C(s)$, then $s$ is called a blockable individual and it is an *R-successor* of $a$. It is not difficult to see that, if the knowledge base does not contain nominals, no tableau derivation rule can connect $s$ with an arbitrary named individual: the individual $s$ can participate only in inferences that derive an assertion of the form $D(s)$ with $D$ a concept, create a new successor of $s$, connect $s$ to an existing predecessor or successor, or, in the presence of (local) reflexivity, connect $s$ to itself. Hence, each ABox $\mathcal{A}'$ obtained from $\mathcal{A}$ can be seen as a "forest" of the form shown in Figure 3: each named individual can be arbitrarily connected to other named individuals and to a tree of blockable successors. The *concept label* $\mathcal{L}_{\mathcal{A}}(s)$ is defined as the set of all concepts $C$ such that $C(s) \in \mathcal{A}$, and the *edge label* $\mathcal{L}_{\mathcal{A}}(s, s')$ as the set of all atomic roles such that $R(s, s') \in \mathcal{A}$.

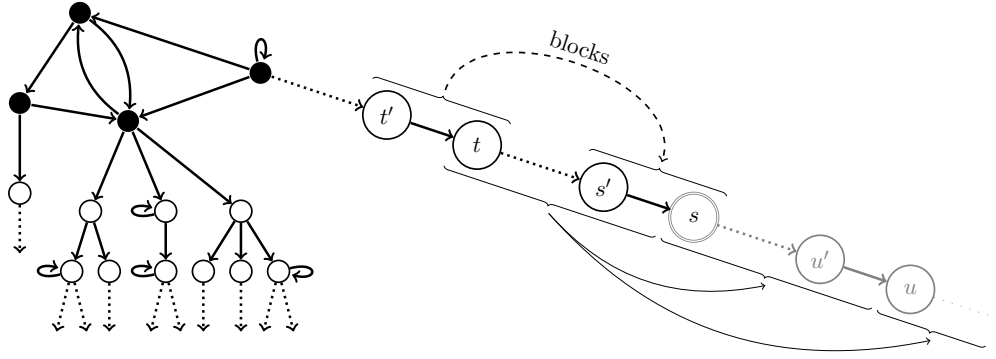

Figure 3: Forest-Like Shape of ABoxes

The forest-like structure of ABoxes enables blocking. Description logics such as $\mathcal{SHIQ}^+$ and $\mathcal{SHOIQ}^+$ allow for inverse roles and number restrictions, which has been handled in the literature by *ancestor pairwise blocking* (Horrocks et al., 2000b): for individuals $s$, $s'$, $t$, and $t'$ occurring in an ABox $\mathcal{A}$ as shown in Figure 3, $t$ blocks $s$ (shown by a double border on $s$) if and only if $\mathcal{L}_{\mathcal{A}}(s) = \mathcal{L}_{\mathcal{A}}(t)$, $\mathcal{L}_{\mathcal{A}}(s') = \mathcal{L}_{\mathcal{A}}(t')$, $\mathcal{L}_{\mathcal{A}}(s, s') = \mathcal{L}_{\mathcal{A}}(t, t')$, and $\mathcal{L}_{\mathcal{A}}(s', s) = \mathcal{L}_{\mathcal{A}}(t', t)$.[4] In tableau algorithms, the $\exists$- and $\geq$-rules are applicable only to nonblocked individuals, which ensures termination: the number of different concept and edge labels is exponential in $|\mathcal{K}|$, so an exponentially long branch in a forest-like ABox must contain a blocked individual, thus limiting the length of each branch in an ABox. Let $\mathcal{A}$ be an ABox as in Figure 3 to which no tableau derivation rule is applicable, and in which $s$ is blocked by $t$. We can construct a model from $\mathcal{A}$ by *unraveling*—that is, by replicating the fragment between $s$ and $t$ infinitely often. Intuitively, blocking ensures that the part of the ABox between $s$ and $s'$ "behaves" just like the part between $t$ and $t'$, so unraveling indeed generates a model. If our logic were able to connect blockable individuals in a non-tree-like way, then unraveling would not generate a model; in fact, the notion of ancestors, descendants, and blocking would itself be ill-defined.

Consider now an "unlucky" run of a tableau algorithm with ancestor pairwise blocking on $\mathcal{K}_2$. The number of elements in $\Pi$ is exponential in $|\mathcal{K}_2|$, so it can happen that blocking comes into effect only after the algorithm constructs an exponentially deep tree; since the tree is binary, it is doubly exponential in total. In a "lucky" run, the algorithm can always pick $B_j$ instead of $C_j$; then, the algorithm constructs a polynomially deep binary tree, so the tree is exponential in total. Thus, the and-branching caused by the ∃- and ≥-rules can lead to unnecessary generation of an ABox that is doubly exponential in the size of the input, which limits the scalability of tableau algorithms in practice.

---

4. Our blocking definition must include both edge labels in both directions because, unlike in some other tableau formalizations, our edge labels include only atomic roles.

## 3.2 The Hypertableau Algorithm at a Glance

In this section we present an informal overview of our hypertableau algorithm that addresses the problems due to or- and and-branching outlined in Section 3.1. We then formalize the algorithm in Section 4.

### 3.2.1 DERIVATION RULES

The hyperresolution calculus (Robinson, 1965) has often been used for first-order theorem proving. It works on *clauses*—implications of the form $\bigwedge_{i=1}^{n} U_i \to \bigvee_{j=1}^{m} V_j$ where $U_i$ and $V_j$ are first-order atoms. The conjunction $\bigwedge_{i=1}^{n} U_i$ is called the *antecedent*, and the disjunction $\bigvee_{j=1}^{m} V_j$ is called the *consequent*; we sometimes omit $\to$ if the antecedent is empty. For $\mathbf{D_i}$ a possibly empty disjunction of literals and $\sigma$ the most general unifier of $(A_1, B_1), \ldots, (A_m, B_m)$, the hyperresolution derivation rule is defined as follows (assuming that the unifier $\sigma$ exists):[5]

$$\frac{A_1 \vee \mathbf{D_1} \qquad \ldots \qquad A_m \vee \mathbf{D_m} \qquad B_1 \wedge \ldots \wedge B_m \to C_1 \vee \ldots \vee C_k}{\mathbf{D_1}\sigma \vee \ldots \vee \mathbf{D_m}\sigma \vee C_1\sigma \vee \ldots \vee C_k\sigma}$$

To make the calculus refutationally complete for first-order logic, one additionally needs a *factoring* derivation rule, which we do not discuss any further.

The hypertableau calculus (Baumgartner et al., 1996) is based on the observation that, if the literals in $C_1\sigma \vee \ldots \vee C_n\sigma$ do not share variables, we can replace the clause with a nondeterministically chosen atom $C_i\sigma$ that we assume to be true. If we assume that all clauses are safe (i.e., that each variable occurring in a clause also occurs in the clause's antecedent), then $A_i \vee \mathbf{D_i}$ and $C_1\sigma \vee \ldots \vee C_n\sigma$ are always ground, so they can always be nondeterministically split into atoms. Such a hypertableau inference is written as

$$\frac{A_1 \qquad \ldots \qquad A_m \qquad B_1 \wedge \ldots \wedge B_m \to C_1 \vee \ldots \vee C_k}{C_1\sigma \quad | \quad \ldots \quad | \quad C_k\sigma}$$

where $\sigma$ is the most general unifier of $(A_1, B_1), \ldots, (A_m, B_m)$ and | represents or-branching. On Horn clauses, each inference is deterministic,[6] and the calculus exhibits a "minimal" amount of don't-known nondeterminism on general clauses.

The hypertableau calculus by Baumgartner et al. (1996) can be easily applied to DLs: GCIs can be translated into first-order formulae (Borgida, 1996), which can then be converted into clauses, as shown in the following example.

$$A \sqsubseteq \exists R.B \quad \rightsquigarrow \quad \forall x : [A(x) \to \exists y : R(x,y) \wedge B(y)] \quad \rightsquigarrow \quad \begin{array}{l} A(x) \to B(f(x)) \\ A(x) \to R(x, f(x)) \end{array}$$

---

5. It is usual in resolution theorem proving to assume that the notation $A_i \vee \mathbf{D_i}$ does not imply that $A_i$ is the left-most disjunct in the disjunction, and we follow this convention.

6. As mentioned before, the order in which inferences are applied is nevertheless don't-care nondeterministic.

Let $\mathcal{A}$ be an ABox containing the assertions $A(a)$, $R(a, b)$, and $B(b)$. The GCI $A \sqsubseteq \exists R.B$ is clearly satisfied in $\mathcal{A}$, so there is no need to perform any inference. The clauses obtained by skolemization, however, are not satisfied in $\mathcal{A}$, so the hypertableau calculus derives $R(a, f(a))$ and $B(f(a))$. Hence, skolemization may make the calculus perform unnecessary inferences, which may be inefficient.

Therefore, instead of working with skolemized clauses, our calculus first preprocesses a $\mathcal{SHOIQ}^+$ knowledge base $\mathcal{K}$ into a pair $\Xi(\mathcal{K}) = (\Xi_{\mathcal{TR}}(\mathcal{K}), \Xi_{\mathcal{A}}(\mathcal{K}))$, where $\Xi_{\mathcal{A}}(\mathcal{K})$ is an ABox and $\Xi_{\mathcal{TR}}(\mathcal{K})$ is a set of *DL-clauses*—implications of the form $\bigwedge_{i=1}^{n} U_i \to \bigvee_{j=1}^{m} V_j$, where $U_i$ are of the form $R(x, y)$ or $A(x)$, and $V_j$ are of the form $R(x, y)$, $A(x)$, $\exists R.C(x)$, $\geq n\, R.C(x)$, or $x \approx y$. The preprocessing step is introduced formally in Section 4.1. The DL-clauses in $\Xi_{\mathcal{TR}}(\mathcal{K})$ are used in the *Hyp-rule*, which is inspired by the hypertableau derivation rule. For example, a GCI $\exists R.\neg A \sqsubseteq B$ is translated into a DL-clause $R(x, y) \to B(x) \lor A(y)$; then, if an ABox contains $R(a, b)$, the *Hyp*-rule derives either $B(a)$ or $A(b)$.

At-most restrictions are translated in our approach into DL-clauses containing equalities; for example, the axiom $A \sqsubseteq\, \leq 2\, R.B$ is translated into the DL-clause

$$A(x) \land R(x, y_1) \land B(y_1) \land R(x, y_2) \land B(y_2) \land R(x, y_3) \land B(y_3) \to y_1 \approx y_2 \lor y_1 \approx y_3 \lor y_2 \approx y_3.$$

While a concept of the form $\leq n\, R.B$ can be encoded using $O(\log n)$ bits, the corresponding DL-clause contains $O(n^2)$ literals; thus, our translation incurs an exponential blowup. We do not believe, however, this issue to be particular to our approach: tableau algorithms deal with at-most restrictions using a specialized $\leq$-rule whose application requires $O(n)$ space; thus, our translation merely makes the exponential space requirement explicit. Consequently, the (hyper)tableau algorithms are unlikely to be able to handle large numbers in number restrictions, and specialized algorithms, such as the one proposed by Faddoul, Farsinia, Haarslev, and Möller (2008), may be required.

Because of the translation described in the previous paragraph, the *Hyp*-rule can derive equalities of the form $s \approx t$. These are then dealt with using the $\approx$-rule: whenever $s \approx t \in \mathcal{A}$ and $s \neq t$, the $\approx$-rule replaces $s$ with $t$ or vice versa in all assertions in $\mathcal{A}$; this is usually called *merging*.

Apart from the *Hyp*- and the $\approx$-rule, our calculus contains the $\geq$-rule from the tableau calculus that deals with existential quantifiers, the $\bot$-rule that detects obvious contradictions (which can be of the form $s \not\approx s$, or $A(s)$ and $\neg A(s)$), and the *NI*-rule that ensures termination in the presence of nominals, number restrictions, and inverse roles. We discuss the *NI*-rule in more detail in Section 3.2.4.

The rules of the algorithm are formalized in Definition 7 on page 193 and Table 5 on page 196, and the reader may find it useful to briefly examine these definitions before continuing.

### 3.2.2 ANYWHERE PAIRWISE BLOCKING

We employ pairwise blocking from Section 3.1.2 to ensure termination of the calculus; to curb and-branching, however, we extend it to *anywhere pairwise blocking*. The key idea is to extend the set of potential blockers for $s$ beyond the ancestors of $s$. In doing so, we must avoid cyclic blocks: if $s$ is allowed to block $t$ and $t$ can block $s$, then neither $s$ nor $t$ is guaranteed to have all its successors constructed, which would render the calculus incomplete. Therefore, we parameterize our algorithm with a strict ordering $\prec$ on individuals that contains the ancestor relation. We allow $t$ to block $s$ only if, in addition to the conditions mentioned in Section 3.1.2, we have $t \prec s$. This version of blocking is formalized in Definition 7 on page 193. Note that, if $\prec$ coincides with the ancestor relation, then anywhere blocking becomes equivalent to ancestor blocking.

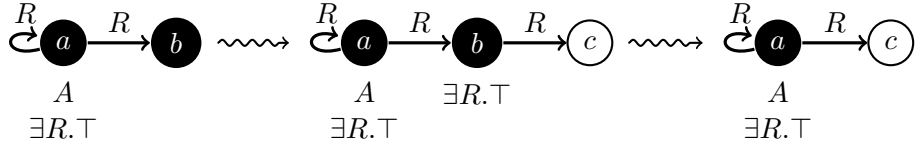

Figure 4: A Yo-Yo Example

Anywhere blocking can reduce and-branching in practice. Consider again the knowledge base $\mathcal{K}_2$ from Section 3.1.2. After we exhaust the exponentially many members of $\Pi$, all subsequently created individuals will be blocked. In the best case, we can always choose $B_j$ instead of $C_j$, so we create a polynomial path in the tree and then use the individuals from that path to block their siblings, as shown in Figure 2b. Hence, there is a derivation for $\mathcal{K}_2$ with anywhere blocking that can be constructed in polynomial time.

### 3.2.3 PROBLEMS DUE TO MERGING

Merging can easily lead to termination problems even for very simple DLs, as shown in the following example. For simplicity, we present the TBox as a set of DL-clauses $\mathcal{C}_3$.

(4)
$$\begin{aligned} \mathcal{A}_3 &= \{\, A(a), \quad \exists R.\top(a), \quad R(a,b), \quad R(a,a) \,\} \\ \mathcal{C}_3 &= \{\, R(x,y_1) \wedge R(x,y_2) \to y_1 \approx y_2, \qquad A(x) \wedge R(x,y) \to \exists R.\top(y) \,\} \end{aligned}$$

Consider now the derivation in our calculus on $\mathcal{A}_3$ and $\mathcal{C}_3$ illustrated in Figure 4: by the second DL-clause, the *Hyp*-rule derives $\exists R.\top(b)$, which the $\exists$-rule expands to $R(b,c)$; then, by the first DL-clause, the *Hyp*-rule derives $b \approx a$, so the $\approx$-rule merges $b$ into $a$. Clearly, the resulting ABox is isomorphic to the original one (that $c$ is a blockable and $b$ a named individual is not relevant here), so we can repeat the same sequence of inferences, which leads to nontermination. To the best of our knowledge, this problem was first identified by Baader and Sattler (2001), and it is commonly known as a "yo-yo."

This problem arises because, due to merging, $a$ can have an unbounded number of blockable $R$-successors: the blockable individual $c$ is created as an $R$-successor of $b$, but merging $b$ into $a$ makes $c$ a blockable $R$-successor of $a$. This, in turn, allows us to apply the DL-clauses from $\mathcal{C}_3$ to $a$ an arbitrary number of times, which leads to nontermination.

This problem can be solved by always merging a descendant $s$ into its ancestor $t$, and *pruning* $s$ before merging—that is, by removing all assertions containing a blockable descendant of $s$ and thus ensuring that $t$ does not "inherit" new successors.[7] Pruning is formally defined in Definition 7 on page 193.

---

7. Horrocks et al. (2000b) do not physically remove successors, but mark them as "not present" by setting the relevant edge labels to $\emptyset$. This has exactly the same effect as pruning.

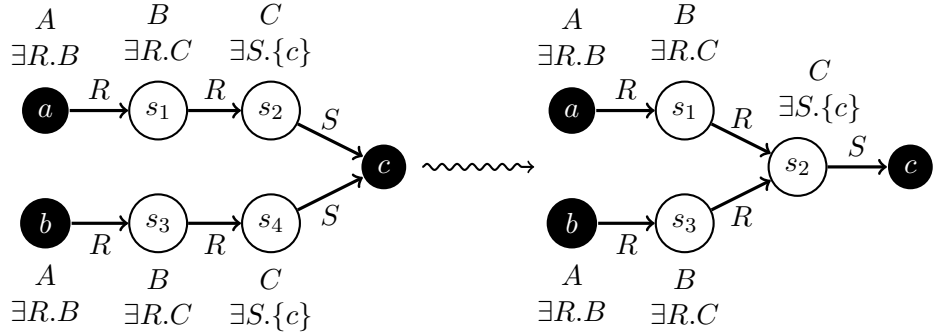

Figure 5: Non-Tree-Like Structures Due to Merging

Thus, before merging $b$ into $a$ in our example, we prune $b$—that is, we remove the assertion $R(b, c)$. Merging then produces an ABox that represents a model of $\mathcal{A}_3$ and $\mathcal{C}_3$, so the algorithm terminates. Note that pruning is well-defined only because our ABoxes are forest-shaped, cf. Figure 3: if connections between individuals were arbitrary and, in particular, cyclic, it would not be clear which part of the ABox should be pruned.

### 3.2.4 NOMINALS

With nominals, it is possible to derive ABoxes that are not forest-like, as the following simple example demonstrates. For presentation purposes, we use the concept $\exists R.\{c\}$ in the DL-clauses even though such concepts would be further decomposed in our algorithm.

$$(5) \quad \begin{aligned} \mathcal{A}_4 &= \{ A(a), \quad A(b) \} \\ \mathcal{C}_4 &= \{ A(x) \to (\exists R.B)(x), \; B(x) \to (\exists R.C)(x), \; C(x) \to (\exists S.\{c\})(x) \} \end{aligned}$$

Successive applications of the *Hyp*- and $\exists$-rules to $\mathcal{A}_4$ and $\mathcal{C}_4$ can produce the ABox $\mathcal{A}_4^1$ shown on the left-hand side of Figure 5. This ABox is clearly not forest-shaped: the two paths of role atoms in $\mathcal{A}_4^1$ start at the named individuals $a$ and $b$ and end in a named individual $c$. Nevertheless, if role relations between blockable individuals remain forest-like, termination of the derivation can be ensured using blocking. Some DLs that include nominals produce only such *extended forest-like* ABoxes (Horrocks & Sattler, 2001).

If a DL includes inverse roles, number restrictions, and nominals, the shape of an ABox becomes much more involved. To this end, assume now that we extend $\mathcal{C}_4$ with the DL-clause $S(y_1, x) \land S(y_2, x) \to y_1 \approx y_2$ (which axiomatizes $S$ to be inverse-functional and effectively introduces number restrictions). On $\mathcal{A}_4^1$, the *Hyp*-rule then derives $s_2 \approx s_4$. Note that both $s_2$ and $s_4$ are blockable individuals; furthermore, neither individual is an ancestor of the other, so we can merge, say, $s_4$ into $s_2$. This produces the ABox $\mathcal{A}_4^2$ shown on the right-hand side of Figure 5, in which the assertion $R(s_3, s_2)$ makes $\mathcal{A}_4^2$ not forest-shaped. By extending the example, it is possible to use nominals, inverse roles, and number restrictions to arrange blockable individuals in cycles. The derived ABoxes are thus not forest-shaped, which makes defining suitable notions of pruning and unraveling difficult and prevents us from using blocking to ensure termination of the calculus.

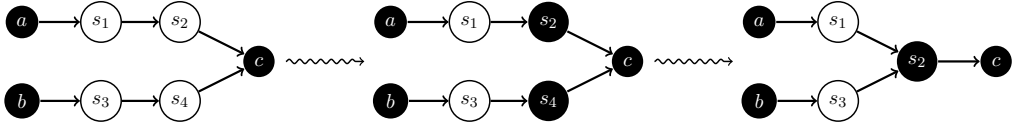

Figure 6: The Introduction of Root Individuals

To solve this problem, we need to extend the arbitrarily interconnected part of $\mathcal{A}_4^2$ by changing the status of $s_2$ from a blockable into a *root individual*—that is, an individual similar to the named ones in that it can be arbitrarily interconnected. Our extended forest-like ABoxes thus consist of a set of arbitrarily interconnected root individuals each of which can be the root of a "tree" (ignoring reflexive connections and connections back to root individuals) that otherwise consists entirely of blockable individuals (see Figure 3 on page 172). Named individuals are just the subset of the root individuals that occur in the input ABox. When we talk about individuals, we mean either root or blockable ones (see Definition 7 on page 193 for a formal definition).

Returning to our example, after changing the status of $s_2$ from a blockable into a root individual, only $s_1$ and $s_3$ are blockable in $\mathcal{A}_4^2$, so the ABox has the extended forest-like shape and we can apply blocking and pruning as usual. This is schematically shown in Figure 6. More generally, we apply the following preliminary version of the *NI-rule*, which we denote with (*) for easier reference:

> We change $s$ into a root individual whenever $\mathcal{A}$ contains assertions $R(s, a)$ and $A(s)$ where $a$ is a root or a named individual, $s$ is a blockable individual that is not a successor of $a$, and $a$ must satisfy an at-most restriction $\leq n\, R^-.A$.

Note that, if $s$ is a successor of $a$, then the part of the ABox involving $s$ and $a$ is forest-shaped, so the *NI*-rule need not be applicable.

This solution, however, introduces another problem: the number of root individuals can now grow arbitrarily, as shown in the following example.

$$
(6) \quad
\begin{aligned}
\mathcal{A}_5 &= \{\, A(b) \,\} \\
\mathcal{C}_5 &= \left\{
\begin{array}{l}
A(x) \to (\exists R.A)(x), \qquad A(x) \to (\exists S.\{a\})(x), \\
S(y_1, x) \land S(y_2, x) \land S(y_3, x) \to y_1 \approx y_2 \lor y_2 \approx y_3 \lor y_1 \approx y_3
\end{array}
\right\}
\end{aligned}
$$

On $\mathcal{A}_5$ and $\mathcal{C}_5$, our calculus can produce the ABox $\mathcal{A}_5^1$ shown on the left-hand side of Figure 7. ABox $\mathcal{A}_5^1$ does not explicitly contain at-most restriction concepts, so the precondition of (*) cannot be checked directly; we shall discuss this issue shortly. For the moment, however, please note that the last DL-clause in $\mathcal{C}_5$ corresponds to the axiom $\top \sqsubseteq\, \leq 2\, S^-.\top$, so individuals $c$ and $d$ can be seen as satisfying the precondition of (*); therefore, we change them into root individuals. Furthermore, the third DL-clause from $\mathcal{C}_5$ is not satisfied, so the *Hyp*-rule derives $c \approx b$, and the $\approx$-rule can merge $c$ into $b$. Since $d$ is now not a blockable individual, we cannot prune it, so we obtain the ABox $\mathcal{A}_5^2$ shown in the middle of Figure 7.[8]

---

8. To reduce clutter, we do not repeat the labels of individuals.

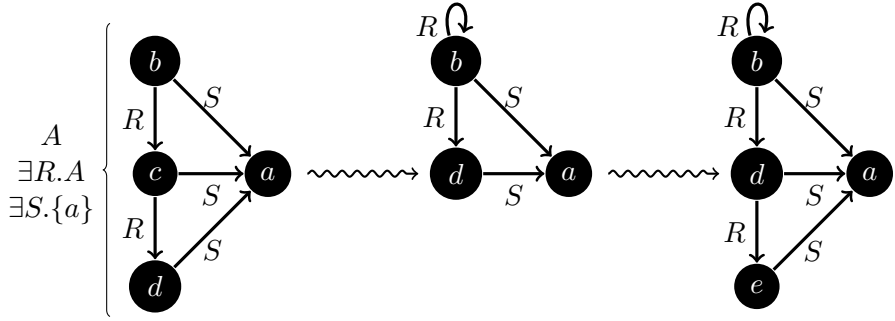

Figure 7: A Yo-Yo With Root Individuals

Since $\exists R.A(d)$ is not satisfied, we can extend $\mathcal{A}_5^2$ with $R(d,e)$, $A(e)$, $\exists R.A(e)$, $\exists S.\{a\}(e)$, and $S(e,a)$ to produce the ABox $\mathcal{A}_5^3$ shown on the right-hand side of Figure 7. Individual $e$ can be seen as satisfying the precondition of (*), so it is changed into a root individual. This ABox is isomorphic to $\mathcal{A}_5^1$, so we can repeat the same inferences forever.

We solve this problem with an *NI-rule* that refines (*). Assume that $\mathcal{A}$ contains an individual $s$ that satisfies the precondition of (*)—that is, $\mathcal{A}$ contains assertions $R(s,a)$ and $A(s)$, where $a$ is a root or a named individual, $s$ is a blockable individual that is not a successor of $a$, and $a$ must satisfy an at-most restriction $\leq n\, R^-.A$. In any model of $\mathcal{A}$, there can be at most $n$ different individuals $b_i$ that participate in assertions of the form $R(b_i,a)$ and $A(b_i)$. Hence, we associate with $a$ a set of $n$ fresh root individuals $\{b_1,\dots,b_n\}$ that represent the $R^-$-neighbors of $a$. We turn $s$ into a root individual by nondeterministically choosing $b_j$ from this set and merging $s$ into $b_j$. In this way, the number of new root individuals that can be introduced as a result of the at-most restriction $\leq n\, R^-.A$ on $a$ is limited to $n$. The complete definition of the *NI-rule* is given in Table 5 on page 196. In the example from Figure 7, the *NI-rule* introduces at most two fresh root individuals. When the *NI-rule* is applied for the third time, instead of introducing $e$, one of the previously introduced root individuals is reused, which ensures termination of the calculus.

When formulating the *NI-rule*, we are faced with a technical problem: at-most restriction concepts are translated in our calculus into DL-clauses, which makes testing the condition from the previous paragraph difficult. For example, an application of the *Hyp*-rule to the third DL-clause in (6) (obtained from the axiom $\top \sqsubseteq\, \leq 2\, S^-.\top$) can produce an equality such as $c \approx b$; this equality alone does not reflect the fact that $a$ must satisfy the at-most restriction $\leq 2\, S^-.\top$. To enable the application of the *NI-rule*, we introduce *annotated equalities* in which the annotations establish an association with the at-most restriction. The third DL-clause from (6) is thus represented in our algorithm as follows:

$$
\begin{aligned}
(7) \qquad & S(y_1,x) \wedge S(y_2,x) \wedge S(y_3,x) \rightarrow \\
& y_1 \approx y_2\ @_{\leq 2\,S^-.\top}^{x} \vee y_2 \approx y_3\ @_{\leq 2\,S^-.\top}^{x} \vee y_1 \approx y_3\ @_{\leq 2\,S^-.\top}^{x}
\end{aligned}
$$

The *Hyp*-rule then derives $c \approx b\ @_{\leq 2\,S^-.\top}^{a}$, which has the same meaning as $c \approx b$; however, the annotation says that, since $a$ must satisfy the at-most restriction $\leq 2\, S^-.\top$, both $b$ and $c$ must also be merged with one of the (two) individuals reserved as $S^-$-neighbors of $a$.

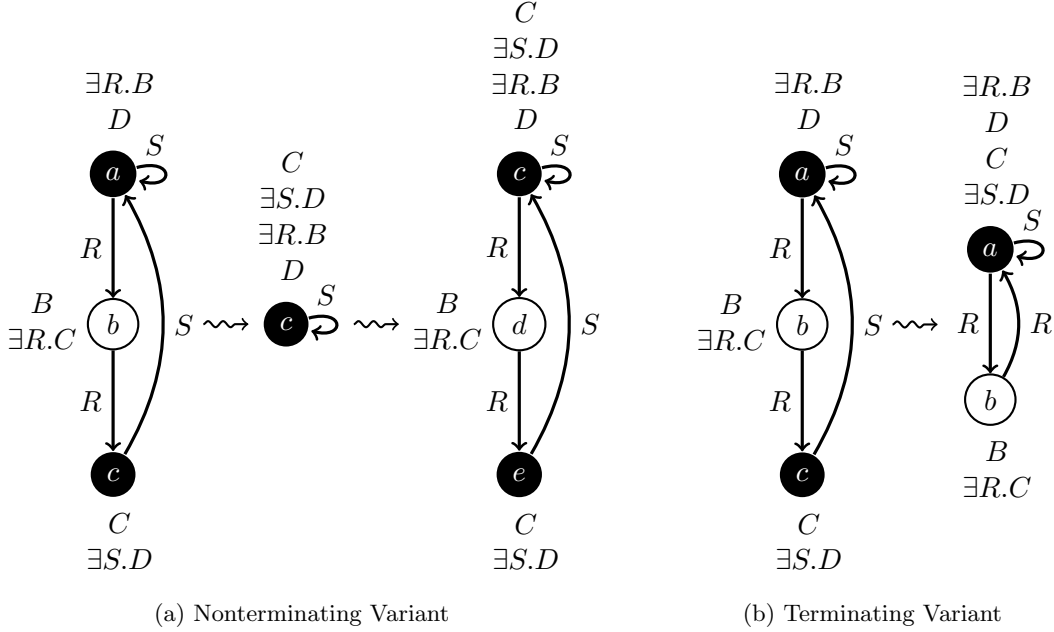

(a) Nonterminating Variant          (b) Terminating Variant

Figure 8: The "Caterpillar" Example

### 3.2.5 Nominals and Merging

The introduction of the *NI*-rule leads to another problem: repeated merging between root individuals can lead to nontermination in a "caterpillar" derivation. Consider, for example, an application of the hypertableau calculus to the following knowledge base:

$$(8) \quad \begin{aligned} \mathcal{A}_6 &= \big\{ \, S(a,a), \quad \exists R.B(a) \, \big\} \\ \mathcal{C}_6 &= \left\{ \begin{array}{ll} B(x) \to \exists R.C(x), & C(x) \to \exists S.D(x), \\ D(x) \to x \approx a, & S(y_1,x) \wedge S(y_2,x) \to y_1 \approx y_2 \, @^x_{\leq 1\,S^-.\top} \end{array} \right\} \end{aligned}$$

The ABox and the first DL-clause cause the introduction of two new blockable individuals $b$ and $c$; the next two DL-clauses connect $c$ with $a$ by the role $S$; the last DL-clause produces $c \approx c\,@^x_{\leq 1\,S^-.\top}$; and an application of the *NI*-rule to this assertion causes $c$ to become a root individual. The ABox $\mathcal{A}_6^1$ resulting from these inferences is shown in the left-hand side of Figure 8a. Since $S$ is inverse-functional, the individuals $a$ and $c$ must be merged. Because individual $c$ is a root, it is no longer a descendant of $a$, so we can choose to merge $a$ into $c$. The blockable individual $b$ is then pruned (in order to avoid the problems outlined in Section 3.2.3), and the resulting ABox is shown in the middle part of Figure 8a. The existential restriction $\exists R.B$ on $c$, however, is not satisfied, so a similar sequence of rule applications constructs the ABox $\mathcal{A}_6^2$ shown in the right-hand side of Figure 8a. This ABox is isomorphic to $\mathcal{A}_6^1$, so the same inferences can be repeated forever.

This problem can be intuitively explained by the following observation. The *NI*-rule introduces fresh root individuals as neighbors of an existing root individual; thus, each root individual in an ABox can be seen as a part of a "chain" showing which individual caused the introduction of which root individual. Each chain is initially anchored at a named individual: such individuals occur in the input ABox and are not introduced by the *NI*-rule. The length of a path of blockable individuals can be used to limit the length of the "chains" of root individuals. If we allow chain anchors to be removed from an ABox, then the chains remain limited in length in any given ABox; however, over the course of derivation, one end of the chain can be extended indefinitely as the other end is shortened.

We solve this problem by allowing named individuals to be merged only into other named individuals, as specified by the postcondition of the ≈-rule in Table 5 on page 196. This ensures that each chain of root individuals always remains anchored at a named individual. In our example, instead of merging $a$ into $c$, we merge $c$ into $a$, which results in the ABox shown in Figure 8b. No derivation rule is applicable to this ABox, so the algorithm terminates.

### 3.2.6 The *NI*-Rule and Unraveling

The *NI*-rule is required not only to ensure that ABoxes are forest shaped, but also to enable the application of blocking and unraveling. Consider, for example, the knowledge base shown in (9), in which we omit the annotations on equalities for the sake of clarity. Intuitively, the axioms of the knowledge base state that the individual $a$ can have no $R^-$-neighbors, and that there is an infinite chain of individuals each of which is an $S^-$-neighbor of $a$.

$$
\begin{aligned}
\mathcal{A}_7 &= \{\, A(a),\ (\exists R.B)(a),\, \} \\
(9) \quad \mathcal{C}_7 &= \left\{
\begin{array}{l}
A(x) \wedge R(y, x) \to \bot, \ \ B(x) \to (\exists R.B)(x), \ \ B(x) \to (\exists S.\{a\})(x), \\
R(y_1, x) \wedge R(y_2, x) \to y_1 \approx y_2, \\
S(y_1, x) \wedge S(y_2, x) \wedge S(y_3, x) \wedge S(y_4, x) \to \\
\quad y_1 \approx y_2 \vee y_1 \approx y_3 \vee y_1 \approx y_4 \vee y_2 \approx y_3 \vee y_2 \approx y_4 \vee y_3 \approx y_4,
\end{array}
\right\}
\end{aligned}
$$

Without the *NI*-rule, an application of our calculus to $\mathcal{A}_7$ and $\mathcal{C}_7$ might produce the ABox $\mathcal{A}_7^1$ shown in Figure 9a. The individual $d$ is blocked in $\mathcal{A}_7^1$ by the individual $c$, so the derivation terminates. Note that the last DL-clause from $\mathcal{C}_7$ (which corresponds to the axiom $\top \sqsubseteq\, \leq 3\,S^-.\top$) is satisfied: $a$ is the only individual in $\mathcal{A}_7^1$ that has $S^-$-neighbors and it has only two such neighbors. To construct a model from $\mathcal{A}_7^1$, we unravel the blocked parts of the ABox—that is, we construct an infinite path that extends past $d$ by "duplicating" the fragment of the model between $c$ and $d$ an infinite number of times. This, however, creates additional $S^-$-neighbors of $a$, which invalidates the last DL-clause from $\mathcal{C}_7$; thus, the unraveled ABox does not define a model of $\mathcal{A}_7$ and $\mathcal{C}_7$.

The *NI*-rule elegantly solves this problem. Since $a$ must satisfy an at-most restriction of the form $\leq 3\,S^-.\top$, as soon as $S(b, a)$, $S(c, a)$, and $S(d, a)$ are derived, the *NI*-rule is applied to turn $b$, $c$, and $d$ into root individuals. This corrects the problems with unraveling: root individuals do not become blocked, so we introduce another fresh blockable individual $e$. This individual is merged with another $S^-$-neighbor of $a$, producing an individual with two $R^-$-neighbors, as illustrated in Figure 9b. $R$ is inverse-functional, however, so the neighbors are merged. Merging continues until $b$ has been merged into $a$, causing $a$ to become its own

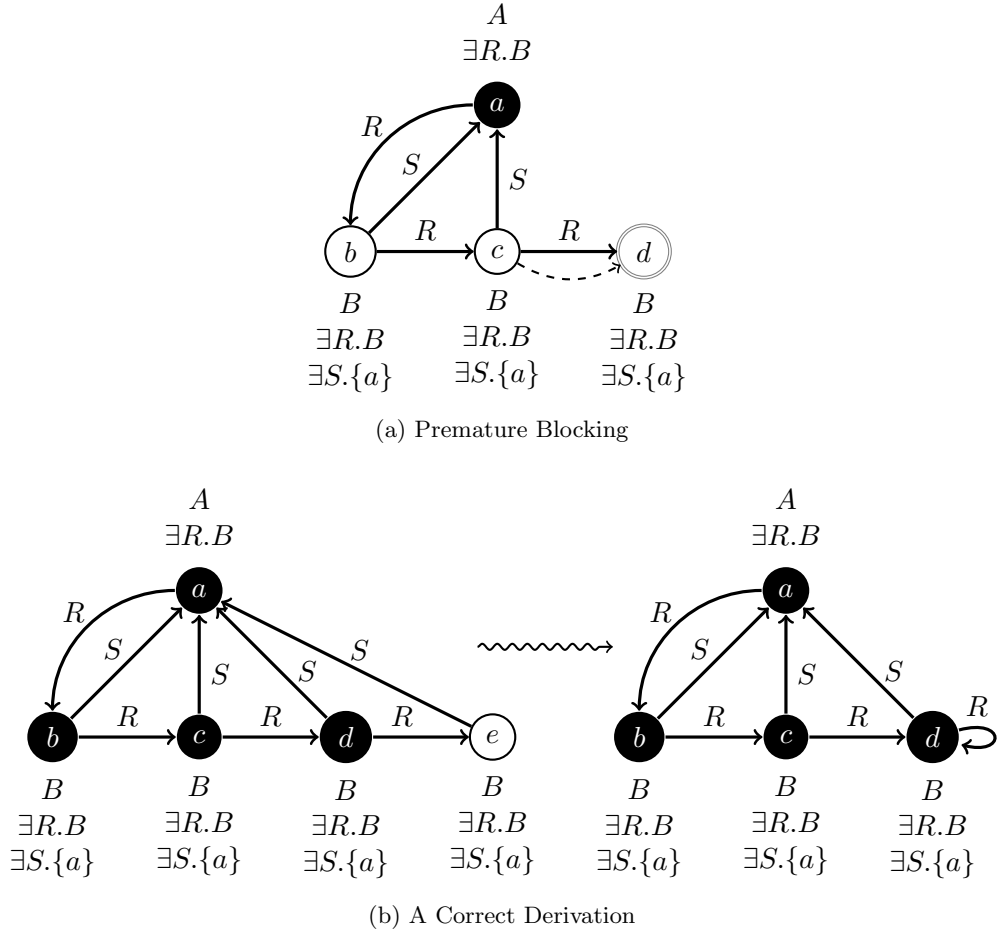

(a) Premature Blocking

(b) A Correct Derivation

Figure 9: The *NI*-rule and Unraveling

*R*-neighbor, at which point our algorithm correctly determines that the knowledge base represented by $\mathcal{A}_7$ and $\mathcal{C}_7$ is unsatisfiable.

## 3.3 Related Work

### 3.3.1 Hypertableau vs. Absorption

*Absorption* has been extensively used in tableau calculi to address the problems with or-branching outlined in Section 3.1.1 (Horrocks, 2007). The basic absorption algorithm tries to rewrite GCIs into the form $A \sqsubseteq C$ where $A$ is an atomic concept. After such preprocessing, instead of deriving $\neg A \sqcup C$ for each individual in an ABox, $C(s)$ is derived only if the ABox contains $A(s)$; thus, the nondeterminism introduced by the absorbed GCIs is localized. This basic technique has been refined and extended in several ways. *Negative absorption* rewrites GCIs into the form $\neg A \sqsubseteq C$ where $A$ is an atomic concept; then, $C(s)$ is derived only if an ABox contains $\neg A(s)$ (Horrocks, 2007). *Role absorption* rewrites GCIs into the form $\exists R.\top \sqsubseteq C$; then, $C(s)$ is derived only if an ABox contains $R(s,t)$ (Tsarkov & Horrocks, 2004). *Binary absorption* rewrites GCIs into the form $A_1 \sqcap A_2 \sqsubseteq C$; then, $C(s)$ is derived only if an ABox contains both $A_1(s)$ and $A_2(s)$ (Hudek & Weddell, 2006).

These techniques have proven indispensable in practice; however, our analysis shows potential for further improvement. For example, the axiom $\exists R.A \sqsubseteq A$ from (2) cannot be absorbed directly, and applying role absorption to (2) produces the axiom $\exists R.\top \sqsubseteq A \sqcup \forall R.\neg A$ containing a disjunction in the consequent. Binary absorption is not directly applicable to (2) since the axiom does not contain two concepts on the left-hand side of $\sqsubseteq$, but the algorithm by Hudek and Weddell (2006) additionally transforms (2) into an absorbable axiom $A \sqsubseteq \forall R^-.A$. Consider, however, the following axiom:

$$(10) \qquad \top \sqsubseteq \forall R.\neg C \sqcup \forall S.D$$

The binary absorption algorithm can process the two disjuncts in (10) in two ways. If $\forall R.\neg C$ is processed before $\forall S.D$, then (10) is transformed into the axioms shown in (11), both of which can be applied deterministically in a tableau algorithm. If, however, $\forall S.D$ is processed before $\forall R.\neg C$, then (10) is transformed into the axioms shown in (12). The first axiom is absorbable, but the second is not, so a tableau algorithm will be nondeterministic.

$$(11) \qquad C \sqsubseteq \forall R^-.Q_1 \qquad\qquad Q_1 \sqsubseteq \forall S.D$$
$$(12) \qquad Q_2 \sqsubseteq \forall R.\neg C \qquad\qquad \top \sqsubseteq D \sqcup \forall S^-.Q_2$$

Heuristics are used in practice to find a "good" absorption (see, e.g., Wu & Haarslev, 2008), but there are no guarantees that the result will incur the "least" amount of nondeterminism; this is so even on Horn knowledge bases, for which reasoning without any nondeterminism is possible in principle (Hustadt et al., 2005). In contrast, our algorithm is guaranteed to preprocesses a Horn knowledge base into Horn DL-clauses that will always result in deterministic derivations. For example, (10) is transformed into a Horn DL-clause (13).

$$(13) \qquad R(x, y_1) \wedge C(y_1) \wedge S(x, y_2) \to D(y_2)$$

Even in the case of inherently nondeterministic knowledge bases, absorption can be further optimized. Consider axiom (14), which is translated into DL-clause (15):

$$(14) \qquad \top \sqsubseteq A \sqcup \forall R.B \sqcup \forall S.C$$
$$(15) \qquad R(x, y_1) \wedge S(x, y_2) \to A(x) \vee B(y_1) \vee C(y_2)$$

The binary absorption algorithm transforms (14) into the following axioms:

$$(16) \qquad Q_1 \sqcap Q_2 \sqsubseteq A$$
$$(17) \qquad \top \sqsubseteq B \sqcup \forall R^-.Q_1$$
$$(18) \qquad \top \sqsubseteq C \sqcup \forall S^-.Q_2$$

Axiom (16) is absorbable; however, (17) and (18) are not, so their application introduces a nondeterministic choice point for each individual occurring in an ABox. This problem can be ameliorated by using role absorption and transforming (17) and (18) into (19) and (20):

$$(19) \qquad \exists R^-.\top \sqsubseteq B \sqcup \forall R^-.Q_1$$

(20) $$\exists S^-.\top \sqsubseteq C \sqcup \forall S^-.Q_2$$

Now (19) can be used to derive $(B \sqcup \forall R^-.Q_1)(b)$ from $R(a, b)$, and (20) can be used to derive $(C \sqcup \forall S^-.Q_2)(d)$ from $S(c, d)$; however, these two disjunctions are derived even if $a \neq c$. In contrast, the DL-clause (15) derives a disjunction only if $a = c$; thus, literals $R(x, y_1)$ and $S(x, y_2)$ in (15) act as "guards." The presence of variables in the antecedent (the shared variable $x$ in this example) makes the guards more selective than if each guard were applied in isolation. Furthermore, if $a = c$, we derive a disjunction $A(a) \lor B(b) \lor C(d)$, which involves three different individuals ($a$, $b$, and $d$ in this case); in contrast, consequences of tableau algorithms typically involve just one individual. Thus, through the usage of variables, DL-clauses can be more global in their effect than tableau rules.

To the best of our knowledge, no known absorption technique can localize the effects of axioms with number restrictions, such as (21).

(21) $$\geq 2\, R.B \sqsubseteq A$$

In order to ensure that only instances of $B$ are counted, tableau algorithms need to include a *choose*-rule that, for each assertion $R(a, b)$, nondeterministically derives $B(b)$ or $\neg B(b)$. In the hypertableau setting, however, (21) is translated into the following DL-clause:

(22) $$R(x, y_1) \land R(x, y_2) \land B(y_1) \land B(y_2) \rightarrow A(x) \lor y_1 \approx y_2$$

No *choose*-rule is needed, as the DL-clause is simply applied to assertions of the form $R(a, b)$, $B(b)$, $R(a, c)$, and $B(c)$; furthermore, the conclusion is a tautology whenever $b = c$. The presence of "guard" atoms in the antecedent of (22) thus significantly reduces the nondeterminism introduced by such number restrictions. Furthermore, on Horn knowledge bases with number restrictions (which includes the common case of functional roles), our calculus exhibits no nondeterminism; in contrast, tableau calculi still need the *choose*-rule, which introduces nondeterminism even if all GCIs have been fully absorbed.

The hypertableau calculus as presented in this paper does not generalize negative absorption directly; for example, the negatively absorbed axiom (23) is translated into a DL-clause (24) which is then applied to all individuals in an ABox.

(23) $$\neg A \sqsubseteq B$$
(24) $$\rightarrow A(x) \lor B(x)$$

Negative absorption can, however, easily be applied in our setting: to negatively absorb an atomic concept $A$, we simply replace in the input ABox and the DL-clauses all occurrences of $A$ with $\neg A'$ where $A'$ is a fresh concept, and then move the literals involving $A'$ to the appropriate side of DL-clauses. In our example, (24) would be thus converted into (25), which can then be applied deterministically.

(25) $$A'(x) \rightarrow B(x)$$

Note that this will transform a DL-clause $A(x) \rightarrow B(x)$ into $\rightarrow A'(x) \lor B(a)$; however, a similar situation arises in tableau calculi, where applying negative absorption to $\neg A \sqsubseteq B$ means that $A \sqsubseteq B$ cannot be absorbed.

To summarize, unlike various absorption techniques that are guided primarily by heuristics, the hypertableau calculus provides a framework that captures all variants of absorption we are aware of, guarantees deterministic behavior whenever the input knowledge base is Horn, eliminates the need for the nondeterministic *choose*-rule, and allows for a more powerful use of "guard" atoms to further localize any remaining nondeterminism. Furthermore, in Section 4.1.3 we show that the our calculus provides a proof-theoretic framework for DLs that can uniformly handle certain useful extensions of $\mathcal{SHOIQ}^+$.

### 3.3.2 Relationship with Caching

Various *caching* optimizations can be used to reduce the sizes of the models constructed during knowledge base classification (Ding & Haarslev, 2006; Horrocks, 2007). In the proposed approaches, caching is used in parallel with blocking—that is, caching alone does not guarantee termination of the calculus, and caching must be carefully integrated with blocking in order not to affect soundness and/or completeness. This integration is particularly problematic in the presence of inverse roles. In contrast, anywhere blocking alone is sufficient to guarantee termination of the calculus. Furthermore, in Section 6.2 we present an optimization of anywhere blocking that can be seen as a very simple but effective form of general caching. Finally, as we discuss in Section 7, an efficient implementation of anywhere blocking can be obtained using very simple techniques. Thus, anywhere blocking achieves many of the effects of caching without much of the added complexity.

Donini and Massacci (2000) have used anywhere blocking with caching of unsatisfiable concepts to obtain a tableau algorithm for the DL $\mathcal{ALC}$ that runs in single exponential time. Goré and Nguyen (2007) have presented an algorithm for the DL $\mathcal{SHI}$ that also runs in exponential time and achieves termination solely by caching both satisfiable and unsatisfiable concepts. These algorithms, however, seem to be incompatible with all absorption variants, and the latter are essential for making tableau algorithms practical. Furthermore, it is unclear how to extend these algorithms to DLs that provide number restrictions, nominals, and inverse roles, such as $\mathcal{SHOIQ}^+$.

### 3.3.3 Relationship with First-Order Calculi

The original hypertableau calculus for first-order logic was subsequently extended with equality and has been implemented in the KRHyper theorem prover (Baumgartner, Furbach, & Pelzer, 2008). The calculus can be used for finite model generation, and it decides function-free clause logic.

Hyperresolution with splitting has been used to decide several description and modal logics (Georgieva et al., 2003; Hustadt & Schmidt, 1999). These approaches, however, rely on skolemization, which, as we have discussed previously, can be inefficient in practice. Furthermore, these approaches deal with logics that are much weaker than $\mathcal{SHOIQ}^+$; in particular, we are not aware of a hyperresolution-based decision procedure that can handle inverse roles, number restrictions, and nominals.

Our hypertableau calculus is related to the Extended Positive (EP) tableau calculus for first-order logic by Bry and Torge (1998). Instead of relying on skolemization, EP satisfies existential quantifiers by introducing new constants, and this is done in a way that makes the calculus complete for finite satisfiability. EP is, however, unlikely to be practical due to a high degree of nondeterminism. Furthermore, EP does not provide a decision procedure for DLs such as $\mathcal{SHOIQ}^+$ that do not enjoy the finite model property (Baader & Nutt, 2007). Consider, for example, the knowledge base whose TBox contains axioms (26) and (27), and whose ABox contains assertion (28):

$$(26) \qquad\qquad\qquad A \sqsubseteq \exists R.A$$

$$(27) \qquad\qquad\qquad \top \sqsubseteq\; \leq 1\, R^-.\top$$

$$(28) \qquad\qquad\qquad (\neg A \sqcap \exists R.A)(a)$$

EP will try to satisfy the existential quantifier on $a$ by "reusing" $a$—that is, by adding assertions $R(a,a)$ and $A(a)$. This leads to a contradiction, so EP will backtrack, introduce a fresh individual $b$, and add assertions $R(a,b)$ and $A(b)$; to satisfy (26), it will then also add $\exists R.A(b)$. To satisfy the existential quantifier in the latter assertion, EP will again try to "reuse" $a$; this will fail, so it will try to "reuse" $b$ by adding an assertion $R(b,b)$. Due to (27), however, $b$ will be merged into $a$, which results in a contradiction; therefore, EP will backtrack, introduce yet another fresh individual $c$ and add the assertions $R(b,c)$, $A(c)$, and $\exists R.A(c)$. By repeating the argument, it is easy to see that EP will generate ever larger models and will not terminate. This is unsurprising since the knowledge base is satisfied only in infinite models. To achieve termination on such knowledge bases, EP would need to be extended with blocking techniques such as the ones described in this paper.

Baumgartner and Schmidt (2006) developed a so-called *blocking* transformation of first-order clauses, which can improve the performance of bottom-up model generation methods. Roughly speaking, the clauses are modified in a way that makes a bottom-up calculus derive $s \approx t$ or $s \not\approx t$ for each term $s$ that is a subterm of $t$; then, an application of paramodulation to $s \approx t$ achieves an effect that is analogous to "reusing" $s$ instead of $t$ in the EP tableau calculus. This transformation, however, does not ensure termination for DLs that do not have the finite model property. For example, for the same reasons as explained in the previous paragraph, hyperresolution with splitting does not terminate on the clauses obtained by an application of the blocking transformation to (the clauses corresponding to) (26)–(28). Furthermore, even for DLs that enjoy the finite model property, an "unlucky" sequence of applications of derivation rules can prevent a bottom-up model generation method with blocking from terminating (please refer to Section 3.2.3 for more details).

## 4. The Satisfiability Checking Algorithm

We now present the hypertableau algorithm that can be used to check the satisfiability of a $\mathcal{SHOIQ}^+$ knowledge base $\mathcal{K}$. Our algorithm consists of two phases: the *preprocessing* phase is described in Section 4.1, and the *hypertableau* phase is described in Section 4.2.

### 4.1 Preprocessing

The goal of the preprocessing phase is to transform a $\mathcal{SHOIQ}^+$ knowledge base $\mathcal{K}$ into an ABox $\Xi_{\mathcal{A}}(\mathcal{K})$ and a set of *DL-clauses* $\Xi_{\mathcal{TR}}(\mathcal{K})$ that are equisatisfiable with $\mathcal{K}$.

**Definition 1** (DL-Clause). *The concepts* $\top$, $\bot$, *and concepts of the form* $A$ *and* $\neg A$ *for* $A$ *an atomic concept are called* literal concepts. *Let* $N_V$ *be a set of* variables *disjoint from the* set of individuals $N_I$. An atom *is an expression of the form* $B(s)$, $\ge n\,S.B(s)$, $R(s,t)$, *or* $s \approx t$, *for $s$ and $t$ individuals or variables, $B$ a literal concept, $R$ an atomic role, $S$ a (not necessarily atomic) role, and $n$ a positive integer. A* DL-clause *is an expression of the form*

$$U_1 \wedge \ldots \wedge U_m \rightarrow V_1 \vee \ldots \vee V_n$$

Table 2: Satisfaction of DL-Clauses in an Interpretation

| | | |
|---|---|---|
| $I, \mu \models C(s)$ | iff | $s^{I,\mu} \in C^I$ |
| $I, \mu \models R(s,t)$ | iff | $\langle s^{I,\mu}, t^{I,\mu} \rangle \in R^I$ |
| $I, \mu \models s \approx t$ | iff | $s^{I,\mu} = t^{I,\mu}$ |
| $I, \mu \models \bigwedge_{i=1}^{m} U_i \rightarrow \bigvee_{j=1}^{n} V_j$ | iff | $I, \mu \models U_i$ for each $1 \le i \le m$ implies |
| | | $I, \mu \models V_j$ for some $1 \le j \le n$ |
| $I \models \bigwedge_{i=1}^{m} U_i \rightarrow \bigvee_{j=1}^{n} V_j$ | iff | $I, \mu \models \bigwedge_{i=1}^{m} U_i \rightarrow \bigvee_{j=1}^{n} V_j$ for all mappings $\mu$ |
| $I \models \mathcal{C}$ | iff | $I \models r$ for each DL-clause $r \in \mathcal{C}$ |

*where $U_i$ and $V_j$ are atoms, $m \ge 0$, and $n \ge 0$. The conjunction $U_1 \wedge \ldots \wedge U_m$ is called the* antecedent, *and the disjunction $V_1 \vee \ldots \vee V_n$ is called the* consequent. *The empty antecedent and the empty consequent of a DL-clause are written as $\top$ and $\bot$, respectively.*

Let $I = (\triangle^I, \cdot^I)$ *be an interpretation and* $\mu : N_V \rightarrow \triangle^I$ *a mapping of variables to elements of the interpretation domain. Let $a^{I,\mu} = a^I$ for an individual $a$ and $x^{I,\mu} = \mu(x)$ for a variable $x$. Satisfaction of an atom, DL-clause, and a set of DL-clauses $\mathcal{C}$ in $I$ and $\mu$ is defined in Table 2.*

### 4.1.1 ELIMINATION OF TRANSITIVITY AXIOMS

Transitivity axioms are handled in tableau algorithms by the $\forall_+$-rule: if $R$ is transitive and an ABox contains $\forall R.C(s)$ and $R(s,t)$, the $\forall_+$-rule derives $\forall R.C(t)$. In our algorithm, however, concepts of the form $\forall R.C$ are translated into DL-clauses, so the $\forall_+$-rule cannot be applied. Therefore, instead of handling transitivity directly, we encode a $\mathcal{SHOIQ}^+$ knowledge base $\mathcal{K}$ into an equisatisfiable $\mathcal{ALCHOIQ}^+$ knowledge base $\Omega(\mathcal{K})$. This encoding eliminates all transitivity axioms, but simulates their effects using additional GCIs.

**Definition 2.** *Given a $\mathcal{SHOIQ}^+$ knowledge base $\mathcal{K} = (\mathcal{R}, \mathcal{T}, \mathcal{A})$, the* concept closure *of $\mathcal{K}$ is the smallest set of concepts* clos($\mathcal{K}$) *such that*

- *if $C \sqsubseteq D \in \mathcal{T}$, then* nnf($\neg C \sqcup D$) $\in$ clos($\mathcal{K}$),

- *if $C(a) \in \mathcal{A}$, then* nnf($C$) $\in$ clos($\mathcal{K}$),

- *if $C \in$ clos($\mathcal{K}$) and $D$ syntactically occurs in $C$, then $D \in$ clos($\mathcal{K}$),*

- *if $\le n\,R.C \in$ clos($\mathcal{K}$), then $\dot{\neg}C \in$ clos($\mathcal{K}$), and*

- *if $\forall R.C \in$ clos($\mathcal{K}$), $S \sqsubseteq_{\mathcal{R}}^* R$, and* Tra($S$) $\in \mathcal{R}$, *then $\forall S.C \in$ clos($\mathcal{K}$).*

*The $\Omega$-encoding of $\mathcal{K}$ is the $\mathcal{ALCHOIQ}^+$ knowledge base $\Omega(\mathcal{K}) = (\mathcal{R}', \mathcal{T}', \mathcal{A})$ where $\mathcal{R}'$ is obtained from $\mathcal{R}$ by removing all transitivity axioms and*

$$\mathcal{T}' = \mathcal{T} \cup \{\forall R.C \sqsubseteq \forall S.(\forall S.C) \mid \forall R.C \in \mathsf{clos}(\mathcal{K}), \ S \sqsubseteq_{\mathcal{R}}^* R, \ \text{and } \mathsf{Tra}(S) \in \mathcal{R}\}.$$

Similar encodings are known for various description (Tobies, 2001) and modal (Schmidt & Hustadt, 2003) logics. Note that, in order to guarantee decidability (Horrocks, Sattler, & Tobies, 2000a), number restrictions and local reflexivity are allowed in $\mathcal{SHOIQ}^+$ only on simple roles—that is, on roles not having transitive subroles; for similar reasons, role disjointness, irreflexivity, and asymmetry axioms are also allowed only on simple roles.

**Lemma 1.** *A $\mathcal{SHOIQ}^+$ knowledge base $\mathcal{K}$ is satisfiable if and only if $\Omega(\mathcal{K})$ is satisfiable, and $\Omega(\mathcal{K})$ can be computed in time polynomial in $|\mathcal{K}|$.*

The full proof of an analogous result for the DL $\mathcal{SHIQ}$ is given by Motik (2006) in Theorem 5.2.3, and the generalization of this result to $\mathcal{SHOIQ}^+$ is straightforward; therefore, we omit the proof of Lemma 1 for the sake of brevity. After the elimination of transitivity axioms, there is no distinction between simple and complex roles. Hence, in the rest of this paper we assume that all roles are simple unless otherwise stated and, without loss of generality, we treat $\exists R.B$ as a syntactic shortcut for $\geq 1\,R.B$.

### 4.1.2 Normalization

Before translation into a set of DL-clauses, a a knowledge base is first brought into a *normalized* form. This is done in order to make all negations explicit, and to ensure that the resulting DL-clauses are compatible with blocking.

To understand the first issue, consider the axiom $\neg A \sqsubseteq \neg(\exists R.\exists R.\exists R.B)$. Converting this axiom into DL-clauses is not straightforward because of the implicit negations; for example, the concept $A$ is seemingly negated but, due to the negation implicit in the implication, $A$ actually occurs positively in the axiom. Therefore, we replace this axiom with the following equivalent axiom. This makes all negations explicit, so the result can be easily translated into a DL-clause.

$$(29) \quad \top \sqsubseteq A \sqcup \forall R.\forall R.\forall R.\neg B \quad \rightsquigarrow \quad R(x, y_1) \wedge R(y_1, y_2) \wedge R(y_2, y_3) \wedge B(y_3) \rightarrow A(x)$$

To understand the second issue, consider the knowledge base $\mathcal{K}_8$, consisting of an ABox $\mathcal{A}_8$ and a TBox that corresponds to the set of DL-clauses $\mathcal{C}_8$.

$$(30) \quad \begin{aligned} \mathcal{A}_8 \ &= \{\ \neg A(a), \quad B(a)\ \} \\ \mathcal{C}_8 \ &= \{\ R(x, y_1) \wedge R(y_1, y_2) \wedge R(y_2, y_3) \wedge B(y_3) \rightarrow A(x), \quad B(x) \rightarrow \exists R.B(x)\ \} \end{aligned}$$

By applying the rules from Section 3.2, our algorithm constructs on $\mathcal{K}_8$ the ABox shown in Figure 10. According to the definition of blocking introduced in Definition 7,[9] $c$ is now blocked by $b$; furthermore, no rule is applicable to the ABox, so the algorithm terminates, leading us to believe that $\mathcal{K}_8$ is satisfiable. The ABox, however, does not represent a model of $\mathcal{K}_8$: if we expand $\exists R.B(c)$ into $R(c, d)$ and $B(d)$, by the first DL-clause in $\mathcal{C}_8$ we can derive

---

9. The version of blocking introduced in Definition 7 differs from the one presented in Section 3.1.2 in that the concept label $\mathcal{L}_{\mathcal{A}}(s)$ of an individual $s$ consists only of atomic concepts $A$ such that $A(s) \in \mathcal{A}$.

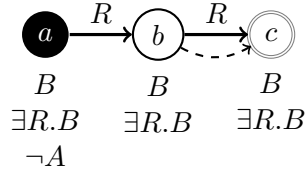

Figure 10: Incorrect Blocking due to Lack of Normalization

$A(a)$, which then contradicts $\neg A(a)$. This problem arises because the antecedent of the first DL-clause in $\mathcal{C}_8$ checks for a path of three $R$-successors, whereas the pairwise blocking condition ensures only that all paths of length two are fully constructed. Intuitively, the antecedents of each DL-clause should check for paths that "fit" into the fully constructed model fragments. We can ensure this by renaming complex concepts into simpler ones. Thus, we transform the culprit DL-clause into the following ones, which check only for paths of length one.

(31)   $\top \sqsubseteq A \sqcup \forall R.\neg Q_1$     $\leadsto$     $R(x,y) \wedge Q_1(y) \to A(x)$

(32)   $\top \sqsubseteq Q_1 \sqcup \forall R.\neg Q_2$     $\leadsto$     $R(x,y) \wedge Q_2(y) \to Q_1(x)$

(33)   $\top \sqsubseteq Q_2 \sqcup \forall R.\neg B$     $\leadsto$     $R(x,y) \wedge B(y) \to Q_2(x)$

The application of these DL-clauses to the ABox shown in Figure 10 would additionally derive $Q_2(a)$, $Q_2(b)$, and $Q_1(a)$, so $c$ would not be blocked. The calculus would then expand $\exists R.B(c)$ and discover a contradiction.

To formalize these ideas, we define a normalized form of DL knowledge bases.

**Definition 3** (Normalized Form). *A GCI is normalized if it is of the form $\top \sqsubseteq \bigsqcup_{i=1}^{n} C_i$, where each $C_i$ is of the form $B$, $\{a\}$, $\forall R.B$, $\exists R.\mathsf{Self}$, $\neg \exists R.\mathsf{Self}$, $\geq n\, R.B$, or $\leq n\, R.B$, for $B$ a literal concept, $R$ a role, and $n$ a nonnegative integer.*

*A TBox $\mathcal{T}$ is normalized if each GCI in it is normalized. An ABox $\mathcal{A}$ is normalized if each concept assertion in $\mathcal{A}$ contains only a literal concept, each role assertion in $\mathcal{A}$ contains only an atomic role, and $\mathcal{A}$ contains at least one assertion. An $\mathcal{ALCHOIQ}^+$ knowledge base $\mathcal{K} = (\mathcal{R}, \mathcal{T}, \mathcal{A})$ is normalized if $\mathcal{T}$ and $\mathcal{A}$ are normalized.*

The following transformation can be used to normalize a knowledge base.

**Definition 4** (Normalization). *For an $\mathcal{ALCHOIQ}^+$ knowledge base $\mathcal{K}$, the knowledge base $\Delta(\mathcal{K})$ is computed as shown in Table 3.*

Normalization can be seen as a variant of the well-known structural transformation (Plaisted & Greenbaum, 1986; Nonnengart & Weidenbach, 2001). An application of the structural transformation to (29) would replace each complex subconcept with a positive atomic concept, eventually producing $\top \sqsubseteq A \sqcup \forall R.Q_1$. This axiom cannot be translated into a Horn DL-clause, whereas (29) can; thus, the structural transformation can destroy

Table 3: The Functions Used in the Normalization

$$\Delta(\mathcal{K}) = \{\top(a)\} \cup \bigcup_{\alpha \in \mathcal{R} \cup \mathcal{A}} \Delta(\alpha) \cup \bigcup_{C_1 \sqsubseteq C_2 \in \mathcal{T}} \Delta(\top \sqsubseteq \mathsf{nnf}(\neg C_1 \sqcup C_2))$$

$$\Delta(\top \sqsubseteq \mathbf{C} \sqcup C') = \Delta(\top \sqsubseteq \mathbf{C} \sqcup \alpha_{C'}) \cup \bigcup_{1 \leq i \leq n} \Delta(\top \sqsubseteq \dot{\neg}\alpha_{C'} \sqcup C_i)$$

$$\text{for } C' \text{ of the form } C' = C_1 \sqcap \ldots \sqcap C_n \text{ and } n \geq 2$$

$$\Delta(\top \sqsubseteq \mathbf{C} \sqcup \forall R.D) = \Delta(\top \sqsubseteq \mathbf{C} \sqcup \forall R.\alpha_D) \cup \Delta(\top \sqsubseteq \dot{\neg}\alpha_D \sqcup D)$$

$$\Delta(\top \sqsubseteq \mathbf{C} \sqcup \geq n\,R.D) = \Delta(\top \sqsubseteq \mathbf{C} \sqcup \geq n\,R.\alpha_D) \cup \Delta(\top \sqsubseteq \dot{\neg}\alpha_D \sqcup D)$$

$$\Delta(\top \sqsubseteq \mathbf{C} \sqcup \leq n\,R.D) = \Delta(\top \sqsubseteq \mathbf{C} \sqcup \leq n\,R.\dot{\neg}\alpha_{\dot{\neg}D}) \cup \Delta(\top \sqsubseteq \dot{\neg}\alpha_{\dot{\neg}D} \sqcup \dot{\neg}D)$$

$$\Delta(\top \sqsubseteq \mathbf{C} \sqcup \neg\{s\}) = \begin{cases} \bot & \text{if } \mathbf{C} \text{ is empty,} \\ \Delta(\mathbf{C}(s)) & \text{otherwise.} \end{cases}$$

$$\Delta(D(s)) = \{\alpha_D(s)\} \cup \Delta(\top \sqsubseteq \dot{\neg}\alpha_D \sqcup \mathsf{nnf}(D))$$

$$\Delta(R^-(s,t)) = \{R(t,s)\}$$

$$\Delta(\beta) = \{\beta\} \text{ for any other axiom } \beta$$

$$\alpha_C = \begin{cases} Q_C & \text{if } \mathsf{pos}(C) = \mathsf{true} \\ \neg Q_C & \text{if } \mathsf{pos}(C) = \mathsf{false} \end{cases}, \text{ where } Q_C \text{ is a fresh atomic concept unique for } C$$

| | |
|---|---|
| $\mathsf{pos}(\top) = \mathsf{false}$ | $\mathsf{pos}(\bot) = \mathsf{false}$ |
| $\mathsf{pos}(A) = \mathsf{true}$ | $\mathsf{pos}(\neg A) = \mathsf{false}$ |
| $\mathsf{pos}(\{s\}) = \mathsf{true}$ | $\mathsf{pos}(\neg\{s\}) = \mathsf{false}$ |
| $\mathsf{pos}(\exists R.\mathsf{Self}) = \mathsf{true}$ | $\mathsf{pos}(\neg\exists R.\mathsf{Self}) = \mathsf{false}$ |
| $\mathsf{pos}(C_1 \sqcap C_2) = \mathsf{pos}(C_1) \vee \mathsf{pos}(C_2)$ | $\mathsf{pos}(C_1 \sqcup C_2) = \mathsf{pos}(C_1) \vee \mathsf{pos}(C_2)$ |
| $\mathsf{pos}(\forall R.C_1) = \mathsf{pos}(C_1)$ | |
| $\mathsf{pos}(\geq n\,R.C_1) = \mathsf{true}$ | $\mathsf{pos}(\leq n\,R.C_1) = \begin{cases} \mathsf{pos}(\dot{\neg}C_1) & \text{if } n = 0 \\ \mathsf{true} & \text{otherwise} \end{cases}$ |

**Note:** $A$ is an atomic concept, $C_{(i)}$ are arbitrary concepts, $\mathbf{C}$ is a possibly empty disjunction of arbitrary concepts, $D$ is not a literal concept, and $a$ is a fresh individual. Note that $\sqcup$ is commutative, so $C'$ in $\mathbf{C} \sqcup C'$ is not necessarily the right-most disjunct.

Horn-ness. To prevent this, we introduce the function $\mathsf{pos}(C)$ (c.f. Table 3) that returns $\mathsf{false}$ if the classification of $C$ does not require adding atoms into the consequent of a DL-clause. We then replace an occurrence of a concept $C$ in a concept $D$ with a negative literal concept $\neg Q_C$ if $\mathsf{pos}(C) = \mathsf{false}$, and with a positive literal concept $Q_C$ if $\mathsf{pos}(C) = \mathsf{true}$. Special care must be taken when replacing a concept $D$ in a concept $\leq n\,R.D$: since $D$ occurs in $\leq n\,R.D$ under an implicit negation, we replace $D$ with $\dot{\neg}\alpha_{\dot{\neg}D}$ in order to preserve Horn-ness. On a Horn knowledge base $\mathcal{K}$ (Hustadt et al., 2005), normalization performs the same replacements as the one presented by Hustadt et al., so $\Delta(\mathcal{K})$ is a Horn knowledge base as well.

**Lemma 2.** *The following properties hold for each $\mathcal{ALCHOIQ}^+$ knowledge base $\mathcal{K}$ and the corresponding knowledge base $\Delta(\mathcal{K})$:*

- *$\mathcal{K}$ is satisfiable if and only if $\Delta(\mathcal{K})$ is satisfiable;*

- *$\Delta(\mathcal{K})$ is normalized; and*

- $\Delta(\mathcal{K})$ can be computed in time polynomial in $|\mathcal{K}|$.

*Proof.* (*Sketch*) Since our transformation can be seen a syntactic variant of the structural transformation, the proof that $\mathcal{K}$ and $\Delta(\mathcal{K})$ are equisatisfiable is completely analogous to the ones by Plaisted and Greenbaum (1986) and Nonnengart and Weidenbach (2001), so we omit it for the sake of brevity. For the second claim, note that $\Delta$ essentially rewrites each GCI into a form $\top \sqsubseteq \bigsqcup_{i=1}^{n} C_i$ and then keeps replacing nested subconcepts of $C_i$ until the GCI becomes normalized; it adds $\top(a)$ to the ABox so that the ABox is not empty; and it replaces all inverse role assertions with equivalent assertions on the atomic roles. Thus, $\Delta(\mathcal{K})$ is normalized. Finally, each occurrence of a concept in $\mathcal{K}$ can be replaced with a new atomic concept at most once, and all necessary syntactic transformations can be performed in polynomial time, so $\Delta(\mathcal{K})$ can be computed in polynomial time. ☐

### 4.1.3 TRANSLATION INTO DL-CLAUSES

We now introduce the notion of HT-clauses—syntactically restricted DL-clauses on which our hypertableau calculus is guaranteed to terminate. In the rest of this paper, we often use the function $\mathsf{ar}$, which, given a role $R$ and variables or constants $s$ and $t$, returns an atom that is semantically equivalent to $R(s,t)$ but that contains an atomic role; that is,

$$\mathsf{ar}(R,s,t) = \begin{cases} R(s,t) & \text{if } R \text{ is an atomic role} \\ S(t,s) & \text{if } R \text{ is an inverse role and } R = S^- \end{cases}.$$

**Definition 5** (HT-Clause). *We assume that, for each individual $a$, the set of atomic concepts $N_C$ contains a unique* nominal guard concept *which we denote as $O_a$; furthermore, we assume that nominal guard concepts do not occur in any input knowledge base.*

*An* annotated equality *is an atom of the form $s \approx t \, @_{\leq n\,S.B}^{u}$, where $s$, $t$, and $u$ are constants or variables, $n$ is a nonnegative integer, $S$ is a role, and $B$ is a literal concept; the part $@_{\leq n\,S.B}^{u}$ of the atom is called the* annotation. *This atom is semantically equivalent to $s \approx t$.[10]*

*An* HT-clause *is a DL-clause $r$ of the following form, for $m \geq 0$ and $n \geq 0$:*

$$(34) \qquad\qquad U_1 \wedge \ldots \wedge U_m \to V_1 \vee \ldots \vee V_n$$

*Furthermore, it must be possible to separate the variables into a* center variable $x$, *a set of* branch variables $y_i$, *and a set of* nominal variables $z_j$ *such that the following properties hold, for $A$ an atomic concept, $B$ a literal concept not containing a nominal guard concept, $O_a$ a nominal guard concept, $R$ an atomic role, and $S$ a role.*

- *Each atom in the antecedent of $r$ is of the form $A(x)$, $R(x,x)$, $R(x,y_i)$, $R(y_i,x)$, $A(y_i)$, or $A(z_j)$.*

- *Each atom in the consequent of $r$ is of the form $B(x)$, $\geq h\,S.B(x)$, $B(y_i)$, $R(x,x)$, $R(x,y_i)$, $R(y_i,x)$, $R(x,z_j)$, $R(z_j,x)$, $x \approx z_j$, or $y_i \approx y_j \, @_{\leq h\,S.B}^{x}$.*

- *Each $y_i$ occurs in the antecedent of $r$ in an atom of the form $R(x,y_i)$ or $R(y_i,x)$.*

---

10. As explained in Section 3.2.4, annotations are only used to ensure termination of the hypertableau phase.

- *Each $z_j$ occurs in the antecedent of $r$ in an atom of the form $O_a(z_j)$.*

- *Each equality $y_i \approx y_j \, @^x_{\leq h\,S.A}$ in the consequent of $r$ occurs in a subclause of $r$ of the form (35) where $y^1, \ldots, y^{h+1}$ are branch variables such that no $y^k$ with $1 \leq k \leq h+1$ occurs elsewhere in $r$.*

$$(35) \qquad \ldots \bigwedge_{k=1}^{h+1} [\mathsf{ar}(S, x, y^k) \wedge A(y^k)] \ldots \rightarrow \ldots \bigvee_{1 \leq k < \ell \leq h+1} y^k \approx y^\ell \, @^x_{\leq h\,S.A} \ldots$$

- *Each equality $y_i \approx y_j \, @^x_{\leq h\,S.\neg A}$ in the consequent of $r$ occurs in a subclause of $r$ of the form (36) where $y^1, \ldots, y^{h+1}$ are branch variables such that no $y^k$ with $1 \leq k \leq h+1$ occurs elsewhere in $r$.*

$$(36) \qquad \ldots \bigwedge_{k=1}^{h+1} \mathsf{ar}(S, x, y^k) \ldots \rightarrow \ldots \bigvee_{k=1}^{h+1} A(y^k) \vee \bigvee_{1 \leq k < \ell \leq h+1} y^k \approx y^\ell \, @^x_{\leq h\,S.\neg A} \ldots$$

HT-clauses are more general than what is strictly needed to capture $\mathcal{ALCHOIQ}^+$ knowledge bases. For example, HT-clauses of the form $R(x, y) \wedge A(y) \rightarrow S(x, y)$ express a form of *relativized* role inclusions, and HT-clauses of the form $R(x, y) \wedge S(y, x) \rightarrow U(x, y) \vee T(y, x)$ capture *safe* role expressions (Tobies, 2001).

We now show how to transform a normalized $\mathcal{ALCHOIQ}^+$ knowledge base into a set of HT-clauses, after which we explain the need for nominal guard concepts.

**Definition 6** (Clausification). *The clausification of a normalized $\mathcal{ALCHOIQ}^+$ knowledge base $\mathcal{K} = (\mathcal{R}, \mathcal{T}, \mathcal{A})$ is the pair $\Xi(\mathcal{K}) = (\Xi_{\mathcal{TR}}(\mathcal{K}), \Xi_{\mathcal{A}}(\mathcal{K}))$ in which $\Xi_{\mathcal{TR}}(\mathcal{K})$ is a set of DL-clauses and $\Xi_{\mathcal{A}}(\mathcal{K})$ is an ABox, both obtained as shown in Table 4.*

By Definition 3, concepts of the form $\neg\{a\}$ are converted to ABox assertions during normalization, so Table 4 need not handle them. Positive nominal concepts are naturally translated into equalities containing constants; for example, $\top \sqsubseteq \neg A \sqcup \{a\}$ corresponds to $A(x) \rightarrow x \approx a$. Such DL-clauses are impractical: given an equality assertion $a \approx b$, the $\approx$-rule would need to replace all occurrences of $a$ with $b$ not only in the assertions, but in the DL-clauses as well; thus, the mentioned DL-clause should be replaced with $A(x) \rightarrow x \approx b$. To avoid the need for changing a set of DL-clauses in a derivation, we "extract" all constants into the ABox; for example, $\top \sqsubseteq \neg A \sqcup \{a\}$ is transformed into the DL-clause $A(x) \wedge O_a(z_{\{a\}}) \rightarrow x \approx z_{\{a\}}$ and the assertion $O_a(a)$. All constants are thus "pushed" into the assertions, so the $\approx$-rule can perform replacements only in the ABox.

**Lemma 3.** *Let $\mathcal{K}$ be a normalized $\mathcal{ALCHIQ}$ knowledge base. Then, $\mathcal{K}$ is equisatisfiable with $\Xi(\mathcal{K}) = (\Xi_{\mathcal{TR}}(\mathcal{K}), \Xi_{\mathcal{A}}(\mathcal{K}))$, and $\Xi_{\mathcal{TR}}(\mathcal{K})$ contains only HT-clauses.*

*Proof.* By inspecting Table 4, $\Xi_{\mathcal{TR}}(KB)$ clearly contains only HT-clauses. The following equivalences between DL concepts and first-order formulae are well known (Borgida, 1996):

$$\begin{aligned}
\forall R.B(x) &\equiv \forall y : \neg R(x, y) \vee B(y) \\
\leq n\,R.B(x) &\equiv \forall y_1, \ldots, y_{n+1} : \bigwedge_{1 \leq i \leq n+1} [R(x, y_i) \wedge B(y_i)] \rightarrow \bigvee_{1 \leq i < j \leq n+1} y_i \approx y_j \\
\{a\}(x) &\equiv x \approx a
\end{aligned}$$

Table 4: Translation of a Normalized Knowledge Base to HT-Clauses

$$\Xi_{\mathcal{T}}(\mathcal{T}) \;=\; \{\bigwedge_{i=1}^{n} \mathsf{lhs}(C_i) \to \bigvee_{i=1}^{n} \mathsf{rhs}(C_i) \;\mid\; \text{for each } \top \sqsubseteq \bigsqcup_{i=1}^{n} C_i \text{ in } \mathcal{T}\}$$

$$\Xi_{\mathcal{R}}(\mathcal{R}) \;=\; \{\mathsf{ar}(R,x,y) \to \mathsf{ar}(S,x,y) \;\mid\; \text{for each } R \sqsubseteq S \text{ in } \mathcal{R}\} \;\cup$$
$$\{\mathsf{ar}(S_1,x,y) \wedge \mathsf{ar}(S_2,x,y) \to \bot \;\mid\; \text{for each } \mathsf{Dis}(S_1,S_2) \in \mathcal{R}\} \;\cup$$
$$\{\top \to \mathsf{ar}(R,x,x) \;\mid\; \text{for each } \mathsf{Ref}(R) \in \mathcal{R}\} \;\cup$$
$$\{\mathsf{ar}(S,x,x) \to \bot \;\mid\; \text{for each } \mathsf{Irr}(S) \in \mathcal{R}\} \;\cup$$
$$\{\mathsf{ar}(R,x,y) \to \mathsf{ar}(R,y,x) \;\mid\; \text{for each } \mathsf{Sym}(R) \in \mathcal{R}\} \;\cup$$
$$\{\mathsf{ar}(S,x,y) \wedge \mathsf{ar}(S,y,x) \to \bot \;\mid\; \text{for each } \mathsf{Asy}(S) \in \mathcal{R}\}$$

$$\Xi_{\mathcal{T}\mathcal{R}}(\mathcal{K}) \;=\; \Xi_{\mathcal{T}}(\mathcal{T}) \cup \Xi_{\mathcal{R}}(\mathcal{R})$$

$$\Xi_{\mathcal{A}}(\mathcal{K}) \;=\; \mathcal{A} \cup \{O_a(a) \mid \text{for each } \{a\} \text{ occurring in } \mathcal{K}\}$$

**Note:** Whenever $\mathsf{lhs}(C_i)$ or $\mathsf{rhs}(C_i)$ is undefined, it is omitted in the HT-clause.

| $C$ | $\mathsf{lhs}(C)$ | $\mathsf{rhs}(C)$ |
|---|---|---|
| $A$ | | $A(x)$ |
| $\neg A$ | $A(x)$ | |
| $\{a\}$ | $O_a(z_C)$ | $x \approx z_C$ |
| $\geq n\,R.A$ | | $\geq n\,R.A(x)$ |
| $\geq n\,R.\neg A$ | | $\geq n\,R.\neg A(x)$ |
| $\exists R.\mathsf{Self}$ | | $\mathsf{ar}(R,x,x)$ |
| $\neg \exists R.\mathsf{Self}$ | $\mathsf{ar}(R,x,x)$ | |
| $\forall R.A$ | $\mathsf{ar}(R,x,y_C)$ | $A(y_C)$ |
| $\forall R.\neg A$ | $\mathsf{ar}(R,x,y_C) \wedge A(y_C)$ | |
| $\leq n\,R.A$ | $\bigwedge_{i=1}^{n+1}[\mathsf{ar}(R,x,y_C^i) \wedge A(y_C^i)]$ | $\bigvee_{1\le i<j\le n+1} y_C^i \approx y_C^j \, @_{\leq n\,R.A}^x$ |
| $\leq n\,R.\neg A$ | $\bigwedge_{i=1}^{n+1}\mathsf{ar}(R,x,y_C^i)$ | $\bigvee_{i=1}^{n+1} A(y_C^i) \vee \bigvee_{1\le i<j\le n+1} y_C^i \approx y_C^j \, @_{\leq n\,R.\neg A}^x$ |

**Note:** Each $y_C^{(i)}$ and $z_C$ is a fresh variable unique for $C$ (and $i$).

Let $\Xi'_{\mathcal{T}\mathcal{R}}(\mathcal{K})$ be the set of HT-clauses defined just like $\Xi_{\mathcal{T}\mathcal{R}}(\mathcal{K})$, but with the difference that $\mathsf{lhs}(\{a\}) = \top$ and $\mathsf{rhs}(\{a\}) = x \approx a$. Then, $(\Xi'_{\mathcal{T}\mathcal{R}}(\mathcal{K}), \Xi_{\mathcal{A}}(\mathcal{K}))$ is obtained from $\mathcal{K}$ by replacing concepts of the form $\forall R.B$, $\leq n\,R.B$ and $\{a\}$ with the equivalent first-order formulae, so $\mathcal{K}$ and $(\Xi'_{\mathcal{T}\mathcal{R}}(\mathcal{K}), \Xi_{\mathcal{A}}(\mathcal{K}))$ are clearly equisatisfiable. We now show that $(\Xi'_{\mathcal{T}\mathcal{R}}(\mathcal{K}), \Xi_{\mathcal{A}}(\mathcal{K}))$ is equisatisfiable with $(\Xi_{\mathcal{T}\mathcal{R}}(\mathcal{K}), \Xi_{\mathcal{A}}(\mathcal{K}))$.

($\Rightarrow$) Each model $I'$ of $(\Xi'_{\mathcal{T}\mathcal{R}}(\mathcal{K}), \Xi_{\mathcal{A}}(\mathcal{K}))$ is extended to a model $I$ of $(\Xi_{\mathcal{T}\mathcal{R}}(\mathcal{K}), \Xi_{\mathcal{A}}(\mathcal{K}))$ by setting $O_a^I = \{a^{I'}\}$ for each nominal guard concept $O_a$.

($\Leftarrow$) Each model $I$ of $\Xi(\mathcal{K})$ is a model of $(\Xi'_{\mathcal{TR}}(\mathcal{K}), \Xi_{\mathcal{A}}(\mathcal{K}))$: for each $\gamma \in \Xi'_{\mathcal{TR}}(\mathcal{K})$, we have $\delta \in \Xi_{\mathcal{TR}}(\mathcal{K})$ and $O_{a_k}(a_k) \in \Xi_{\mathcal{A}}(\mathcal{K})$, where $\gamma$ and $\delta$ are of the form shown below.

$$\gamma = \quad \bigwedge U_i \to \bigvee V_j \vee \bigvee_{k=1}^{n} x_k \approx a_k$$
$$\delta = \quad \bigwedge U_i \wedge \bigwedge_{k=1}^{n} O_{a_k}(z_{\{a_k\}}) \to \bigvee V_j \vee \bigvee_{k=1}^{n} x_k \approx z_{\{a_k\}}$$

Now if the disjunction $\bigvee_{k=1}^{n} x_k \approx a_k$ in some $\gamma$ were not true in $I$ for some values of $x_1, \ldots, x_n$, then clearly $\delta$ would not be true in $I$ for the same values of $x_1, \ldots, x_n$. $\qquad \square$

## 4.2 The Hypertableau Calculus for HT-Clauses

We now present the hypertableau calculus for deciding the satisfiability of an ABox $\mathcal{A}$ and a set of HT-clauses $\mathcal{C}$. As explained in Section 3, our algorithm uses several types of individuals. Each individual is either root or blockable as summarized next; when we refer simply to an individual, we mean either a root or a blockable one.

- *Root* individuals are those that either occur in the input ABox, or are introduced by the *NI*-rule. Their important characteristic is that they can be connected in arbitrary, and not just tree-like, ways.

    - Root individuals that occur in the input ABox are called *named* individuals.
    - Root individuals that are introduced by the *NI*-rule are defined as finite strings of the form $a.\gamma_1.\ldots.\gamma_n$ where $a$ is a named individual, each $\gamma_\ell$ is of the form $\langle R.B.i \rangle$, and $n \geq 0$. Root individuals introduced by applying the *NI*-rule to an assertion $s \approx t \, @^u_{\leq n \, R.B}$ are all of the form $u.\langle R.B.i \rangle$ with $1 \leq i \leq n$.

- *Blockable* individuals are introduced by the $\geq$-rule, and make up the tree-like parts of a model. The set of blockable individuals is disjoint from the set of root individuals. Blockable individuals are defined as finite strings of the form $s.i_1.i_2.\ldots.i_n$ where $s$ is a root individual, each $i_\ell$ is an integer, and $n \geq 1$. This string representation naturally induces the parent–child relationship between individuals; for example, $s.2$ is the second child of the individual $s$, which can be either blockable or root.

We now introduce our algorithm.

**Definition 7** (Hypertableau Algorithm).

   **Individuals.** *Given a set of* named *individuals* $N_I$, *the set of* root individuals $N_O$ *is the smallest set such that* $N_I \subseteq N_O$ *and, if* $x \in N_O$, *then* $x.\langle R, B, i \rangle \in N_O$ *for each role* $R$, *literal concept* $B$, *and positive integer* $i$. *The set of* all *individuals* $N_A$ *is the smallest set such that* $N_O \subseteq N_A$ *and, if* $x \in N_A$, *then* $x.i \in N_A$ *for each positive integer* $i$. *The individuals in* $N_A \setminus N_O$ *are blockable individuals. A blockable individual* $x.i$ *is a* successor *of* $x$, *and* $x$ *is a* predecessor *of* $x.i$. Descendant *and* ancestor *are the transitive closures of successor and predecessor, respectively.*

   **ABoxes.** *The hypertableau algorithm operates on ABoxes that are obtained by extending the standard definition from Section 2 as follows.*

- *In addition to assertions from Section 2, an ABox can contain annotated equality assertions and a special assertion* $\perp$ *that is false in all interpretations. Furthermore, assertions can refer to the individuals from* $N_A$ *and not only from* $N_I$.

- *Each (in)equality $s \approx t$ ($s \not\approx t$) also stands for the symmetric (in)equality $t \approx s$ ($t \not\approx s$). The same is true for annotated equalities.*

- *An ABox $\mathcal{A}$ can contain* renamings *of the form $a \mapsto b$ where $a$ and $b$ are root individuals. Let $\mapsto^*$ be the reflexive-transitive closure of $\mapsto$ in $\mathcal{A}$. An individual $b$ is the* canonical name *of a root individual $a$ in $\mathcal{A}$, written $b = \|a\|_{\mathcal{A}}$, if $b$ is the only individual such that $a \mapsto^* b$ and there exists no individual $c \neq b$ such that $b \mapsto^* c$; if no such individual exists, then $\|a\|_{\mathcal{A}} = a$.[11]*

*An* input ABox *is an ABox containing only named individuals, no annotated equalities, and no renamings, and in which all concepts are literal and all roles are atomic.*

*Satisfaction of such ABoxes in an interpretation is obtained by a straightforward generalization of the definitions in Section 2: all individuals are interpreted as elements of the interpretation domain $\triangle^I$, and $I \models a \mapsto b$ iff $a^I = b^I$.*

**Pairwise Anywhere Blocking.** *The* labels *of an individual $s$ and of an individual pair $\langle s, t \rangle$ in an ABox $\mathcal{A}$ are defined as follows:*

$$\mathcal{L}_{\mathcal{A}}(s) = \{ A \mid A(s) \in \mathcal{A} \text{ and } A \text{ is an atomic concept} \}$$
$$\mathcal{L}_{\mathcal{A}}(s, t) = \{ R \mid R(s, t) \in \mathcal{A} \}$$

*Let $\prec$ be a strict ordering (i.e., a transitive and irreflexive relation) on $N_A$ containing the ancestor relation—that is, if $s'$ is an ancestor of $s$, then $s' \prec s$. By induction on $\prec$, we assign to each individual $s$ in $\mathcal{A}$ a status as follows:*

- *a blockable individual $s$ is* directly blocked *by a blockable individual $t$ if and only if the following conditions are satisfied, for $s'$ and $t'$ the predecessors of $s$ and $t$, respectively:*

  - *$t$ is not blocked,*
  - *$t \prec s$,*
  - *$\mathcal{L}_{\mathcal{A}}(s) = \mathcal{L}_{\mathcal{A}}(t)$ and $\mathcal{L}_{\mathcal{A}}(s') = \mathcal{L}_{\mathcal{A}}(t')$, and*
  - *$\mathcal{L}_{\mathcal{A}}(s, s') = \mathcal{L}_{\mathcal{A}}(t, t')$ and $\mathcal{L}_{\mathcal{A}}(s', s) = \mathcal{L}_{\mathcal{A}}(t', t)$;*

- *$s$ is* indirectly blocked *iff it has a predecessor that is blocked; and*

- *$s$ is* blocked *iff it is either directly or indirectly blocked.*

**Pruning.** *The ABox $\mathsf{prune}_{\mathcal{A}}(s)$ is obtained from $\mathcal{A}$ by removing all assertions containing a descendant of $s$.*

**Merging.** *The ABox $\mathsf{merge}_{\mathcal{A}}(s \to t)$ is obtained from $\mathsf{prune}_{\mathcal{A}}(s)$ by replacing the individual $s$ with the individual $t$ in all assertions and their annotations (but not in renamings) and, if both $s$ and $t$ are root individuals, adding the renaming $s \mapsto t$.*

**Derivation Rules.** *Table 5 specifies derivation rules that, given an ABox $\mathcal{A}$ and a set of HT-clauses $\mathcal{C}$, derive one or more ABoxes $\mathcal{A}_1, \dots, \mathcal{A}_n$. In the Hyp-rule, $\sigma$ is a mapping*

---

11. As we show in Lemma 4, the derivation rules of our calculus ensure that $\mapsto$ is a functional and acyclic relation, so an individual $b$ satisfying the definition always exists. The second part of the definition of $\|a\|_{\mathcal{A}}$ is thus just a technical aid necessary to make the definition complete.

*from the set of variables $N_V$ to the individuals occurring in the assertions of $\mathcal{A}$, and $\sigma(U)$ is the result of replacing each variable $x$ in the atom $U$ with $\sigma(x)$.*

**Rule Precedence.** *The $\approx$-rule can be applied to a (possibly annotated) equality $s \approx t$ in an ABox $\mathcal{A}$ only if $\mathcal{A}$ does not contain an equality $s \approx t @^u_{\leq n\,R.B}$ to which the NI-rule is applicable (with the same $s$ and $t$).*

**Clash.** *An ABox $\mathcal{A}$ contains a clash iff $\perp \in \mathcal{A}$; otherwise, $\mathcal{A}$ is clash-free.*

**Derivation.** *For a set of HT-clauses $\mathcal{C}$ and an input ABox $\mathcal{A}$, a derivation is a pair $(T, \lambda)$ where $T$ is a finitely branching tree and $\lambda$ is a function that labels the nodes of $T$ with ABoxes such that the following properties hold for each node $t \in T$:*

- *$\lambda(t) = \mathcal{A}$ if $t$ is the root of $T$;*

- *$t$ is a leaf of $T$ if $\perp \in \lambda(t)$ or no derivation rule is applicable to $\lambda(t)$ and $\mathcal{C}$;*

- *$t$ has children $t_1, \ldots, t_n$ such that $\lambda(t_1), \ldots, \lambda(t_n)$ are exactly the results of applying one (arbitrarily chosen, but respecting the rule precedence) applicable rule to $\lambda(t)$ and $\mathcal{C}$ in all other cases.*

We stress several important aspects of Definition 7. If the preconditions of the *NI*-rule are satisfied for an annotated equality $s \approx t @^u_{\leq n\,R.B}$, then the rule must be applied even if $s = t$; hence, such an equality plays a role in a derivation even though it is a logical tautology. Furthermore, even though the *NI*-rule is not applied to $s \approx t @^u_{\leq n\,R.B}$ if $u$ is a blockable individual, the equality cannot be eagerly simplified into $s \approx t$ because $u$ can subsequently be merged into a root individual so the annotation might become important. Finally, if $\mathcal{C}$ has been obtained by a normalization of a DL knowledge base that does not use nominals, inverse roles, or number restrictions, then the precondition of the *NI*-rule will never be satisfied, so we need not keep track of annotations at all.

Renamings are used to keep track of root individuals that are merged into other root individuals, which is necessary to make the *NI*-rule sound. For example, if a root individual $a.\langle R, B, 2 \rangle$ is merged into a named individual $b$, then the *NI*-rule must use $b$ instead of $a.\langle R, B, 2 \rangle$ in all future inferences.

The proof of Lemma 6 shows that assertions containing at least one indirectly blocked individual are not used to construct a model from an ABox labeling a leaf in a derivation. All derivation rules are therefore applicable only to individuals that are either directly blocked or not blocked, as this is sufficient for completeness. Since all rules are sound, however, one may choose to disregard this restriction if that makes implementation easier.

We next introduce a notion of HT-ABoxes, which formalizes the idea of forest-shaped ABoxes introduced in Section 3.1.2.

**Definition 8** (HT-ABoxes)**.** *An ABox $\mathcal{A}$ is an* HT-ABox *if it satisfies the following conditions, for $R$ an atomic role, $S$ a role, $B$ a literal concept not containing a nominal guard concept, $O_a$ a nominal guard concept, $s, t, u \in N_A$, $a \in N_O$, $b \in N_I$, and $i, j$ integers.*

1. *Each role assertion in $\mathcal{A}$ is of the form $R(a, s)$, $R(s, a)$, $R(s, s.i)$, $R(s.i, s)$, or $R(s, s)$.*

Table 5: Derivation Rules of the Hypertableau Calculus

| | |
|---|---|
| *Hyp*-rule | If 1. $r \in \mathcal{C}$, where $r = U_1 \wedge \ldots \wedge U_m \rightarrow V_1 \vee \ldots \vee V_n$, and |
| | 2. a mapping $\sigma$ from the variables in $r$ to the individuals of $\mathcal{A}$ exists such that |
| | 2.1 there is no $x \in N_V$ such that $\sigma(x)$ is indirectly blocked, |
| | 2.2 $\sigma(U_i) \in \mathcal{A}$ for each $1 \leq i \leq m$, and |
| | 2.3 $\sigma(V_j) \notin \mathcal{A}$ for each $1 \leq j \leq n$, |
| | then $\mathcal{A}_1 := \mathcal{A} \cup \{\bot\}$ if $n = 0$; |
| | $\mathcal{A}_j := \mathcal{A} \cup \{\sigma(V_j)\}$ for $1 \leq j \leq n$ otherwise. |
| $\geq$-rule | If 1. $\geq n\,R.B(s) \in \mathcal{A}$, |
| | 2. $s$ is not blocked in $\mathcal{A}$, and |
| | 3. $\mathcal{A}$ does not contain individuals $u_1, \ldots, u_n$ such that |
| | 3.1 $\{\mathsf{ar}(R, s, u_i), B(u_i) \mid 1 \leq i \leq n\} \cup \{u_i \not\approx u_j \mid 1 \leq i < j \leq n\} \subseteq \mathcal{A}$, and |
| | 3.2 for each $1 \leq i \leq n$, either $u_i$ is a successor of $s$ or $u_i$ is not blocked in $\mathcal{A}$, |
| | then $\mathcal{A}_1 := \mathcal{A} \cup \{\mathsf{ar}(R, s, t_i),\ B(t_i) \mid 1 \leq i \leq n\} \cup \{t_i \not\approx t_j \mid 1 \leq i < j \leq n\}$ |
| | where $t_1, \ldots, t_n$ are fresh distinct successors of $s$. |
| $\approx$-rule | If 1. $s \approx t \in \mathcal{A}$ (the equality can possibly be annotated), |
| | 2. $s \neq t$, and |
| | 3. neither $s$ nor $t$ is indirectly blocked |
| | then $\mathcal{A}_1 := \mathsf{merge}_{\mathcal{A}}(s \rightarrow t)$ if $t$ is a named individual, or $t$ is a root individual and $s$ is not a named individual, or $s$ is a descendant of $t$; |
| | $\mathcal{A}_1 := \mathsf{merge}_{\mathcal{A}}(t \rightarrow s)$ otherwise. |
| $\bot$-rule | If $s \not\approx s \in \mathcal{A}$ or $\{A(s), \neg A(s)\} \subseteq \mathcal{A}$ where $s$ is not indirectly blocked |
| | then $\mathcal{A}_1 := \mathcal{A} \cup \{\bot\}$. |
| *NI*-rule | If 1. $s \approx t\,@^a_{\leq n\,R.B} \in \mathcal{A}$ (the symmetry of $\approx$ applies as usual), |
| | 2. $u$ is a root individual, |
| | 3. $s$ is a blockable individual that is not a successor of $u$, |
| | 4. $t$ is a blockable individual, and |
| | 5. neither $s$ nor $t$ is indirectly blocked |
| | then $\mathcal{A}_i := \mathsf{merge}_{\mathcal{A}}(s \rightarrow \|u.\langle R, B, i\rangle\|_{\mathcal{A}})$ for each $1 \leq i \leq n$. |

2. *Each equality in $\mathcal{A}$ is either of the form $s \approx t\,@^a_{\leq n\,R.B}$ with $s$ a blockable individual that is not a successor of $a$ and $t$ a blockable individual, or it is a possibly annotated equality of the form $s.i \approx s.j$, $s.i \approx s$, $s.i.j \approx s$, $s \approx s$, or $s \approx a$. (The symmetry of $\approx$ applies in all these cases as usual.)*

3. *Each concept assertion in $\mathcal{A}$ is of the form $B(s)$, $\geq n\,S.B(s)$, or $O_a(b)$.*

4. *If $\mathcal{A}$ contains $s \approx t\,@^u_{\leq n\,R.B}$, then $\mathcal{A}$ also contains $\mathsf{ar}(R, u, s)$ and $\mathsf{ar}(R, u, t)$.*

5. *If $\mathcal{A}$ contains a blockable individual $s.i$ in some assertion, then $\mathcal{A}$ must contain an assertion of the form $R(s, s.i)$ or $R(s.i, s)$.*

6. *$\mathcal{A}$ contains at least one assertion.*

Table 6: Cases in an Application of the *Hyp*-Rule to Role Assertions

| $\mathsf{ar}(R, u, s)$ | $\mathsf{ar}(R, u, t)$ | $s \approx t \, @^{u}_{\leq_k R.B}$ |
|---|---|---|
| $\mathsf{ar}(R, v, a)$ | $\mathsf{ar}(R, v, b)$ | $a \approx b \, @^{v}_{\leq_k R.B}$ |
| $\mathsf{ar}(R, v, a)$ | $\mathsf{ar}(R, v.n)$ | $a \approx v.n \, @^{v}_{\leq_k R.B}$ |
| $\mathsf{ar}(R, v, a)$ | $\mathsf{ar}(R, v, v)$ | $a \approx v \, @^{v}_{\leq_k R.B}$ |
| $\mathsf{ar}(R, v.n, a)$ | $\mathsf{ar}(R, v.n, v)$ | $a \approx v \, @^{v.n}_{\leq_k R.B}$ |
| $\mathsf{ar}(R, v, v.m)$ | $\mathsf{ar}(R, v, v.n)$ | $v.m \approx v.n \, @^{v}_{\leq_k R.B}$ |
| $\mathsf{ar}(R, v, v.m)$ | $\mathsf{ar}(R, v, v)$ | $v.m \approx v \, @^{v}_{\leq_k R.B}$ |
| $\mathsf{ar}(R, v.n, v.n.m)$ | $\mathsf{ar}(R, v.n, v)$ | $v.n.m \approx v \, @^{v.n}_{\leq_k R.B}$ |
| $\mathsf{ar}(R, v, v)$ | $\mathsf{ar}(R, v, v)$ | $v \approx v \, @^{v}_{\leq_k R.B}$ |
| $\mathsf{ar}(R, v.n, v.n)$ | $\mathsf{ar}(R, v.n, v)$ | $v.n \approx v \, @^{v.n}_{\leq_k R.B}$ |
| $\mathsf{ar}(R, v.n, v)$ | $\mathsf{ar}(R, v.n, v)$ | $v \approx v \, @^{v.n}_{\leq_k R.B}$ |

7. *The relation $\mapsto$ in $\mathcal{A}$ is acyclic, $\mathcal{A}$ contains at most one renaming $a \mapsto b$ for an individual $a$, and, if $\mathcal{A}$ contains $a \mapsto b$, then $a$ does not occur in any assertion in $\mathcal{A}$.*

Clearly, each input ABox is an HT-ABox. We now prove that, given an HT-ABox, our calculus produces only HT-ABoxes.

**Lemma 4** (HT-Preservation). *For $\mathcal{C}$ a set of HT-clauses and $\mathcal{A}$ an HT-ABox, each ABox $\mathcal{A}'$ obtained by applying a derivation rule to $\mathcal{C}$ and $\mathcal{A}$ is an HT-ABox.*

*Proof.* Let $\mathcal{C}$, $\mathcal{A}$, and $\mathcal{A}'$ be as stated in the lemma. We now analyze each derivation rule from Table 5 and show that $\mathcal{A}'$ satisfies the remaining conditions of HT-ABoxes.

(*Hyp*-rule) Consider an application of the *Hyp*-rule to an HT-clause $r$ of type (34) with a mapping $\sigma$, deriving an assertion $\sigma(V)$.

Assume that $V$ is of the form $y_i \approx y_j \, @^{x}_{\leq_k R.B}$, so $\sigma(V)$ is of the form $s \approx t \, @^{u}_{\leq_k R.B}$. By Definition 5, the antecedent of $r$ then contains atoms of the form $\mathsf{ar}(R, x, y_i)$ and $\mathsf{ar}(R, x, y_j)$ so, by the precondition of the *Hyp*-rule, $\mathcal{A}$ contains assertions $\mathsf{ar}(R, u, s)$ and $\mathsf{ar}(R, u, t)$. If $u$ is a root individual and either $s$ or $t$ is a blockable individual that is not a successor of $u$, then $\sigma(V)$ clearly satisfies Property (2) of HT-ABoxes. Otherwise, since $\mathcal{A}$ satisfies Property (1) of HT-ABoxes, we have the possibilities shown in Table 6, for $v$ a blockable individual, and $a$ and $b$ root individuals. For brevity, we omit the symmetric combinations where the roles of $\mathsf{ar}(R, u, s)$ and $\mathsf{ar}(R, u, t)$ are exchanged. Clearly, $\sigma(V)$ satisfies Property (2) of HT-ABoxes. Finally, $\sigma(V)$ obviously satisfies Property (4) of HT-ABoxes.

Assume that $V$ is of the form $x \approx z_j$, so $\sigma(V)$ is of the form $s \approx t$. By Definition 5, the antecedent of $r$ then contains an atom $O_a(z_j)$, so either $O_a(s) \in \mathcal{A}$ or $O_a(t) \in \mathcal{A}$. By Property (3) of HT-ABoxes, either $s$ or $t$ is a named individual, so $\sigma(V)$ satisfies Property (2) of HT-ABoxes.

Assume that $V$ is of the form $R(x, x)$. Then, $\sigma(V)$ is of the form $R(s, s)$, and it satisfies Property (1) of HT-ABoxes.

Assume that $V$ is of the form $R(x, y_i)$ or $R(y_i, x)$, so $\sigma(V)$ is of the form $R(s, t)$. By Definition 5, the antecedent of $r$ then contains an atom of the form $S(x, y_i)$ or $S(y_i, x)$, and either $S(s, t) \in \mathcal{A}$ or $S(t, s) \in \mathcal{A}$; these assertions satisfy Property (1) of HT-ABoxes, so $R(s, t)$ satisfies it as well.

Assume that $V$ is of the form $R(x, z_j)$ or $R(z_j, x)$, so $\sigma(V)$ is of the form $R(s, t)$. By Definition 5, the antecedent of $r$ then contains an atom of the form $O_a(z_j)$ for $O_a$ a nominal guard concept, and either $O_a(s) \in \mathcal{A}$ or $O_a(t) \in \mathcal{A}$; by Property (3) of HT-ABoxes, either $s$ or $t$ is a named individual, so $R(s, t)$ satisfies Property (1) of HT-ABoxes.

Assume that $V$ is of the form $B(x)$, $\geq n\,S.B(x)$, or $B(y_i)$, so $\sigma(V)$ is of the form $B(s)$ or $\geq n\,S.B(s)$. By Definition 5, $B$ is a literal but not a nominal guard concept, so $\sigma(V)$ satisfies Property (3) of HT-ABoxes.

($\geq$-rule) Consider an application of the $\geq$-rule to an assertion $\geq n\,R.B(s)$. By Property (3) of HT-ABoxes, $B$ is not a nominal guard concept, so all assertions $B(t_i)$ introduced by the rule satisfy Property (3) of HT-ABoxes. Furthermore, all $t_i$ introduced by the rule are fresh blockable successors of $s$, and all role assertions introduced by the rule are of the form $R(s, t_i)$ or $R(t_i, s)$, so they satisfy Properties (1) and (5) of HT-ABoxes. The inequalities introduced by the rule trivially satisfy the properties of HT-ABoxes.

($\approx$-rule) Consider an application of the $\approx$-rule to a possibly annotated equality $s \approx t$, where $s$ is merged into $t$ (the annotation of the equality plays no role here). By the conditions on the $\mapsto$ relation of $\mathcal{A}$, the ABox $\mathcal{A}$ contains no renaming for $s$ or $t$, so the renaming $s \mapsto t$ is the only renaming for $s$ in $\mathcal{A}'$, and adding this renaming to $\mathcal{A}$ does not introduce a cycle in $\mapsto$. Merging replaces all occurrences of $s$ in $\mathcal{A}$, so no assertion of $\mathcal{A}'$ contains $s$. Hence, the $\mapsto$ relation in $\mathcal{A}'$ satisfies Property (7) of HT-ABoxes.

The $NI$-rule is not applicable to $s \approx t$ by the rule precedence, so, by the preconditions of the $NI$-rule and Property (2) of HT-ABoxes, $s \approx t$ can be of the form $v \approx a$, $v.i \approx v.j$, $v.i \approx v$, or $v.i.j \approx v$ for $a \in N_O$ and $v \in N_A$; we denote this property with (*). Since pruning and replacements are applied to all assertions of $\mathcal{A}$ uniformly, $\mathcal{A}'$ clearly satisfies Property (4) of HT-ABoxes. Furthermore, pruning removes all successors of $s$, so $\mathcal{A}'$ satisfies Property (5) of HT-ABoxes. We next consider the types of assertions of $\mathcal{A}$ that change when $s$ is merged into $t$.

Consider a role assertion $R(s, u) \in \mathcal{A}$ that is changed into $R(t, u) \in \mathcal{A}'$. If either $t$ or $u$ is a root individual, then $R(t, u)$ clearly satisfies Property (1) of HT-ABoxes, so assume that $t$ and $u$ are both blockable individuals. Then, $u$ is not a successor of $s$, since the $\approx$-rule prunes all assertions that contain a descendant of the merged individual. But then, by (*) and since $R(s, u)$ satisfies Property (1) of HT-ABoxes, we have the possibilities shown in Table 7. The cases when $R(u, s) \in \mathcal{A}$ is changed into $R(u, t) \in \mathcal{A}'$ by merging are analogous.

We now consider the form of equalities that can be derived from other equalities via merging. An equality $u \approx v\,@^s_{\leq n\,R.C}$ can be changed into $u \approx v\,@^t_{\leq n\,R.C}$, but the resulting equality always satisfies Property (2) of HT-ABoxes. Furthermore, for $a$ a root individual, $s \approx u\,@^a_{\leq n\,R.C}$ can be changed into $t \approx u\,@^a_{\leq n\,R.C}$, and $s \approx a$ can be changed into $t \approx a$; however, in both cases, the resulting equality satisfies Property (2) of HT-ABoxes. For the remaining cases, assume that a possibly annotated equality $s \approx u$ is changed into a possibly annotated equality $t \approx u$. If $s$ is a root individual, then $t$ is a root individual as well (the $\approx$-rule never merges a root individual into a blockable one), so $t \approx u$ satisfies Property (2) of HT-ABoxes. Assume that $s$ is a blockable individual. Since the ≈-rule prunes all assertions that contain a descendant of the merged individual, $u$ is not a successor of $s$. By (*), Property (2) of HT-ABoxes, and the fact that the *NI*-rule is not applicable to $\mathcal{A}$, we have the possibilities shown in Table 8. In all cases, the resulting assertion satisfies Property (2) of HT-ABoxes. Furthermore, replacing $s$ with $t$ in $s \approx t \in \mathcal{A}$ results in $t \approx t \in \mathcal{A}'$, so $\mathcal{A}'$ satisfies Property (6) of HT-ABoxes.

Table 7: Cases in an Application of the ≈-Rule to Role Assertions

| $R(s, u)$ | $s \approx t$ | $R(t, u)$ |
|---|---|---|
| $R(v.i, v)$ | $v.i \approx v.j$ | $R(v.j, v)$ |
| $R(v.i, v)$ | $v.i \approx v$ | $R(v, v)$ |
| $R(t.j.i, t.j)$ | $t.j.i \approx t$ | $R(t, t.j)$ |
| $R(v.i, v.i)$ | $v.i \approx v.j$ | $R(v.j, v.j)$ |
| $R(v.i, v.i)$ | $v.i \approx v$ | $R(v, v)$ |
| $R(t.j.i, t.j.i)$ | $t.j.i \approx t$ | $R(t, t)$ |

Table 8: Cases in an Application of the ≈-Rule to Equalities

| $s \approx u$ | $s \approx t$ | $t \approx u$ |
|---|---|---|
| $v.i \approx v.k$ | $v.i \approx v.j$ | $v.j \approx v.k$ |
| $v.i \approx v$ | $v.i \approx v.j$ | $v.j \approx v$ |
| $u.k.i \approx u$ | $u.k.i \approx u.k.j$ | $u.k.j \approx u$ |
| $v.i \approx v.k$ | $v.i \approx v$ | $v \approx v.k$ |
| $v.i \approx v$ | $v.i \approx v$ | $v \approx v$ |
| $u.k.i \approx u$ | $u.k.i \approx u.k$ | $u.k \approx u$ |
| $t.j.i \approx t.j.k$ | $t.j.i \approx t$ | $t \approx t.j.k$ |
| $t.j.i \approx t.j$ | $t.j.i \approx t$ | $t \approx t.j$ |
| $t.j.i \approx t$ | $t.j.i \approx t$ | $t \approx t$ |

Consider an assertion $C(s) \in \mathcal{A}$ that is changed into $C(t) \in \mathcal{A}'$. The only nontrivial case is when $C$ is a nominal guard concept $O_a$. By Property (3) of HT-ABoxes, $s$ is then a named individual. The ≈-rule replaces named individuals only with other named individuals, so $t$ is a named individual as well. Thus, $C(t)$ satisfies Property (3) of HT-ABoxes.

(*NI*-rule) Consider an application of the *NI*-rule to an equality $s \approx t \,@_{\leq n\,R.B}^u$ that merges $s$ into a root individual $\|u.\langle R, B, i\rangle\|_{\mathcal{A}}$. The individual $s$ is blockable, so no renaming is added to $\mathcal{A}$ and the $\mapsto$ relation in $\mathcal{A}'$ satisfies Property (7) of HT-ABoxes. Since $s$ is replaced by a root individual in role and equality assertions, all resulting assertions satisfy Properties (1) and (2) of HT-ABoxes. Since $s$ is not a named individual, no assertion involving a nominal guard concept is affected by merging, so $\mathcal{A}'$ satisfies Property (3). Since pruning and replacements are applied to all assertions of $\mathcal{A}$ uniformly, $\mathcal{A}'$ clearly satisfies Property (4) of HT-ABoxes. Pruning removes all successors of $s$, so $\mathcal{A}'$ satisfies Property (5) of

HT-ABoxes. Finally, $\mathcal{A}'$ is clearly not empty, so it satisfies Property (6). □

We next prove soundness and completeness of our calculus. We use these notions as is customary in resolution-based theorem proving: a calculus is sound if its derivation rules preserve satisfiability of a theory, and it is complete if, whenever the calculus terminates without detecting a contradiction, the theory is indeed satisfiable.

**Lemma 5** (Soundness). *Let $\mathcal{C}$ be a set of HT-clauses and $\mathcal{A}$ an input ABox such that $(\mathcal{C}, \mathcal{A})$ is satisfiable. Then, each derivation for $\mathcal{C}$ and $\mathcal{A}$ contains a branch such that $\lambda(t)$ is clash-free for each node $t$ on the branch.*

*Proof.* We say that a model $I$ of an ABox $\mathcal{A}_0$ is *NI-compatible* with $\mathcal{A}_0$ if the following conditions are satisfied:

- For each root individual $a$ occurring in $\mathcal{A}_0$, each concept $\leq n\,R.B$, and each $\alpha \in \triangle^I$ such that $a^I \in (\leq n\,R.B)^I$, $\langle a^I, \alpha \rangle \in R^I$, and $\alpha \in B^I$, we have $\alpha = (a.\langle R, B, i \rangle)^I$ for some $1 \leq i \leq n$.[12]

- If $s \approx t\,@^u_{\leq n\,R.B} \in \mathcal{A}_0$, then we have $\langle u^I, s^I \rangle \in R^I$, $\langle u^I, t^I \rangle \in R^I$, $s^I \in B^I$, $t^I \in B^I$, and $u^I \in (\leq n\,R.B)^I$.[13]

To prove this lemma, we first show the following property (*): if $(\mathcal{C}, \mathcal{A}_0)$ is satisfiable in a model that is *NI-compatible* with $\mathcal{A}_0$ and $\mathcal{A}_1, \ldots, \mathcal{A}_n$ are ABoxes obtained by applying a derivation rule to $\mathcal{C}$ and $\mathcal{A}_0$, then some $(\mathcal{C}, \mathcal{A}_i)$ is satisfiable in a model that is *NI-compatible* with $\mathcal{A}_i$. Let $I$ be a model of $(\mathcal{C}, \mathcal{A}_0)$ that is *NI-compatible* with $\mathcal{A}_0$, and consider all possible derivation rules that can derive $\mathcal{A}_1, \ldots, \mathcal{A}_n$ from $\mathcal{A}_0$ and $\mathcal{C}$.

(*Hyp*-rule) Consider an application of the *Hyp*-rule to an HT-clause $r$ of the form (34). Since $\sigma(U_i) \in \mathcal{A}_0$, we have $I \models \sigma(U_i)$ for all $1 \leq i \leq m$. But then, $I \models \sigma(V_j)$ for some $1 \leq j \leq n$. Since $\mathcal{A}_j := \mathcal{A}_0 \cup \{\sigma(V_j)\}$, we have $I \models (\mathcal{C}, \mathcal{A}_j)$.

If $I \models \sigma(V_j)$ for some atom $V_j$ not of the form $\psi = y_k \approx y_\ell\,@^x_{\leq h\,R.B}$, then $I$ is clearly *NI-compatible* with $\mathcal{A}_j$. Furthermore, for each $V_j$ of the form $\psi$, clearly $\langle \sigma(x)^I, \sigma(y_k)^I \rangle \in R^I$, $\langle \sigma(x)^I, \sigma(y_\ell)^I \rangle \in R^I$, $\sigma(y_k)^I \in B^I$, and $\sigma(y_\ell)^I \in B^I$. Let (**) denote these two properties.

Assume that $I$ is not *NI-compatible* with $\mathcal{A}_j$ for each $1 \leq j \leq n$. By (**), then $I \not\models \sigma(V_j)$ for each $V_j$ not of the form $\psi$, and $\sigma(x)^I \notin (\leq h\,R.B)^I$ for each $V_j$ of the form $\psi$. Let $\mu : N_V \to \triangle^I$ be a variable mapping such that $\mu(x) = \sigma(x)^I$ and $\mu(y_k) = \sigma(y_k)^I$ for each branch variable $y_k$ not occurring in an atom of the form $\psi$; furthermore, for each set of branch variables $y_1, \ldots, y_{h+1}$ occurring in an atom of the form $\psi$, we set $\mu(y_1), \ldots, \mu(y_{h+1})$ to arbitrarily chosen domain elements that verify $\sigma(x)^I \notin (\leq h\,R.B)^I$. Clearly, $I, \mu \not\models V_j$ for each $V_j$ not occurring in a subset (35) or (36) of $r$; furthermore, by the definition of $\mu$, we have that $I, \mu \not\models V_j$ for each $V_j$ occurring in a subset of (35) or (36) of $r$. But then, we conclude $I, \mu \not\models (\mathcal{C}, \mathcal{A}_0)$, which is a contradiction.

($\geq$-rule) Since $\geq n\,R.B(s) \in \mathcal{A}_0$, we have $I \models \geq n\,R.B(s)$, which implies that domain elements $\alpha_1, \ldots, \alpha_n \in \triangle^I$ exist where $\langle s^I, \alpha_i \rangle \in R^I$ and $\alpha_i \in B^I$ for $1 \leq i \leq n$, and $\alpha_i \neq \alpha_j$ for $1 \leq i < j \leq n$. Let $I'$ be an interpretation obtained from $I$ by setting $t_i^{I'} = \alpha_i$. Clearly, $I' \models \mathsf{ar}(R, s, t_i)$, $I' \models B(t_i)$, and $I' \models t_i \not\approx t_j$ for $i \neq j$, so $I' \models (\mathcal{C}, \mathcal{A}_1)$. The individuals $t_i$ are not root individuals, so $I'$ is *NI*-compatible with $\mathcal{A}_1$.

---

12. Intuitively, this condition ensures that each root individual $a.\langle R, B, i \rangle$ is interpreted as an appropriate "neighbor" of $a^I$.

13. Intuitively, this condition ensures that $u$, $s$, and $t$ are interpreted in $I$ in accordance with the annotation.

($\approx$-rule) Assume that the $\approx$-rule is applied to the assertion $s \approx t \in \mathcal{A}_0$ and $s$ is merged into $t$. Since $I \models s \approx t$, we have $s^I = t^I$. Pruning removes assertions, so $I$ is a model of the pruned ABox by monotonicity. Merging simply replaces an individual with a synonym, so $I \models (\mathcal{C}, \mathcal{A}_1)$. Furthermore, by Property (7) of HT-ABoxes, $\mathcal{A}$ does not contain renamings for $s$ and $t$, so $\|s\|_{\mathcal{A}_1} = t$; hence, $I$ is *NI*-compatible with $\mathcal{A}_1$.

($\bot$-rule) This rule is never applicable if $(\mathcal{C}, \mathcal{A}_0)$ is satisfiable.

(*NI*-rule) Assume that the *NI*-rule is applied to some $s \approx t @^u_{\leq n \, R.B} \in \mathcal{A}_0$ and $s$ is merged into a root individual. Since $I$ is *NI*-compatible with $\mathcal{A}_0$, we have $u^I \in (\leq n \, R.B)^I$, $\langle u^I, s^I \rangle \in R^I$, $s^I \in B^I$, and $s^I = (u.\langle R, B, i \rangle)^I$ for some $1 \leq i \leq n$. Let $v_i = \|u.\langle R, B, i \rangle\|_{\mathcal{A}_0}$; since $I$ is *NI*-compatible, we have $(u.\langle R, B, i \rangle)^I = v_i^I$. Thus, the *NI*-rule replaces $s$ by its synonym $v_i$, so $I \models (\mathcal{C}, \mathcal{A}_i)$ just like in the case of the $\approx$-rule. If $v_i$ does not occur in $\mathcal{A}_0$, the interpretation $I$ may not be *NI*-compatible with $\mathcal{A}_i$ because it does not interpret $v_i.\langle S, C, \ell \rangle$ correctly. We then extend $I$ to $I'$ as follows. For each $m$, $S$, and $C$ such that $v_i^I \in (\leq m \, S.C)^I$, let $\alpha_1, \ldots, \alpha_k$ be the elements of $\triangle^I$ such that $\langle v_i^I, \alpha_j \rangle \in S^I$ and $\alpha_j \in C^I$; clearly, $k \leq m$. We then set $(v_i.\langle S, C, \ell \rangle)^{I'} = \alpha_\ell$ for $1 \leq \ell \leq k$. Since none of $v_i.\langle S, C, \ell \rangle$ occurs in $\mathcal{A}_i$, we have $I' \models (\mathcal{C}, \mathcal{A}_i)$, so $I'$ is *NI*-compatible with $\mathcal{A}_j$.

This completes the proof of (*). To prove the main claim of this lemma, let $\mathcal{A}$ be an input ABox. Similarly as for the *NI*-rule in the proof of (*), we can extend $I$ to a model $I'$ of $(\mathcal{C}, \mathcal{A})$. Since $\mathcal{A}$ does not contain annotated equalities, $I'$ is *NI*-compatible with $\mathcal{A}$. The claim of this lemma then follows by a straightforward inductive application of (*). □

**Lemma 6** (Completeness). *If a derivation for a set of HT-clauses $\mathcal{C}$ and an input ABox $\mathcal{A}$ exists in which some leaf node is labeled with a clash-free ABox $\mathcal{A}'$, then $(\mathcal{C}, \mathcal{A})$ is satisfiable.*

*Proof.* We prove the lemma by constructing from $\mathcal{A}'$ a model of $(\mathcal{C}, \mathcal{A})$. Since our logic does not have the finite model property, we obtain this model by *unraveling* $\mathcal{A}'$ as intuitively explained in Section 3.1.2. As usual, elements of the unraveled model are paths (Horrocks & Sattler, 2001, 2007), as defined next.

Given an individual $s$ that is directly blocked in $\mathcal{A}'$, let the *blocker* of $s$ be an arbitrarily chosen but fixed individual $t$ such that $s$ is directly blocked by $t$.

A *path* is finite sequence of pairs of individuals $p = [\frac{s_0}{s_0'}, \ldots, \frac{s_n}{s_n'}]$. Let $\mathsf{tail}(p) = s_n$ and $\mathsf{tail}'(p) = s_n'$. Furthermore, let $q = [p \mid \frac{s_{n+1}}{s_{n+1}'}]$ be the path $[\frac{s_0}{s_0'}, \ldots, \frac{s_n}{s_n'}, \frac{s_{n+1}}{s_{n+1}'}]$; we say that $q$ is a *successor* of $p$, and $p$ is a *predecessor* of $q$. The set of all paths $\mathcal{P}(\mathcal{A}')$ is defined inductively as follows:

- $[\frac{a}{a}] \in \mathcal{P}(\mathcal{A}')$ for each root individual $a$ occurring in $\mathcal{A}'$;

- $[p \mid \frac{s'}{s'}] \in \mathcal{P}(\mathcal{A}')$ if $p \in \mathcal{P}(\mathcal{A}')$, $s'$ is a successor of $\mathsf{tail}(p)$, $s'$ occurs in $\mathcal{A}'$, and $s'$ is not blocked in $\mathcal{A}'$; and

- $[p \mid \frac{s}{s'}] \in \mathcal{P}(\mathcal{A}')$ if $p \in \mathcal{P}(\mathcal{A}')$, $s'$ is a successor of $\mathsf{tail}(p)$, $s'$ occurs in $\mathcal{A}'$, $s'$ is directly blocked in $\mathcal{A}'$, and $s$ is the blocker of $s'$ in $\mathcal{A}'$.

Table 9: The Construction of an Interpretation from $\mathcal{A}'$

$$
\begin{aligned}
\triangle^I &= \mathcal{P}(\mathcal{A}') \\
a^I &= \left[\tfrac{a}{a}\right] \text{ for each root individual } a \text{ that occurs in an assertion in } \mathcal{A}' \\
a^I &= b^I \text{ if } a \neq b \text{ and } \|a\|_{\mathcal{A}'} = b \\
A^I &= \{p \in \triangle^I \mid A(\mathsf{tail}(p)) \in \mathcal{A}'\} \\
R^I &= \{\langle[\tfrac{a}{a}], p\rangle \in \triangle^I \times \triangle^I \quad \mid a \text{ is a root individual and } R(a, \mathsf{tail}(p)) \in \mathcal{A}'\} \cup \\
&\quad\ \{\langle p, [\tfrac{a}{a}]\rangle \in \triangle^I \times \triangle^I \quad \mid a \text{ is a root individual and } R(\mathsf{tail}(p), a) \in \mathcal{A}'\} \cup \\
&\quad\ \{\langle p, [p \mid \tfrac{s}{s'}]\rangle \in \triangle^I \times \triangle^I \quad \mid R(\mathsf{tail}(p), s') \in \mathcal{A}'\} \cup \\
&\quad\ \{\langle[p \mid \tfrac{s}{s'}], p\rangle \in \triangle^I \times \triangle^I \quad \mid R(s', \mathsf{tail}(p)) \in \mathcal{A}'\} \cup \\
&\quad\ \{\langle p, p\rangle \in \triangle^I \times \triangle^I \quad \mid R(\mathsf{tail}(p), \mathsf{tail}(p)) \in \mathcal{A}'\}
\end{aligned}
$$

Let $I$ be the interpretation constructed from $\mathcal{A}'$ as shown in Table 9. $\mathcal{A}'$ is an HT-ABox, so $\triangle^I$ is not empty. We now show that, for each $p_s$ of the form $[\tfrac{s}{s'}]$ or $[q_s \mid \tfrac{s}{s'}]$ and each individual $w$, the following claims hold (*):

- $R(s, s) \in \mathcal{A}'$ (resp. $A(s) \in \mathcal{A}'$) iff $\langle p_s, p_s \rangle \in R^I$ (resp. $p_s \in A^I$): Immediate by the definition of $I$.

- If $B(w) \in \mathcal{A}'$ and $\mathcal{L}_{\mathcal{A}'}(w) = \mathcal{L}_{\mathcal{A}'}(s')$ for $B$ a literal concept, then $p_s \in B^I$: The proof is immediate if $B$ is atomic. If $B = \neg A$, since the $\bot$-rule is not applicable to $\mathcal{A}'$, we have $A(w) \notin \mathcal{A}'$; but then, we have $A(s') \notin \mathcal{A}'$ and $A(s) \notin \mathcal{A}'$, which by the case for atomic concepts implies $p_s \notin A^I$.

- If $\geq n\,R.B(s) \in \mathcal{A}'$, then $p_s \in (\geq n\,R.B)^I$: By the definition of paths, $s$ is not blocked; since the $\geq$-rule is not applicable to $\geq n\,R.B(s)$, individuals $u_1, \ldots, u_n$ exist such that $\mathsf{ar}(R, s, u_i) \in \mathcal{A}'$ and $B(u_i) \in \mathcal{A}'$ for $1 \leq i \leq n$, and $u_i \not\approx u_j \in \mathcal{A}'$ for $1 \leq i < j \leq n$. Each assertion $\mathsf{ar}(R, s, u_i)$ satisfies Property (1) of HT-ABoxes, so each $u_i$ can be of one of the following forms.

  - $u_i = s$. Let $p_{u_i} = p_s$. But then, by the previous two cases we conclude that $\mathsf{ar}(R, s, u_i) \in \mathcal{A}'$ and $B(u_i) \in \mathcal{A}'$ imply $\langle p_s, p_{u_i} \rangle \in R^I$ and $p_{u_i} \in B^I$.

  - $u_i$ is a successor of $s$. If $u_i$ is directly blocked by the blocker $v_i$, let $p_{u_i} = [p_s \mid \tfrac{v_i}{u_i}]$; otherwise, $u_i$ is not blocked because $s$ is not blocked, and let $p_{u_i} = [p_s \mid \tfrac{u_i}{u_i}]$. Either way, we have $\mathsf{ar}(R, \mathsf{tail}(p_s), u_i) \in \mathcal{A}'$, which, by the definition of $I$, implies $\langle p_s, p_{u_i} \rangle \in R^I$. Furthermore, $B(u_i) \in \mathcal{A}'$ and $\mathcal{L}_{\mathcal{A}'}(u_i) = \mathcal{L}_{\mathcal{A}'}(\mathsf{tail}(p_{u_i}))$ imply $p_{u_i} \in B^I$.

  - $u_i$ is a blockable predecessor of $s$. Since $s$ is blockable, we have $p_s = [q_s \mid \tfrac{s}{s'}]$; hence, let $p_{u_i} = q_s$. If $s'$ is not blocked, then $s = s'$ and $\mathsf{tail}(p_{u_i}) = u_i$, so we have $\mathsf{ar}(R, s', \mathsf{tail}(p_{u_i})) \in \mathcal{A}'$. If $s'$ is blocked by the blocker $s$, then by the definition of pairwise blocking $\mathcal{L}_{\mathcal{A}'}(\mathsf{tail}(p_{u_i}), s') = \mathcal{L}_{\mathcal{A}'}(u_i, s)$ and $\mathcal{L}_{\mathcal{A}'}(s', \mathsf{tail}(p_{u_i})) = \mathcal{L}_{\mathcal{A}'}(s, u_i)$, so we again have $\mathsf{ar}(R, s', \mathsf{tail}(p_{u_i})) \in \mathcal{A}'$. Either way, we have $\langle p_s, p_{u_i} \rangle \in R^I$ by the definition of $I$. Furthermore, $B(u_i) \in \mathcal{A}'$ and $\mathcal{L}_{\mathcal{A}'}(u_i) = \mathcal{L}_{\mathcal{A}'}(\mathsf{tail}(p_{u_i}))$ imply $p_{u_i} \in B^I$.

- $u_i$ and $s$ do not satisfy any of the previous three conditions. If $s$ is a blockable individual, then $u_i$ is a root individual, so let $p_{u_i} = \lceil \frac{u_i}{u_i} \rceil$. If $s$ is a root individual, then $u_i$ is not blocked in $\mathcal{A}'$ by Condition 3.2 of the $\geq$-rule, so some $p_{u_i} \in \triangle^I$ exists that has the form $p_{u_i} = [p \mid \frac{u_i}{u_i}]$. Either way, we have $\mathsf{ar}(R, s, u_i) \in \mathcal{A}'$ and $B(u_i) \in \mathcal{A}'$, which imply $\langle p_s, p_{u_i} \rangle \in R^I$ and $p_{u_i} \in B^I$.

Consider now each $1 \leq i < j \leq n$. If $\mathsf{tail}'(p_{u_i}) \not\approx \mathsf{tail}'(p_{u_j}) \in \mathcal{A}'$, since $\perp \notin \mathcal{A}'$ and the $\perp$-rule is not applicable, we have $\mathsf{tail}'(p_{u_i}) \neq \mathsf{tail}'(p_{u_j})$, so $p_{u_i} \neq p_{u_j}$. Furthermore, if $\mathsf{tail}'(p_{u_i}) \not\approx \mathsf{tail}'(p_{u_j}) \notin \mathcal{A}'$, this is because $\mathsf{tail}'(p_{u_i}) \neq u_i$, which is possible only if $s'$ is directly blocked by the blocker $s$ and $u_i = s$ or $u_i$ is a blockable predecessor of $s$. Note, however, that $s$ can have at most one blockable predecessor, and that there can be at most one $u_i$ such that $u_i = s$. Therefore, we have $u_i \neq u_j$, which implies $p_{u_i} \neq p_{u_j}$, and we conclude $p_s \in (\geq n\,R.B)^I$.

For an assertion $\alpha' \in \mathcal{A}'$ of the form $a \approx b$ and $a \not\approx b$ with $a$ and $b$ named individuals, it is straightforward to see that $I \models \alpha'$. Furthermore, if $\alpha'$ is of the form $R(a, b)$ or $B(a)$, or $\geq n\,R.B(a)$ with $a$ a named individual, (*) implies $I \models \alpha'$. Consider now each $\alpha \in \mathcal{A}$. By induction on the application of the derivation rules, it is straightforward to show that, if $\alpha \notin \mathcal{A}'$, then $\mathcal{A}'$ contains renamings that, when applied to $\alpha$, produce an assertion $\alpha' \in \mathcal{A}'$. But then, since $I \models \alpha'$, we have $I \models \alpha$ by the definition of $I$.

It remains to be shown that $I \models \mathcal{C}$. Consider each HT-clause $r \in \mathcal{C}$ containing atoms of the form $A_i(x)$, $U_k(x, x)$, $\mathsf{ar}(R_i, x, y_i)$, $B_i(y_i)$, and $C_j(z_j)$ in the antecedent. Furthermore, consider a variable mapping $\mu$ such that the antecedent of $r$ is true in $I$ and $\mu$—that is, $p_x \in A_i^I$, $\langle p_x, p_x \rangle \in U_k^I$, $\langle p_x, p_{y_i} \rangle \in R_i^I$, $p_{y_i} \in B_i^I$, and $p_{z_j} \in C_j^I$ for $p_x = \mu(x)$, $p_{y_i} = \mu(y_i)$, and $p_{z_j} = \mu(z_j)$. Let $s = \mathsf{tail}(p_x)$, $s' = \mathsf{tail}'(p_x)$, and $t_i' = \mathsf{tail}'(p_{y_i})$. By the definition of $I$ and the fact that $\mathcal{L}_{\mathcal{A}'}(s_i') = \mathcal{L}_{\mathcal{A}'}(s_i)$, we have $A_i(s) \in \mathcal{A}'$, $U_k(s, s) \in \mathcal{A}'$, and $B_i(t_i') \in \mathcal{A}'$. Depending on the relationship between $p_x$ and $p_{y_i}$, we define $t_i$ as follows.

- $p_{y_i}$ is a successor of $p_x$ or $p_{y_i} = p_x$. Let $t_i = t_i'$. Clearly, $B_i(t_i) \in \mathcal{A}'$; furthermore, the definition of $I$ and $\langle p_x, p_{y_i} \rangle \in R_i^I$ imply $\mathsf{ar}(R_i, s, t_i') \in \mathcal{A}'$, so we have $\mathsf{ar}(R_i, s, t_i) \in \mathcal{A}'$.

- $p_{y_i}$ is a predecessor of $p_x$. We have the following cases.

  - $s$ directly blocks $s'$. Let $t_i$ be the predecessor of $s$; such $t_i$ exists since $s$ is blockable. The definition of $I$ and $\langle p_x, p_{y_i} \rangle \in R_i^I$ imply $\mathsf{ar}(R_i, s', \mathsf{tail}(p_{y_i})) \in \mathcal{A}'$ and $B(\mathsf{tail}(p_{y_i})) \in \mathcal{A}'$, and by the definition of pairwise blocking we conclude that $\mathsf{ar}(R_i, s, t_i) \in \mathcal{A}'$ and $B_i(t_i) \in \mathcal{A}'$.

  - $s'$ is not blocked. Let $t_i = t_i'$. By the definition of $I$, we have $B_i(t_i) \in \mathcal{A}'$ and $\mathsf{ar}(R_i, s, t_i) \in \mathcal{A}'$.

- $p_{y_i}$ and $p_x$ do not match any of the conditions mentioned thus far. By the definition of $I$, then either $p_x$ or $p_{y_i}$ is of the form $\lceil \frac{a}{a} \rceil$. Let $t_i = \mathsf{tail}(p_{y_i})$. By $\langle p_x, p_{y_i} \rangle \in R_i^I$ and the definition of $I$, we conclude that $B_i(t_i) \in \mathcal{A}'$ and $\mathsf{ar}(R_i, s, t_i) \in \mathcal{A}'$.

By Definition 5, the antecedent of $r$ contains an atom of the form $O_a(z_j)$ for each nominal variable $z_j$. Thus, by the definition of $I$ and Property (3) of HT-ABoxes, we have $p_{z_j}$ is of the form $\left[\frac{u_j}{u_j}\right]$ for $u_j$ a named individual; furthermore, $C_j(u_j) \in \mathcal{A}'$.

Let $\sigma$ be a mapping such that $\sigma(x) = s$, $\sigma(y_i) = t_i$, and $\sigma(z_j) = u_j$. Clearly, neither $s$ nor $t_i$ are indirectly blocked, and $\sigma(U_j) \in \mathcal{A}'$ for each atom $U_j$ in the antecedent of $r$. The $Hyp$-rule is not applicable to $r$, $\mathcal{A}'$, and $\sigma$, so $r$ contains an atom $V_i$ in the consequent such that $\sigma(V_i) \in \mathcal{A}'$. Depending on the type of $V_i$, we have the following possibilities.

Assume that $V_i$ is of the form $y_i \approx y_j @^x_{\leq k\,S.B}$; thus, we have $t_i \approx t_j @^s_{\leq k\,S.B} \in \mathcal{A}'$. Since the $\approx$-rule is not applicable to $\mathcal{A}'$, we have $t_i = t_j$. By Definition 5, $r$ contains a subclause of the form (35) or (36), so the antecedent of $r$ contains atoms $\mathsf{ar}(S, x, y_i)$ and $\mathsf{ar}(S, x, y_j)$; therefore, $\langle p_x, p_{y_i} \rangle \in S^I$ and $\langle p_x, p_{y_j} \rangle \in S^I$. The $NI$-rule is not applicable to $t_i \approx t_j @^s_{\leq k\,S.B}$ so, by the preconditions of the $NI$-rule, if $s$ is a root individual, then $t_i$ ($t_j$) is either a root individual or a successor of $s$. This rules out the possibility when $p_x$ is of the form $\left[\frac{a}{a}\right]$ and $p_{y_i}$ ($p_{y_j}$) is neither a successor of $p_x$ nor of the form $\left[\frac{b}{b}\right]$. Hence, by the construction of $I$, we have that $p_{y_i}$ ($p_{y_j}$) is either a successor of $p_x$, equal to $p_x$, the predecessor of $p_x$, or is of the form $\left[\frac{a}{a}\right]$. We now consider the following cases (w.l.o.g. we omit the symmetric cases obtained by swapping $p_{y_i}$ and $p_{y_j}$):

- $p_{y_i}$ is of the form $\left[\frac{a}{a}\right]$. Then, $t_i = t_j$ implies $p_{y_i} = p_{y_j}$ by the definition of paths.

- $p_{y_i}$ is a successor of $p_x$. Then, $p_{y_i} = [p_x \mid \frac{u_i}{t_i}]$ for $u_i = t_i$ if $t_i$ is not blocked or $u_i$ the blocker of $t_i$. Either way, $t_i$ is different from $s$ and the predecessor of $s$ (if the latter exists). We have the following possibilities for $p_{y_j}$:

  - $p_{y_j}$ is a successor of $p_x$. Then, $p_{y_j} = [p_x \mid \frac{u_j}{t_j}]$, so $t_i = t_j$ clearly implies $p_{y_i} = p_{y_j}$.

  - $p_{y_j} = p_x$ or $p_{y_j}$ is the predecessor of $p_x$. Then $t_j = s$ or $t_j$ is the predecessor of $s$, which contradicts the fact that $t_i \neq t_j$.

- $p_{y_i} = p_x$. Then $t_i = s$. The only nontrivial case is if $p_{y_j}$ is the predecessor of $p_x$; but then, $t_j \neq s$, which contradicts the fact that $t_i \neq t_j$.

- $p_{y_i}$ is the predecessor of $p_x$. The only remaining possibility is for $p_{y_j}$ to be the predecessor of $p_x$. Since $p_x$ can have at most one predecessor, we have $p_{y_i} = p_{y_j}$.

Thus, we conclude that $I, \mu \models r$.

Assume that $V_i$ is of the form $x \approx z_j$; thus, we have $s \approx u_j \in \mathcal{A}$. Since the $\approx$-rule is not applicable to $\mathcal{A}'$, we have $s = u_j$. Since $u_j$ is a named individual, it cannot block other individuals, so $s' = s$, which implies $p_x = p_{z_j}$. Thus, $I, \mu \models r$.

Assume that $V_i$ is of the form $T_i(x, x)$; thus, we have $T_i(s, s) \in \mathcal{A}'$. By (*), we then have $\langle p_x, p_x \rangle \in R_i^I$. Thus, we have $I, \mu \models r$.

Assume that $V_i$ is of the form $D_i(x)$ for $D_i$ a literal concept or of the form $\geq n\,T.B$; thus, we have $D_i(s) \in \mathcal{A}'$. By (*), we then have $p_x \in D_i^I$. Thus, we have $I, \mu \models r$.

Assume that $V_i$ is of the form $E_i(y_i)$ for $E_i$ a literal concept; thus, we have $E_i(t_i) \in \mathcal{A}'$. We have already established that $\mathcal{L}_{\mathcal{A}'}(t_i) = \mathcal{L}_{\mathcal{A}'}(t_i')$; by (*), we then have $p_{y_i} \in E_i^I$. Thus, we have $I, \mu \models r$.

Assume that $V_i$ is of the form $\mathsf{ar}(S_i, x, y_i)$, so $\mathsf{ar}(S_i, s, t_i) \in \mathcal{A}'$. By the definition of blocking, we have $\mathsf{ar}(S_i, s', t_i') \in \mathcal{A}'$. Finally, by the definition of $I$, we have $\langle p_x, p_{y_i} \rangle \in S_i^I$. Thus, we have $I, \mu \models r$.

Assume that $V_i$ is of the form $\mathsf{ar}(S_j, x, z_j)$, so $\mathsf{ar}(S_j, s, u_j) \in \mathcal{A}'$. Since $u_j$ is a named individual, by the definition of $I$ we have $\langle p_x, p_{z_j} \rangle \in S_j^I$. Thus, we have $I, \mu \models r$. □

We next prove termination of the hypertableau calculus.

**Lemma 7** (Termination). *For a set of HT-clauses $\mathcal{C}$ and an input ABox $\mathcal{A}$, let $|\mathcal{C}, \mathcal{A}|$ be the sum of the size of $\mathcal{A}$, of the number of concepts and roles in $\mathcal{C}$, and of $\lceil \log n \rceil$ for each integer $n$ occurring in $\mathcal{C}$ in an atom of the form $\geq n\,R.B$ and $y_i \approx y_j @_{\leq n\,R.B}^x$. The total number of individuals introduced on each path in each derivation for $\mathcal{C}$ and $\mathcal{A}$ is at most doubly exponential in $|\mathcal{C}, \mathcal{A}|$, and each derivation for $\mathcal{C}$ and $\mathcal{A}$ is finite.*

*Proof.* We prove the claim by showing that *(i)* each derivation rule can be applied at most once to a fixed set of individuals on a derivation path, and *(ii)* the number of new individuals introduced on each derivation path is at most doubly exponential in $|\mathcal{C}, \mathcal{A}|$. The supply of blockable individuals is infinite, so we can assume that no blockable individual is introduced twice on a derivation path. Furthermore, if the root individual $s$ is removed from an ABox $\mathcal{A}'$ due to merging, then a renaming is added to $\mathcal{A}'$ that ensures $\|s\|_{\mathcal{A}'} \neq s$. Once a renaming is added to $\mathcal{A}'$, all ABoxes occurring below $\mathcal{A}'$ in a derivation will contain this renaming as well, so no subsequent application of the *NI*-rule can reintroduce $s$.

Next, we prove *(i)* by considering each derivation rule.

- An application of the *Hyp*-rule to an HT-clause $r$ of the form (34) and a mapping $\sigma$ introduces an assertion $\sigma(V_i)$, which prevents a subsequent reapplication of the *Hyp*-rule to the same $r$ and $\sigma$. Merging and pruning can remove $\sigma(V_i)$ in subsequent derivation steps, but this also removes at least one individual occurring in $\sigma$ from the set of potential premises of the *Hyp*-rule, thus preventing the reuse of the same $\sigma$ in a future application of the *Hyp*-rule to $r$.

- An application of the $\geq$-rule to an assertion $\geq n\,R.B(s)$ introduces $t_1, \ldots, t_n$ as fresh successors of $s$ and the assertions $B(t_i)$, $\mathsf{ar}(R, s, t_i)$, and $t_i \not\approx t_j$ for $1 \leq i < j \leq n$. Thus, the individuals $u_1, \ldots, u_n$ from Condition 3 of the $\geq$-rule can be matched to $t_1, \ldots, t_n$. Furthermore, if $s$ is a root individual, none of $t_i$ can become blocked and Condition 3.2 is always satisfied for $t_i$; moreover, if $s$ is blockable, Condition 3.2 is trivially satisfied for $t_i$. If some $t_i$ is merged into another individual $v$, then $B(v)$, $\mathsf{ar}(R, s, v)$, and $v \not\approx t_j$ are added to the ABox, so the ABox still contains individuals that can be matched to Condition 3 of the $\geq$-rule. Finally, if some $t_i$ becomes indirectly blocked, then $s$ is blocked and the $\geq$-rule is not applicable to $s$.

- An application of the $\approx$-rule to $s \approx t$ removes either $s$ or $t$, so the rule cannot be reapplied to the same $s$ and $t$

- An application of the $\bot$-rule produces an ABox that labels a derivation leaf.

- An application of the *NI*-rule to an equality $s \approx t @_{\leq n\,R.B}^u$ removes $s$, so the rule cannot be reapplied to the same $\leq n\,R.B$, $s$ and $u$.

Next, we prove *(ii)*—that is, that the total number of individuals introduced on a derivation path is at most doubly exponential in $|\mathcal{C}, \mathcal{A}|$. A *path of length* $n$ between individuals $s$ and $t$ in an ABox $\mathcal{A}'$ is a sequence of individuals $u_0, u_1, \ldots, u_n$ such that $u_0 = s$, $u_n = t$, and, for each $0 \leq i \leq n-1$, either $R(u_i, u_{i+1}) \in \mathcal{A}'$ or $R(u_{i+1}, u_i) \in \mathcal{A}'$ for $R$ an atomic role.

A *root path* for a root individual $t$ in an ABox $\mathcal{A}'$ is a path between $t$ and a named individual $s$ such that all intermediate individuals $u_i$, $1 \leq i \leq n-1$, are root individuals. The *level* $\mathsf{lev}(t)$ of $t$ is the length of the shortest root path for $t$. Thus, $\mathsf{lev}(t) = 0$ if $t$ is a named individual.

The *depth* $\mathsf{dep}(t)$ of an individual $t$ is the number of ancestors of $t$. Thus, $\mathsf{dep}(t) = 0$ if $t$ is a root individual. Due to Property (5) of HT-ABoxes, if an individual $t$ occurs in an ABox $\mathcal{A}'$, then $\mathcal{A}'$ contains a path of length $\mathsf{dep}(t)$ between a root individual $s$ and $t$ such that the individuals $u_i$, $0 \leq i \leq n-1$, are all ancestors of $t$; since each individual has at most one predecessor, these $u_i$ are also the only ancestors of $t$.

We now show that the maximum level of a root individual and the maximum depth of every individual are both at most exponential in the size of $\mathcal{C}$ and $\mathcal{A}$.

An application of an derivation rule never increases the level of an individual. This is because a named individual is never pruned and can be merged only into another named individual,[14] and a root individual can be merged only into another root individual. Such rule applications can only make a root path shorter, and not longer.

Let $m$ be the number of atomic concepts and $n$ the number of atomic roles that occur in $\mathcal{A}$ and $\mathcal{C}$, let $\wp = 2^{2m+2n} + 1$, and let $\mathcal{A}'$ be an ABox labeling a node of a derivation for $\mathcal{A}$ and $\mathcal{C}$. We next show that (1) $\mathsf{dep}(t) \leq \wp$ for each individual $t$ occurring in $\mathcal{A}'$, and (2) if $t$ is a root individual, then $\mathsf{lev}(t) \leq \wp$.

(Claim 1) For a pair of individuals $s$ and $t$ occurring in $\mathcal{A}'$, there are $2^m$ different possible labels $\mathcal{L}_{\mathcal{A}'}(s)$ and $2^n$ different possible labels $\mathcal{L}_{\mathcal{A}'}(s, t)$. Thus, if $\mathcal{A}'$ contains at least $\wp = 2^m \cdot 2^m \cdot 2^n \cdot 2^n + 1$ predecessor-successor pairs of blockable individuals, then $\mathcal{A}'$ must contain two pairs $\langle s, s.i \rangle$ and $\langle t, t.j \rangle$ such that the following conditions are satisfied:

$$\begin{aligned}
\mathcal{L}_{\mathcal{A}'}(s.i) &= \mathcal{L}_{\mathcal{A}'}(t.j) & \mathcal{L}_{\mathcal{A}'}(s) &= \mathcal{L}_{\mathcal{A}'}(t) \\
\mathcal{L}_{\mathcal{A}'}(s, s.i) &= \mathcal{L}_{\mathcal{A}'}(t, t.j) & \mathcal{L}_{\mathcal{A}'}(s.i, s) &= \mathcal{L}_{\mathcal{A}'}(t.j, t)
\end{aligned}$$

Since $\prec$ contains the ancestor relation, a path in $\mathcal{A}'$ containing $\wp$ blockable individuals must include at least one blocked individual, so a blockable individual of depth $\wp$ must be blocked. The $\geq$-rule is applied only to individuals that are not blocked, so the rule cannot introduce an individual $u$ such that $\mathsf{dep}(u) > \wp$.

(Claim 2) We show that the following stronger claim (*) holds for each root individual $s$ occurring in an assertion in $\mathcal{A}'$ (the symmetry of $\approx$ applies as usual):

1. $\mathsf{lev}(s) \leq \wp$;

2. if $R(s, t) \in \mathcal{A}'$ or $R(t, s) \in \mathcal{A}'$ or $t \approx u \,@^s_{\leq n\,R.B} \in \mathcal{A}'$ with $t$ a blockable nonsuccessor of $s$, then $\mathsf{lev}(s) + \mathsf{dep}(t) \leq \wp$; and

3. if $s \approx t \in \mathcal{A}'$ with $t$ a blockable nonsuccessor of $s$ (where the equality can be annotated), then $\mathsf{lev}(s) + \mathsf{dep}(t) \leq \wp + 1$.

---

14. If a derivation rule replaced a named individual with an individual that is not named, the levels of other root individuals could increase.

This claim is clearly true for the input ABox $\mathcal{A}$ labeling the root of a derivation, which contains only named individuals. We now assume that (*) holds for some ABox $\mathcal{A}'$ and consider all possible derivation rules that can be applied to $\mathcal{A}'$.

- Assume that the *Hyp*-rule derives an assertion $R(s, t)$ or $R(t, s)$, where $s$ is a root individual and $t$ is a blockable nonsuccessor of $s$. Let $R(x, y)$ or $R(y, x)$ be the atom from the consequent of an HT-clause $r$ that is instantiated by the derivation rule. We have the following two possibilities for the antecedent of $r$.

    - The antecedent of $r$ contains an atom of the form $S(x, y)$ or $S(y, x)$ that is matched to an assertion of the form $S(s, t)$ or $S(t, s)$ in $\mathcal{A}'$. Since $\mathcal{A}'$ satisfies (*), the resulting ABox satisfies (*) as well.
    - The antecedent of $r$ contains an atom of the form $O_a(x)$ or $O_a(y)$ that is matched to an assertion of the form $O_a(s)$ in $\mathcal{A}'$ (since $t$ is blockable, $\mathcal{A}'$ cannot contain $O_a(t)$ by Property 3 of HT-ABoxes). Then $\mathsf{dep}(t) \leq \wp$ and $\mathsf{lev}(s) = 0$, so the resulting ABox satisfies (*) as well.

- Assume that the *Hyp*-rule derives an assertion $t \approx u \, @^s_{\leq_n R.B}$, where $s$ is a root individual and $t$ is a blockable nonsuccessor of $s$. By Definition 5, the antecedent of the HT-clause then contains atoms of the form $\mathsf{ar}(R, x, y_i)$ and $\mathsf{ar}(R, x, y_j)$ that are matched to assertions $\mathsf{ar}(R, s, t)$ and $\mathsf{ar}(R, s, u)$ in $\mathcal{A}'$. Since $\mathcal{A}'$ satisfies (*), we have $\mathsf{lev}(s) + \mathsf{dep}(t) \leq \wp$, so the resulting equality satisfies Item 2 of (*). To show that $t \approx u \, @^s_{\leq_n R.B}$ satisfies Item 3 of (*), assume that $u$ is a root individual and $t$ is a nonsuccessor of $u$. Since $\mathcal{A}'$ contains $\mathsf{ar}(R, s, u)$, we have that $\mathsf{lev}(u) \leq \mathsf{lev}(s) + 1$; but then, $\mathsf{lev}(u) + \mathsf{dep}(t) \leq \wp + 1$, as required.

- If the *Hyp*-rule derives an assertion $s \approx t$, where $s$ is a root individual and $t$ is a blockable nonsuccessor of $s$, the only remaining possibility is that the consequent of the HT-clause then contains the equality $x \approx z_j$. By Definition 5, the antecedent then contains $O_a(z_j)$ that is matched to an assertion $O_a(s)$ in $\mathcal{A}'$, where $s$ is a named individual. Then $\mathsf{dep}(t) \leq \wp$ and $\mathsf{lev}(s) = 0$, so the resulting ABox satisfies (*).

- Assume that the $\geq$-rule introduces an assertion of the form $R(s, t)$ or $R(t, s)$ where $t$ is a fresh individual. Individual $t$ is always a successor of $s$, so the resulting ABox trivially satisfies (*).

- Assume that the $\approx$-rule is applied to an assertion of the form $u \approx s$ and that $u$ is merged into $s$. By the definition of merging, we have that $\mathsf{dep}(u) \geq \mathsf{dep}(s)$ and $u$ is pruned. If $s$ is a blockable individual, then $u$ is blockable as well, and the resulting ABox satisfies (*) because $u$ is replaced with an individual of equal or smaller depth. Therefore, we assume that $s$ is a root individual and consider the types of assertions that can be added to $\mathcal{A}'$ as a result of merging.

    - If $R(u, u)$ is changed into $R(s, s)$, the resulting ABox clearly satisfies (*).
    - Assume that $R(u, t)$ where $t$ is a root individual is changed into $R(s, t)$. This inference can make root paths to $s$ and $t$ only shorter and not longer, so the levels of $s$ and $t$ can only decrease rather than increase. Thus, the resulting ABox satisfies Item 1 of (*).

- Assume that $R(u, t)$, where $t$ is a predecessor of $u$, is changed into $R(s, t)$; the only nontrivial case is when $t$ is a blockable nonsuccessor of $s$. Since $t$ is a predecessor of $u$, we have $\mathsf{dep}(t) + 1 = \mathsf{dep}(u)$; since $\mathcal{A}'$ satisfies (*), we have $\mathsf{lev}(s) + \mathsf{dep}(u) \leq \wp + 1$; but then, $\mathsf{lev}(s) + \mathsf{dep}(t) \leq \wp$ as required.

- The cases when $R(t, u)$ is changed into $R(t, s)$ are analogous.

- Assume that a possibly annotated equality $v \approx u$ is changed into $v \approx s$. The only nontrivial case is when $v$ is a blockable nonsuccessor of $s$. If $u$ is a root individual, then the level of $s$ after merging is bounded by $\min(\mathsf{lev}(s), \mathsf{lev}(u))$ before merging, so (*) is preserved. If $u$ and $v$ are both blockable individuals, then by Property (2) of HT-ABoxes, either $u$ is an ancestor of $v$, or $u$ and $v$ are siblings, or $v$ is an ancestor of $u$. If $u$ is an ancestor of $v$, then pruning $u$ removes $v \approx u$ from $\mathcal{A}'$. If $v$ is a sibling or an ancestor of $u$, then $u$ must be a nonsuccessor of $s$, so $\mathsf{lev}(s) + \mathsf{dep}(u) \leq \wp + 1$; but then, $\mathsf{dep}(v) \leq \mathsf{dep}(u)$, so $\mathsf{lev}(s) + \mathsf{dep}(v) \leq \wp + 1$ and (*) is preserved.

- Assume that $v \approx v' @^u_{\leq n R.B}$ is changed into $v \approx v' @^s_{\leq n R.B}$ or $v \approx s @^s_{\leq n R.B}$. The only nontrivial case is when $v$ is a blockable nonsuccessor of $s$. Since $u$ is pruned before merging, by Properties (2) and (4) of HT-ABoxes $v$ must be a predecessor of $u$, so $\mathsf{dep}(v) + 1 = \mathsf{dep}(u)$. Furthermore, by the same properties $u$ must be a blockable nonsuccessor of $s$, so $\mathsf{lev}(s) + \mathsf{dep}(u) \leq \wp + 1$. But then, $\mathsf{lev}(s) + \mathsf{dep}(v) \leq \wp$, as required.

- An application of the $\perp$-rule trivially preserves (*).

- Assume that the *NI*-rule is applied to an assertion $s \approx t @^u_{\leq n R.B}$ replacing $s$ with a root individual $v = \|u.\langle R, B, i \rangle\|_{\mathcal{A}'}$. If $v$ already occurs in an assertion in $\mathcal{A}'$, then $v$ satisfies Item 1 of (*). If, however, $v$ is fresh, by Property (4) of HT-ABoxes $v$ will be connected to $u$ by a role assertion, so $\mathsf{lev}(v) \leq \mathsf{lev}(u) + 1$. Furthermore, since $s$ is a blockable nonsuccessor of $u$, we have $\mathsf{lev}(u) + \mathsf{dep}(s) \leq \wp$. Finally, since $s$ is blockable, $\mathsf{dep}(s) \geq 1$, so $\mathsf{lev}(u) \leq \wp - 1$. As a consequence, we conclude that $\mathsf{lev}(v) \leq \wp$, which proves Item 1 of (*). The proof that the assertions introduced through merging satisfy (*) is analogous to the case for the $\approx$-rule.

We now complete the proof of claim *(ii)*—that is, that the total number of individuals introduced by derivation rules is at most doubly exponential in $|\mathcal{C}, \mathcal{A}|$.

All named individuals are of level 0 and are never introduced by the derivation rules. An application of the *NI*-rule to a root individual $u$ of level $\ell$ can introduce at most $n$ root individuals of level $\ell + 1$ for each concept $\leq n R.B$ that occurs in $\mathcal{C}$. Thus, for each named individual, the derivation rules can create a tree of root individuals. The maximum depth of the tree is $\wp$, which is exponential in $|\mathcal{C}, \mathcal{A}|$. Furthermore, the maximum branching factor $b$ is equal to the sum of all numbers occurring in $\mathcal{C}$ in atoms of the form $y_i \approx y_j @^x_{\leq n R.B}$. Clearly, $b$ is exponential in $|\mathcal{C}, \mathcal{A}|$, so each such tree is doubly exponential in $|\mathcal{C}, \mathcal{A}|$.[15]

Similarly, each root individual can become the root of a tree of blockable individuals of depth $\wp$. Each blockable individual is introduced by applying the $\geq$-rule to its predecessor.

---

15. If numbers were coded in unary, then the branching factor would be polynomial, but each such tree would still be doubly exponential in $|\mathcal{C}, \mathcal{A}|$.

Furthermore, the $\geq$-rule can be applied to an individual $s$ at most once for each concept of the form $\geq n\,R.B$. Thus, the branching factor is exponential assuming binary coding of numbers, and each such tree is at most doubly exponential in $|\mathcal{C}, \mathcal{A}|$.

Thus, the total number of individuals appearing in a derivation is at most doubly exponential in $|\mathcal{C}, \mathcal{A}|$. Since the branching factor in the derivation is exponentially bounded by $|\mathcal{C}, \mathcal{A}|$, each derivation is finite. □

We now state the main theorem of this section.

**Theorem 1.** *The satisfiability of a $\mathcal{SHOIQ}^+$ knowledge base $\mathcal{K}$ can be decided by computing $\mathcal{K}' = \Delta(\Omega(\mathcal{K}))$ and then checking whether some derivation for $\Xi(\mathcal{K}')$ contains a leaf node labeled with a clash-free ABox. Such an algorithm can be implemented such that it runs in* 2NExpTime *in $|\mathcal{K}|$.*

*Proof.* The first part of the theorem follows immediately from Lemmas 1, 2, 5, and 6. By Lemma 7, the total number of individuals is doubly exponential in $|\Xi_{\mathcal{A}}(\mathcal{K}'), \Xi_{\mathcal{TR}}(\mathcal{K}')|$. Since the structural transformation is polynomial, the total number of individuals is doubly exponential in $|\mathcal{K}|$. Thus, the existence of a leaf derivation node labeled with a clash-free ABox can be checked by nondeterministically applying the hypertableau derivation rules to construct an ABox that is at most doubly exponential in $|\mathcal{K}|$. □

## 5. Discussion

In this section we discuss the possibilities of optimizing the blocking condition to single and subset blocking; furthermore, we argue that modifying the algorithm to make it optimal w.r.t. worst-case complexity might be difficult.

### 5.1 Single Blocking

For DLs such as $\mathcal{SHOQ}^+$ that do not provide for inverse roles, pairwise blocking can be weakened to atomic single blocking, defined as follows.

**Definition 9** (Atomic Single Blocking). *Atomic single blocking is obtained from pairwise blocking (see Definition 7) by changing the notion of direct blocking: a blockable individual $s$ is directly blocked by a blockable individual $t$ if and only if $t$ is not blocked, $t \prec s$, and $\mathcal{L}_{\mathcal{A}}(s) = \mathcal{L}_{\mathcal{A}}(t)$ for $\mathcal{L}_{\mathcal{A}}(s)$ as in Definition 7.*[16]

In some cases, this simpler blocking condition can make the hypertableau algorithm construct smaller ABoxes, which can lead to increased efficiency. We next formalize the notion of HT-clauses to which atomic single blocking is applicable.

**Definition 10** (Simple HT-Clause). *An HT-clause $r$ is simple if it satisfies the following restrictions, for $x$ a center variable, $y_i$ a branch variable, $z_j$ a nominal variable, $B$ a literal concept, and $R$ an atomic role:*

- *Each atom in the antecedent of $r$ is of the form $A(x)$, $R(x,x)$, $R(x,y_i)$, $A(y_i)$, or $A(z_j)$.*

---

16. The name "atomic" reflects the fact that $\mathcal{L}_{\mathcal{A}}(s)$ contains only atomic concepts.

- *Each atom in the consequent of $r$ is of the form $B(x)$, $\geq h\,R.B(x)$, $B(y_i)$, $R(x,x)$, $R(x,y_i)$, $R(x,z_j)$, $x \approx z_j$, or $y_i \approx y_j$.*

It is straightforward to see that, if $\mathcal{K}$ is a $\mathcal{SHOQ}^+$ knowledge base, then $\Xi_{\mathcal{TR}}(\mathcal{K})$ contains only simple HT-clauses. The completeness of the hypertableau algorithm with atomic single blocking on simple HT-clauses is straightforward to show.

**Lemma 8.** *Let $\mathcal{C}$ be a set of simple HT-clauses, and $\mathcal{A}$ an input ABox. If a derivation with atomic single blocking for $\mathcal{C}$ and $\mathcal{A}$ exists in which a leaf node is labeled with a clash-free ABox $\mathcal{A}'$, then $(\mathcal{C}, \mathcal{A})$ is satisfiable.*

*Proof.* By slightly modifying the proof of Lemma 4, it is possible to show the following property (*): each atom in $\mathcal{A}'$ involving an atomic role is of the form $R(s,a)$, $R(s,s)$, or $R(s,s.i)$, for $a$ a named individual and $s$ any individual.

Let $I$ be a model constructed in the same way as in Lemma 6, but by using single blocking. Due to (*), whenever $\langle p_1, p_2 \rangle \in R^I$, then $p_2$ is either of the form $\left[\frac{a}{a}\right]$ for $a$ a named individual, it is a successor of $p_1$, or $p_2 = p_1$. The proof that $I$ is a model of $(\mathcal{C}, \mathcal{A})$ is a straightforward consequence of the following observations about the proof of Lemma 6:

- In the proof that $\geq n\,R.B(s) \in \mathcal{A}'$ implies $p_s \in (\geq n\,R.B)^I$, individual $u_i$ can never be a blockable predecessor of $s$. Thus, labels $\mathcal{L}_{\mathcal{A}'}(s,u_i)$, $\mathcal{L}_{\mathcal{A}'}(u_i,s)$, and $\mathcal{L}_{\mathcal{A}'}(u_i)$ are never relevant.

- In the proof that $I \models \mathcal{C}$, it is not possible that $p_{y_i}$ is a predecessor of $p_x$. Thus, labels $\mathcal{L}_{\mathcal{A}'}(s', \mathsf{tail}(p_{y_i}))$, $\mathcal{L}_{\mathcal{A}'}(\mathsf{tail}(p_{y_i}), s')$, and $\mathcal{L}_{\mathcal{A}'}(\mathsf{tail}(p_{y_i}))$ are never relevant.

The proof that $I$ is a model of $(\mathcal{A}, \mathcal{C})$ thus requires only $\mathcal{L}_{\mathcal{A}'}(s) = \mathcal{L}_{\mathcal{A}'}(t)$ to hold when $s$ is blocked by the blocker $t$; hence, $I$ is a model of $(\mathcal{A}, \mathcal{C})$ even if atomic single blocking is used. $\qquad\square$

The following variant of single blocking can also be applied to DLs with inverse roles but no number restrictions, such as $\mathcal{SHOI}$.

**Definition 11** (Full Single Blocking). *Full single blocking is obtained from atomic single blocking (see Definition 9) by changing the definition of $\mathcal{L}_{\mathcal{A}}(s)$ as follows:*

$$\mathcal{L}_{\mathcal{A}}(s) = \{\, C \mid C(s) \in \mathcal{A} \text{ where } C \text{ is of the form } A \text{ or } \geq 1\,R.B$$
$$\text{with } A \text{ an atomic and } B \text{ a literal concept} \,\}$$

For $t$ to directly block $s$ in $\mathcal{A}$ under atomic single blocking, it suffices if $s$ and $t$ occur in the same atomic concepts in $\mathcal{A}$. Intuitively, this is because the model construction from Lemma 6 "copies" all nonatomic concepts from $t$ to $s$; hence, assertions of the form $C(s)$ where $C$ is not atomic are not relevant. In contrast, in full single blocking, $s$ and $t$ must occur in $\mathcal{A}$ in *exactly* the same concepts (apart from negated atomic concepts). Intuitively, given a clash-free ABox $\mathcal{A}'$ to which no derivation rule is applicable, a model for $(\mathcal{A}, \mathcal{C})$ is constructed from $\mathcal{A}'$ by replacing $s$ with $t$; for the result to be a model, the two individuals must occur in exactly the same concepts.

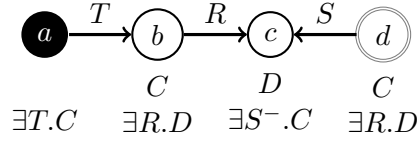

Figure 11: Problems with Single Blocking

Full single blocking must be applied with care in the hypertableau setting. Consider the following knowledge base $\mathcal{K}_9$, consisting of an ABox $\mathcal{A}_9$ and a set of HT-clauses $\mathcal{C}_9$.

$$
\begin{aligned}
(37) \qquad \mathcal{A}_9 \;&=\; \{\, \exists T.C(a) \,\} \\
\mathcal{C}_9 \;&=\; \{ C(x) \to \exists R.D(x),\; D(x) \to \exists S^-.C(x),\; R(x,y_1) \wedge S(x,y_2) \to \bot \}
\end{aligned}
$$

On $\mathcal{K}_9$, the hypertableau algorithm with full single blocking produces the ABox shown in Figure 11. The individual $d$ is blocked by $b$, so the algorithm terminates; an expansion of $\exists R.D(d)$, however, would reveal that $\mathcal{K}_9$ is unsatisfiable. The problem arises because the HT-clause $R(x,y_1) \wedge S(x,y_2) \to \bot$ contains two role atoms, which allows the HT-clause to examine both the successor and the predecessor of $x$. Full single blocking, however, does not ensure that both predecessors and successors of $x$ have been fully built. We can correct this problem by requiring the normalized GCIs to contain at most one $\forall R.C$ concept. For example, if we replace our HT-clause with $R(x,y_1) \to Q(x)$ and $Q(x) \wedge S(x,y_2) \to \bot$, then the first HT-clause would additionally derive $Q(b)$, so $d$ would not be blocked by $b$.

We can apply full single blocking to the DL $\mathcal{SHOI}$ provided that each HT-clause contains at most one role atom in the antecedent. We can always ensure this by suitably renaming complex concepts with atomic ones.

**Lemma 9.** *Let $\mathcal{A}$ be an ABox and $\mathcal{C}$ a set of HT-clauses such that, for each $r \in \mathcal{C}$, (i) $r$ contains no atoms of the form $R(x,x)$, (ii) the antecedent of $r$ contains at most one role atom, and (iii) all at-least restriction concepts are of the form $\geq 1\, S.B$ for $S$ a role and $B$ a literal concept. If a derivation with full single blocking for $\mathcal{C}$ and $\mathcal{A}$ exists in which a leaf node is labeled with a clash-free ABox $\mathcal{A}'$, then $(\mathcal{C}, \mathcal{A})$ is satisfiable.*

*Proof.* Let $\mathcal{A}''$ be obtained from $\mathcal{A}'$ by removing each assertion containing an indirectly blocked individual. Since no derivation rule is applicable to indirectly blocked individuals, no derivation rule is applicable to $\mathcal{A}''$ and $\mathcal{C}$. For an individual $s$ occurring in $\mathcal{A}''$, let $[s]_{\mathcal{A}''} = s$ if $s$ is not blocked in $\mathcal{A}''$, and let $[s]_{\mathcal{A}''} = s'$ if $s$ is blocked in $\mathcal{A}''$ by the blocker $s'$.

Note the following useful property (*): if $\neg A(s) \in \mathcal{A}''$, then $A(s) \notin \mathcal{A}''$ since the $\bot$-rule is not applicable to $\mathcal{A}''$; but then, $A([s]_{\mathcal{A}''}) \notin \mathcal{A}''$ as well.

We now construct an interpretation $I$ from $\mathcal{A}''$ as follows.

$$
\begin{aligned}
\triangle^I \;&=\; \{ s \mid s \text{ occurs in } \mathcal{A}'' \text{ and it is not blocked in } \mathcal{A}'' \} \\
s^I \;&=\; [s]_{\mathcal{A}''} \text{ for each individual } s \text{ occurring in } \mathcal{A}'' \\
A^I \;&=\; \{ [s]_{\mathcal{A}''} \mid A(s) \in \mathcal{A}'' \} \\
R^I \;&=\; \{ \langle [s]_{\mathcal{A}''}, [t]_{\mathcal{A}''} \rangle \mid R(s,t) \in \mathcal{A}'' \}
\end{aligned}
$$

It is straightforward to see that $I \models \mathcal{A}''$. Consider now each HT-clause $r \in \mathcal{C}$ that contains in the antecedent one atom of the form $R(x, y)$, as well as atoms of the form $A_i(x)$, $B_i(y)$, $C_i(z_i)$. Let $\sigma$ be a mapping from the variables of $r$ to the individuals in $\mathcal{A}''$ such that $I \models \sigma(U_i)$ for each atom $U_i$ from the antecedent of $r$. By the definition of $I$, individuals $s$ and $t$ then exist such that $R(s, t) \in \mathcal{A}''$, $\sigma(x) = [s]_{\mathcal{A}''}$, and $\sigma(y) = [t]_{\mathcal{A}''}$. By the definition of full single blocking, then $A_i(s) \in \mathcal{A}''$ and $B_i(t) \in \mathcal{A}''$ as well. Furthermore, since each $z_i$ occurs in a nominal guard concept, $\sigma(z_i)$ is a named individual. Let $\sigma'$ be such that $\sigma'(x) = s$, $\sigma'(y) = t$, and $\sigma'(z_i) = \sigma(z_i)$. Since the $Hyp$-rule is not applicable to $\mathcal{C}$ and $\mathcal{A}''$ for $\sigma'$, we have $\sigma'(V_j) \in \mathcal{A}''$ for some atom $V_j$ from the consequent of $r$. Consider now the possible forms that $V_j$ can have.

- If $V_j = S(x, y)$, then $I \models S(\sigma(x), \sigma(y))$ by the definition of $I$. The case $V_j = S(y, x)$ is analogous.

- If $V_j = A(x)$ for $A$ an atomic concept, then $A([s]_{\mathcal{A}''}) \in \mathcal{A}''$ by the definition of full single blocking; but then, $I \models A(\sigma(x))$ by the definition of $I$. The case when $V_j = A(y)$ is analogous.

- If $V_j = \neg A(x)$, then $A([s]_{\mathcal{A}''}) \notin \mathcal{A}''$ by (*); but then, by the definition of $I$ we have $I \models \neg A(\sigma(x))$. The case when $V_j = \neg A(y)$ is analogous.

- If $V_j = D(x)$ for $D = \geq 1\,R.B$, then $D([s]_{\mathcal{A}''}) \in \mathcal{A}''$ by the definition of full single blocking. Since the $\geq$-rule is not applicable to $[s]_{\mathcal{A}''}$, an individual $t$ exists such that $\mathsf{ar}(R, s, t) \in \mathcal{A}''$ and if $B$ is atomic, then $B(t) \in \mathcal{A}''$, and if $B = \neg A$, then $A(t) \notin \mathcal{A}''$. By the definition of full single blocking, if $B$ is atomic, then $B([t]_{\mathcal{A}''}) \in \mathcal{A}''$, and if $B = \neg A$, then $A([t]_{\mathcal{A}''}) \notin \mathcal{A}''$. By the definition of $I$, we have $\langle [s]_{\mathcal{A}''}, [t]_{\mathcal{A}''} \rangle \in R^I$, and $[t]_{\mathcal{A}''} \in B^I$; therefore, $I \models D(\sigma(x))$. The case when $V_j = D(y)$ is analogous.

- If $V_j = x \approx z_i$, then $\sigma'(x) \approx \sigma'(z_i) \in \mathcal{A}'$; since the $\approx$-rule is not applicable to $\mathcal{A}''$, we have $\sigma'(x) = \sigma'(z_i)$. But then, since named individuals cannot block other individuals, we have $\sigma(x) = \sigma'(x)$; hence, $I \models \sigma(x) \approx \sigma(z_i)$.

Thus, in all cases we have $I \models \sigma(V_j)$. The case when $r$ does not contain a role atom $R(x, y)$ in the antecedent is analogous, so $I \models (\mathcal{A}, \mathcal{C})$. $\qquad\square$

## 5.2 Subset Blocking

In tableau algorithms for DLs without inverse roles, full single blocking condition from Definition 11 can be further weakened to *full subset blocking* (Baader et al., 1996).

**Definition 12** (Full Subset Blocking). Full subset blocking *is obtained from full single blocking (see Definition 11) by changing the notion of direct blocking: a blockable individual $s$ is directly blocked by an individual $t$ if and only if $t$ is not blocked, $t \prec s$, and $\mathcal{L}_{\mathcal{A}}(s) \subseteq \mathcal{L}_{\mathcal{A}}(t)$.*

Full subset blocking is problematic in the hypertableau setting. Consider the knowledge base that consists of an ABox $\mathcal{A}_{10}$ and a TBox corresponding to the HT-clauses $\mathcal{C}_{10}$.

$$(38) \quad \begin{aligned} \mathcal{A}_{10} &= \{\, \exists T.C(a) \,\} \\ \mathcal{C}_{10} &= \left\{ \begin{array}{ll} C(x) \to \exists R.C(x), & C(x) \to \exists S.D(x), \\ S(x, y) \wedge D(y) \to E(x), & R(x, y) \wedge E(y) \to \bot \end{array} \right\} \end{aligned}$$

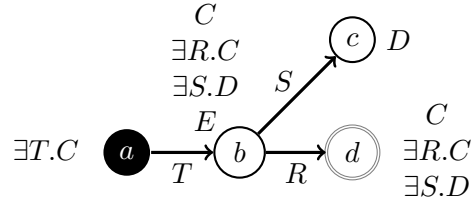

Figure 12: Problems with Full Subset Blocking

On $\mathcal{K}_{10}$, our algorithm can produce the ABox shown in Figure 12, in which $d$ is blocked by $b$. If, however, we expand $\exists S.D(d)$ into $S(d, e)$ and $D(e)$, we can derive $E(d)$; together with $R(b, d)$ and the HT-clause $R(x, y) \land E(y) \to \bot$, we get a contradiction.

The problem arises because, in the hypertableau setting, the syntactic distinction between atomic and inverse roles is lost: an atom $R^-(x, y)$ is transformed (by the function ar) into the semantically equivalent atom $R(y, x)$. The HT-clause $S(x, y) \land D(y) \to E(x)$ can be seen as including an implicit inverse role, because it examines a successor of $x$ in the antecedent in order to derive new information about $x$ in the consequent, thus mimicking the behavior of tableau algorithms with the semantically equivalent GCI $D \sqsubseteq \forall S^-.E$.

The semantically equivalent but inverse-free GCI $\exists S.D \sqsubseteq E$ would, in our hypertableau algorithm, be transformed into exactly the same HT-clause. In the tableau setting, however, this GCI would be treated very differently: it would result in the $\sqsubseteq$-rule deriving $(E \sqcup \forall S.\neg D)(s)$ for all individuals $s$. A similar effect could be achieved in the hypertableau setting by translating $\exists S.D \sqsubseteq E$ into two HT-clauses: $\top \to E(x) \lor Q(x)$ and $Q(x) \land S(x, y) \land D(y) \to \bot$. This introduces nondeterminism, but solves the problem with full subset blocking by deriving either $E(c)$ or $Q(c)$, the first of which leads to an immediate contradiction, and the second of which delays blocking.

In general, it is easy to see that full subset blocking could be used in the hypertableau setting by modifying the preprocessing phase so as to ensure that HT-clauses do not include such implicit inverses. It is not clear, however, if this would be very useful: it would result in (possibly) smaller ABoxes, but at the cost of (possibly) larger derivation trees.

## 5.3 The Number of Blockable Individuals

Buchheit et al. (1993) presented a tableau algorithm for the DL $\mathcal{ALCNR}$ which, due to anywhere blocking, runs in NExpTime instead of 2NExpTime, and Donini et al. (1998) presented a similar result for the basic DL $\mathcal{ALC}$. It is interesting to compare these algorithms to ours to see whether anywhere blocking can improve the worst-case complexity of our algorithm when $\mathcal{K}$ is a $\mathcal{SHIQ}^+$ knowledge base. In such a case, no HT-clause in $\Xi(\mathcal{K})$ contains a nominal guard concept, which prevents the derivation of assertions satisfying the preconditions of the NI-rule; hence, no new root individuals are introduced in a derivation, which eliminates a significant source of complexity.

The following example shows that, unfortunately, anywhere blocking does not improve the worst-case complexity; in fact, we identify a tension between and- and or-branching. In the example, we use the well-known encoding of binary numbers by concepts $B_0, B_1, \ldots, B_{k-1}$: we assign to each individual $s$ in an ABox $\mathcal{A}$ a binary number $\ell_{\mathcal{A}}(s) = b_{k-1} \ldots b_1 b_0$ such that $b_i = 1$ if and only if $B_i(s) \in \mathcal{A}$. Using $k$ concepts, we can thus encode $2^k$ different binary numbers. Furthermore, for any atomic role $R$, using the well-known $R$-*successor counting formula* (Tobies, 2000), we can ensure that, whenever an individual $t$ is an $R$-successor of $s$ in $\mathcal{A}$, then $\ell_{\mathcal{A}}(t) = (\ell_{\mathcal{A}}(s) + 1) \bmod 2^k$; we omit this formula for the sake of brevity. Let $\mathcal{K}_{11}$ be the following knowledge base. For the sake of brevity, we omit the HT-clauses corresponding to the axioms in $\mathcal{K}_{11}$.

(39) $$C(a)$$

(40) $$C \sqsubseteq \exists L.C \sqcap \exists R.C$$

(41) $$(\textit{The } R\textit{-successor formula for } B_0, \ldots, B_{k-1})$$

(42) $$(\textit{The } L\textit{-successor formula for } B_0, \ldots, B_{k-1})$$

(43) $$B_0 \sqcap \ldots \sqcap B_{k-1} \sqsubseteq A$$

(44) $$\exists L.A \sqcap \exists R.A \sqsubseteq A$$

Figure 13 schematically presents a derivation on $\mathcal{K}_{11}$ in which a doubly exponential number of blockable individuals is introduced.[17] For simplicity of presentation, we use single anywhere blocking. Due to (39)–(42), our algorithm can create individuals $a.1$, $a.2$, $a.1.1$, $a.1.2$, $a.1.1.1$, $a.1.1.2$, and so on, such that $s.1$ is an $L$-successor of $s$, and $s.2$ is an $R$-successor of $s$. After creating the individuals of the form $a.1^{2^k-1}.1$ and $a.1^{2^k-1}.2$ where $1^{2^k-1}$ is a string of $2^k-1$ ones, each individual $x.1$ blocks $x.2$ (c.f. Figure 13a). But then, due to (43), $a.1^{2^k-1}.1$ and $a.1^{2^k-1}.2$ become instances of $A$. By (44), $a.1^{2^k-1}$ is made an instance of $A$ as well, so it does not block its sibling $a.1^{2^k-2}.2$ any more; hence, $a.1^{2^k-2}.2$ is now expanded to exponential depth (c.f. Figure 13b). By repeating this process, the algorithm derives that $a.1^{2^k-2}$ is an instance of $A$, but then it does not block its sibling $a.1^{2^k-3}.2$ (c.f. Figure 13d). Eventually, the algorithm constructs a binary tree of exponential depth, thus creating a doubly-exponential number of blockable nodes in total (c.f. Figure 13d).

Buchheit et al. and Donini et al. obtained the nondeterministic exponential behavior by applying the $\sqcap$-, $\sqcup$-, $\forall$-, and $\sqsubseteq$-rules exhaustively before applying the $\exists$-rule. Such a strategy ensures that the label of an individual $s$ is fully constructed before introducing a successor of $s$, which prevents individuals from being indirectly blocked. On $\mathcal{K}_{11}$, this means that the GCI (44) is applied to each individual $s$ *before* introducing its successors. Thus, before the existentials on $s$ are expanded, the assertion $(\forall L.\neg A \sqcup \forall R.\neg A \sqcup A)(s)$ is introduced and one disjunct is chosen nondeterministically. The choices $(\forall L.\neg A)(s)$ and $(\forall R.\neg A)(s)$ will lead to a clash, so the algorithm eventually derives $A(s)$, before it expands the existentials on $s$ and introduces $s.1$ and $s.2$. Thus, while generating at most exponential models, this algorithm incurs a massive amount of nondeterminism.

Nondeterministic exponential behavior can be guaranteed in the hypertableau algorithm by nondeterministically fixing the label of each individual before applying the $\leq$-rule to it. This technique is similar to the one used by Tobies (2001) in order to obtain a PSPACE

---

17. Initially, we suggested informally that our algorithm should run in NEXPTIME on $\mathcal{SHIQ}$ (Motik, Shearer, & Horrocks, 2007). As this example shows, this is not the case.

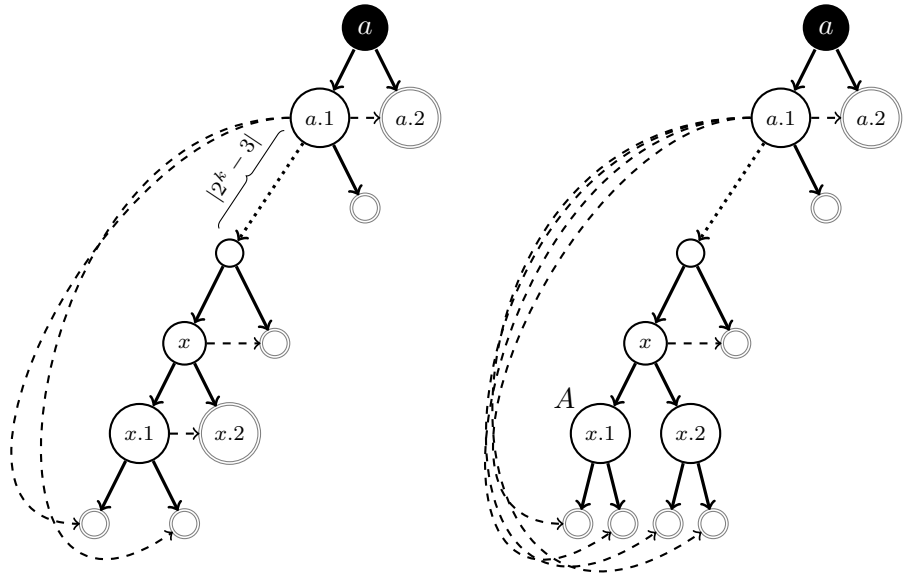

(a) An exponential path is constructed with each blockable individual blocking its sibling. No individual contains $A$ in its label.

(b) Adding $A$ to the label of $x.1$ unblocks $x.2$.

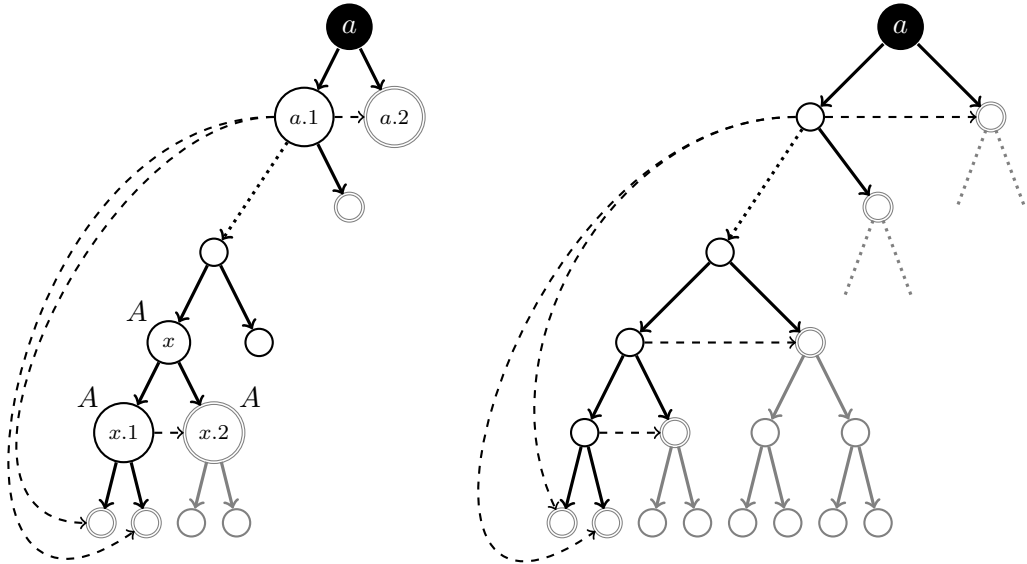

(c) Adding $A$ to the label of $x.2$ makes $x.2$ blocked, and forces the addition of $A$ to the label of $x$. This unblocks the sibling of $x$ so another subtree is created.

(d) Derivation terminates with an exponential number of unblocked individuals, but a doubly-exponential number of indirectly blocked individuals.

Figure 13: Creation of an Exponentially Deep Binary Tree of Blockable Individuals

decision procedure for concept satisfiability in a DL with inverse roles but without GCIs. The performance results in Section 7, however, seem to suggest that this might not be beneficial in practice. Still, it might be worth exploring whether nondeterministically adding concepts to labels of individuals can be used as an optimization that would detect "early blocks" and thus prevent the construction of large models.

## 5.4 The Number of Root Individuals

$\mathcal{SHOIQ}$ is NEXPTIME-complete (Tobies, 2000), and it is straightforward to extend this result to $\mathcal{SHOIQ}^+$. Thus, one might wonder whether the complexity result in Theorem 1 can be sharpened to obtain a worst-case optimal decision procedure. This, unfortunately, is not the case: we present an example on which our algorithm generates a doubly-exponential number of root individuals. We construct $\mathcal{K}_{12}$ by extending $\mathcal{K}_{11}$ (axioms (39)–(44)) with the following two axioms:

$$(45) \qquad\qquad B_0 \sqcap \ldots \sqcap B_{k-1} \sqsubseteq \{b\}$$

$$(46) \qquad\qquad A \sqsubseteq \, \leq 2\, L^-.\top \sqcap \, \leq 2\, R^-.\top$$

As shown in Section 5.3, the axioms of $\mathcal{K}_{11}$ can cause our algorithm to construct a binary tree of blockable individuals with exponential depth. Axiom (45) of $\mathcal{K}_{12}$, however, merges the leaves of this tree into the single named individual $b$, and axiom (46) ensures that the $NI$-rule is applied to each of the remaining blockable individuals, beginning with the neighbors of $b$. If, at each application of the $NI$-rule, we always merge blockable individuals into root individuals as shown in Figure 14a, then our algorithm constructs the ABox shown in Figure 14b, which contains two binary trees of root individuals of depth $2^{k/2}$. Unlike the case with $\mathcal{K}_{11}$, fully constructing individual labels does not avoid double-exponential behavior, since the promotion of blockable individuals to root individuals prevents blocking.

## 6. Algorithm Optimizations

DL reasoning algorithms are often used in practice to compute a *classification* of a knowledge base $\mathcal{K}$—that is, to determine whether $\mathcal{K} \models A \sqsubseteq B$ for each pair of atomic concepts $A$ and $B$ occurring in $\mathcal{K}$. A naïve classification algorithm would involve a quadratic number of calls to the subsumption checking algorithm, each of which can potentially be highly expensive. To obtain acceptable levels of performance, various optimizations have been developed that reduce the number of subsumption checks and the time required for each check (Baader, Hollunder, Nebel, Profitlich, & Franconi, 1994). The well-known dependency-directed back-tracking optimization (Horrocks, 2007) can readily be used with the hypertableau calculus. Furthermore, we have developed two simple optimizations that, to the best of our knowledge, have not been considered previously in the literature.

### 6.1 Reading Classification Relationships from Concept Labels

Let $(\mathcal{A}, \mathcal{C})$ be an ABox and a set of HT-clauses obtained by clausifying a knowledge base $\mathcal{K}$, and let $A$ and $B$ be atomic concepts for which we want to check whether $\mathcal{K} \models A \sqsubseteq B$; since $A$ and $B$ are atomic, this is the case if and only if $(\mathcal{A}', \mathcal{C})$ is unsatisfiable where

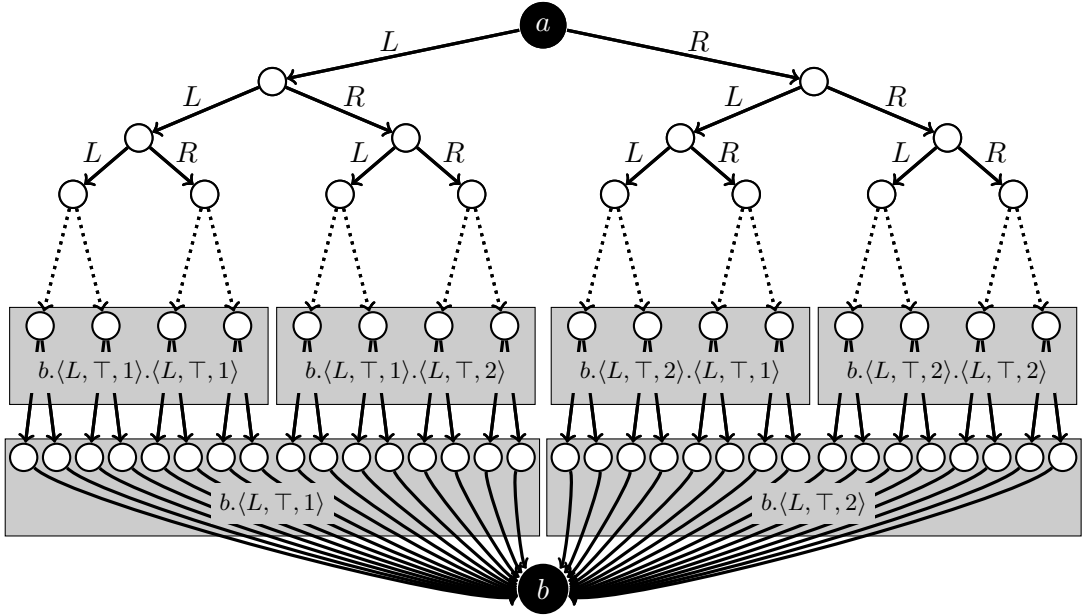

(a) A root introduction strategy for the *NI*-rule

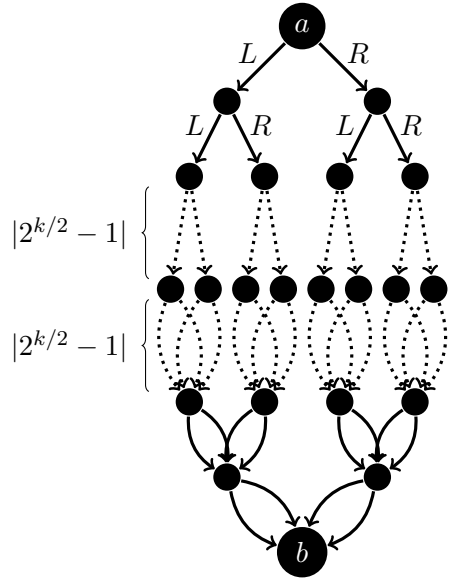

(b) The resulting tree, containing a doubly-exponential number of root individuals

$\mathcal{A}' = \mathcal{A} \cup \{A(a), \neg B(a)\}$ and $a$ is a fresh individual. Let $\mathcal{A}_1$ be a clash-free ABox labeling a leaf in a derivation from $(\mathcal{A}', \mathcal{C})$. We can use $\mathcal{A}_1$ to learn the following things about subsumption in $\mathcal{K}$. The proofs of these claims are straightforward.

1. If $C(a) \in \mathcal{A}_1$ for some concept $C$ and the derivation of $C(a)$ does not depend on a nondeterministic choice, then $\mathcal{K} \models A \sqsubseteq C$.

2. If $\mathcal{A}_1$ has been obtained from $\mathcal{A}'$ deterministically, then $\mathcal{K} \models A \sqsubseteq C$ only if $C(a) \in \mathcal{A}_1$.

3. If $C(b) \in \mathcal{A}_1$ but $D(b) \notin \mathcal{A}_1$ for $C$ and $D$ concepts and $b$ an individual that is not blocked, then $\mathcal{K} \not\models C \sqsubseteq D$.

Thus, if $\mathcal{K}$ is deterministic, we can classify it using a *linear* number of calls to the hypertableau algorithm: for each atomic concept $A$, we check the satisfiability of $(\mathcal{A} \cup \{A(a)\}, \mathcal{C})$; if the algorithm produces a clash-free ABox $\mathcal{A}_1$, the set of subsumers of $A$ are contained in $\mathcal{L}_{\mathcal{A}_1}(a)$. These optimizations are applicable in the case of tableau algorithms as well; however, they might be less effective due to increased or-branching.

## 6.2 Caching Blocking Labels

Let $\mathcal{T}$ and $\mathcal{R}$ be a $\mathcal{SHIQ}^+$ TBox and RBox, respectively, and let $\mathcal{C} = \Xi_{\mathcal{TR}}(\mathcal{T} \cup \mathcal{R})$; since $\mathcal{T}$ does not contain nominals, no assertions involving nominal guard concepts are needed. Furthermore, assume that the classification of $\mathcal{T} \cup \mathcal{R}$ involves $n$ calls to the hypertableau algorithm for $(\{A_i(a_i), \neg B_i(a_i)\}, \mathcal{C})$. Then, if a derivation for $(\{A_i(a_i), \neg B_i(a_i)\}, \mathcal{C})$ contains a leaf node labeled with a clash-free ABox $\mathcal{A}_i$, we can use the nonblocked individuals from $\mathcal{A}_i$ as blockers in all subsequent satisfiability checks of $(\{A_j(a_j), \neg B_j(a_j)\}, \mathcal{C})$ for $j > i$.

This is a simple consequence of the following fact. Let $I_1$ and $I_2$ be two models of $\mathcal{T} \cup \mathcal{R}$ such that $\triangle^{I_1} \cap \triangle^{I_2} = \emptyset$; furthermore, let $I$ be defined as $\triangle^I = \triangle^{I_1} \cup \triangle^{I_2}$, $A^I = A^{I_1} \cup A^{I_2}$, and $R^I = R^{I_1} \cup R^{I_2}$, for each atomic concept $A$ and each atomic role $R$. Then, by a simple induction on the structure of axioms in $\mathcal{T} \cup \mathcal{R}$, it is trivial to show that $I \models \mathcal{T} \cup \mathcal{R}$. This property does not hold in the presence of nominals, which can impose a bound on the number of elements in the interpretation of a concept; the bound could be satisfied in $I_1$ and $I_2$ individually, but violated in $I$.

Our optimization is correct because, instead of $(\{A_i(a_i), \neg B_i(a_i)\}, \mathcal{C})$, we can check the satisfiability of $(\mathcal{A}_i \cup \{A_i(a_i), \neg B_i(a_i)\}, \mathcal{C})$, and in doing so we can use the individuals from $\mathcal{A}_i$ as potential blockers due to anywhere blocking. This optimization can be seen as a very simple form of model caching (Horrocks, 2007), and it has been key to obtaining the results that we present in Section 7. For example, on GALEN only one subsumption test is costly because it computes a substantial part of a model of the TBox; all subsequent subsumption tests reuse large parts of that model.

In practice, we do not need to keep the entire ABox $\mathcal{A}_i$ around; rather, for each nonblocked blockable individual $t$ with a predecessor $t'$, we simply need to retain the sets $\mathcal{L}_{\mathcal{A}_i}(t)$, $\mathcal{L}_{\mathcal{A}_i}(t')$, $\mathcal{L}_{\mathcal{A}_i}(t, t')$, and $\mathcal{L}_{\mathcal{A}_i}(t', t)$.

## 7. Implementation and Evaluation

Based on the calculus from Section 4, we have implemented a prototype DL reasoner called HermiT. In order to estimate how well our calculus performs in practice, we have compared HermiT with two state-of-the-art tableau reasoners on several practical problems. The objective of this evaluation was *not* to establish the superiority of HermiT, but to compare the behavior of our calculus with that of the tableau calculi used in many existing systems, and to demonstrate the usefulness of our calculus on realistic problems.

It is important to understand that HermiT is a prototype, and as such does not always outperform the well-established reasoners. In particular, HermiT may be uncompetitive on ontologies where specialized optimizations are needed for good performance. For example, HermiT cannot process the SNOMED CT ontology due to the very large number of concepts, while many other reasoners can classify the ontology easily. These reasoners, however, employ techniques that are quite different from the standard tableau algorithm; for example, on an $\mathcal{EL}++$ ontology such as SNOMED CT, Pellet uses the reasoning algorithm by Baader, Brandt, and Lutz (2005), and other reasoners employ specialized techniques as well (Haarslev, Möller, & Wandelt, 2008). Similarly, artificial test problems such as those used in the TANCS comparison at the Tableaux'98 conference (Balsiger & Heuerding, 1998; Balsiger, Heuerding, & Schwendimann, 2000) and the DL'98 workshop (Horrocks & Patel-Schneider, 1998b) are often either easy for reasoners employing particular optimizations or are only difficult due to the fact that they encode large propositional satisfiability problems (Horrocks & Patel-Schneider, 1998a). Since our goal was to demonstrate the usefulness of the hypertableau calculus on realistic problems, we have chosen to ignore such ontologies and test problems, as they mainly test specialized calculi and optimizations that are applicable to various sublanguages of $\mathcal{SHOIQ}^+$. Instead, we focus our evaluation on practical ontologies in which the main difficulty is due to nontrivial reasoning problems encountered during classification.

In addition to the hypertableau calculus described in Section 4, HermiT also implements the optimizations from Section 6 and the well-known dependency directed backtracking optimization (Horrocks, 2007). Thus, HermiT fully supports $\mathcal{SHOIQ}^+$ and it can perform both satisfiability and subsumption testing as well as knowledge base classification. An extensive discussion of implementation techniques is beyond the scope of this paper; we only comment briefly on the implementation of anywhere blocking. In the DL community, it is commonly understood that anywhere blocking is more costly than ancestor blocking because, to determine the blocking status of an individual, one may need to examine all individuals in an ABox and not just the individual's ancestors. Our implementation avoids this problem by maintaining a hash table in which individuals are indexed by their four blocking labels. The table is created by scanning all individuals in $\mathcal{A}$ in the increasing sequence of the ordering $\prec$. For each individual $s$ in $\mathcal{A}$, if the parent of $s$ is blocked, then $s$ is indirectly blocked; otherwise, the algorithm queries the hash table for an individual whose blocking labels are equal to those of $s$. If the hash table contains such an individual $t$, then $s$ is directly blocked in $\mathcal{A}$ by $t$; otherwise, $s$ is not blocked in $\mathcal{A}$ so it is added into the hash table. The blocking status of all individuals in $\mathcal{A}$ can thus be determined with a linear number of hash table lookups.

We used Pellet 2.0.0rc4 (Parsia & Sirin, 2004) and FaCT++ 1.2.2 (Tsarkov & Horrocks, 2006) as reference implementations of the $\mathcal{SHOIQ}$ tableau algorithm (Horrocks & Sattler, 2007). Pellet employs ancestor blocking, while FaCT++ has recently been extended with anywhere blocking. At the time of writing, however, the implementation of anywhere blocking in FaCT++ was known to be incorrect,[18] so we switched this feature off and used FaCT++ with ancestor blocking as well. To measure the effects of ancestor vs. anywhere blocking, we also used HermiT-Anc—a version of HermiT with ancestor blocking.

We used a collection of 392 test ontologies that we assembled from three independent sources.

---

18. Personal communication with Dmitry Tsarkov.

- The Gardiner ontology suite (Gardiner, Horrocks, & Tsarkov, 2006) is a collection of OWL ontologies gathered from the Web and includes many of the most commonly-used OWL ontologies.

- The Open Biological Ontologies (OBO) Foundry[19] is a collection of biology and life science ontologies.

- GALEN (Rector & Rogers, 2006) is a large and complex biomedical ontology which has proven notoriously difficult to classify with existing reasoners.

We have preprocessed all ontologies to resolve ontology imports and eliminate some trivial syntactic errors. Thus, each test ontology can be parsed as a single file using the OWL API. All test ontologies are available online.[20]

We measured the time needed to classify each test ontology using the mentioned reasoners. All tests were performed on a 2.2 GHz MacBook Pro with 2 GB of physical memory. A classification attempt was aborted if it exhausted all available memory (Java tools were allowed to use 1 GB of heap space), or if it exceeded a timeout of 30 minutes.

The three reasoners exhibited negligible differences in performance on most of the test ontologies. Therefore, we discuss next only the test results for "interesting" ontologies—that is, ontologies that can be classified by at least one of the tested reasoners, and that are either not trivial or on which the tested reasoners exhibited a significant difference in performance. These include several ontologies from the OBO corpus (Molecule Role, XP Uber Anatomy, XP Plant Anatomy, Cellular Component, Gazetteer, CHEBI), two versions of the National Cancer Institute (NCI) Thesaurus (Hartel et al., 2005), two versions of the GALEN medical terminology ontology, two versions of the Foundational Model of Anatomy (FMA) (Golbreich et al., 2006), the Wine ontology from the OWL Guide,[21] two SWEET ontologies developed at NASA,[22] and a version of the DOLCE ontology developed at the Institute of Cognitive Science and Technology of the Italian National Research Council.[23] Basic statistical information about these ontologies is summarized in Table 10.

We noticed that, for all three reasoners, classification times may vary from run to run. For Pellet and HermiT, this is due to Java's collection library: the order of iteration over collections often depends on the objects' hash codes, and these may vary from run to run; that, in turn, may change the order in which the derivation rules are applied, and some orders may be better than others. We conjecture that FaCT++ is susceptible to similar variations. While the times may vary, we have not noticed a case where an ontology might be successfully classified in one run, but not in another. Therefore, in Table 11 we present the classification times for the "interesting" ontologies that we obtained on one particular run; these times can be taken as being "typical." We identified four groups of ontologies, which we delineate in Tables 10 and 11 by horizontal lines.

---

19. `http://obofoundry.org/`
20. `http://hermit-reasoner.com/2009/JAIR_benchmarks/`
21. `http://www.w3.org/TR/owl-guide/`
22. `http://sweet.jpl.nasa.gov/ontology/`
23. `http://www.loa-cnr.it/DOLCE.html`

Table 10: Statistics of "Interesting" Ontologies

| | | | | Number of Axioms | | | |
|---|---|---|---|---|---|---|---|
| Ontology Name | Classes | Roles | Individuals | TBox | RBox | ABox | Expressivity |
| Molecule Role | 8849 | 2 | 128056 | 9243 | 1 | 128056 | $\mathcal{ALE}+$ |
| XP Uber Anatomy | 11427 | 82 | 88955 | 14669 | 80 | 88955 | $\mathcal{ALEHIF}+$ |
| XP Plant Anatomy | 19145 | 82 | 86099 | 35770 | 87 | 86099 | $\mathcal{SHIF}$ |
| XP Regulators | 25520 | 4 | 155169 | 42896 | 3 | 155169 | $\mathcal{SH}$ |
| Cellular Component | 27889 | 4 | 163244 | 47345 | 3 | 163244 | $\mathcal{SH}$ |
| NCI-1 | 27652 | 70 | 0 | 46800 | 140 | 0 | $\mathcal{ALE}$ |
| Gazetteer | 150979 | 2 | 214804 | 167349 | 2 | 214804 | $\mathcal{ALE}+$ |
| GALEN-doctored | 2748 | 413 | 0 | 3937 | 799 | 0 | $\mathcal{ALEHIF}+$ |
| GALEN-undoctored | 2748 | 413 | 0 | 4179 | 800 | 0 | $\mathcal{ALEHIF}+$ |
| CHEBI | 20977 | 9 | 243972 | 38375 | 2 | 243972 | $\mathcal{ALE}+$ |
| FMA-Lite | 75141 | 2 | 46225 | 119558 | 3 | 46225 | $\mathcal{ALEI}+$ |
| SWEET Phenomena | 1728 | 145 | 171 | 2419 | 239 | 491 | $\mathcal{SHOIN}(\mathcal{D})$ |
| SWEET Numerics | 1506 | 177 | 113 | 2184 | 305 | 340 | $\mathcal{SHOIN}(\mathcal{D})$ |
| Wine | 138 | 17 | 206 | 355 | 40 | 494 | $\mathcal{SHOIN}(\mathcal{D})$ |
| DOLCE-Plans | 118 | 264 | 27 | 265 | 948 | 68 | $\mathcal{SHOIN}(\mathcal{D})$ |
| NCI-2 | 70576 | 189 | 0 | 100304 | 290 | 0 | $\mathcal{ALCH}(\mathcal{D})$ |
| FMA-Constitutional | 41648 | 168 | 85 | 122695 | 395 | 86 | $\mathcal{ALCOIF}(\mathcal{D})$ |

On the ontologies in the first group, HermiT performs similarly to HermiT-Anc, which suggests little impact of anywhere blocking on the performance. Consequently, we believe that HermiT outperforms the other reasoners mainly due to the reduced nondeterminism of the hypertableau calculus. As shown in Table 10, Molecule Role, XP Uber Anatomy, and NCI-1 do not use disjunctions, so HermiT classifies them deterministically using a linear number of calls to the hypertableau algorithm. FaCT++ outperforms HermiT on NCI-1 because this ontology can be classified using the *completely defined concepts* optimization (Tsarkov & Horrocks, 2005a), which FaCT++ implements but HermiT does not. This optimization enables FaCT++ to use simpler structural reasoning techniques on ontologies that satisfy certain syntactic constraints.

On the ontologies in the second group, HermiT-Anc is significantly slower than HermiT. This suggests that anywhere blocking significantly improves the performance since it prevents the construction of large models. Pellet runs out of memory on all ontologies in this group; furthermore, FaCT++ cannot process two of them and is significantly slower than HermiT on CHEBI. FaCT++, however, is faster than HermiT-Anc on CHEBI and GALEN-doctored, and we conjecture that this is mainly due to the ordering heuristics (Tsarkov & Horrocks, 2005b) used by FaCT++. The superior performance of HermiT on the ontologies in this group is mainly due to the fact that all of these ontologies can be classified deterministically using a linear number of concept satisfiability tests. Furthermore, HermiT's classification time is in most cases dominated by only the first such test, as the caching of blocking labels described in Section 6.2 makes subsequent tests easy.

On the ontologies in the third group, HermiT is significantly slower than the other reasoners. As Table 10 shows, all ontologies in this group contain nominals, which prevents HermiT from caching blocking labels. Furthermore, due to nominals, the ABox must be

Table 11: Results of Performance Evaluation

| Ontology Name | Classification Times (seconds) | | | |
|---|---|---|---|---|
| | HermiT | HermiT-Anc | Pellet | FaCT++ |
| Molecule Role | 3.3 | 3.4 | 25.7 | 304.5 |
| XP Uber Anatomy | 5.4 | 4.9 | — | 86.0 |
| XP Plant Anatomy | 12.8 | 11.2 | 87.2 | 22.9 |
| XP Regulators | 14.1 | 17.1 | 35.4 | 66.6 |
| Celular Component | 18.6 | 18.0 | 40.5 | 76.7 |
| NCI-1 | 14.1 | 14.4 | 23.2 | 3.0 |
| Gazetteer | 131.9 | 132.3 | — | — |
| GALEN-doctored | 8.8 | 456.3 | — | 15.9 |
| GALEN-undoctored | 126.3 | — | — | — |
| CHEBI | 24.2 | — | — | 397.0 |
| FMA-Lite | 107.2 | — | — | — |
| SWEET Phenomena | 13.5 | 11.2 | — | 0.2 |
| SWEET Numerics | 76.7 | 72.6 | 3.7 | 0.2 |
| Wine | 343.7 | 524.6 | 19.5 | 162.1 |
| DOLCE-Plans | 1075.1 | — | 105.1 | — |
| NCI-2 | — | — | 172.0 | 60.7 |
| FMA-Constitutional | — | — | — | 616.7 |

**Note:** entry — means that reasoner was unable to classify the ontology either due to time out or memory exhaustion.

taken into account during classification, and HermiT currently reapplies the hypertableau rules to the entire ABox in each run. Effectively, HermiT does not reuse any computation between different hypertableau runs. The other two reasoners, however, use the *completion graph caching* optimization (Sirin, Cuenca Grau, & Parsia, 2006), in which the tableau rules are first applied to the entire ABox, and the resulting completion graph is used as a starting point in each subsequent run.

HermiT was unable to classify the two ontologies in the fourth group. Both NCI-2 and FMA-Constitutional use disjunctions, so they cannot be classified using a linear number of concept satisfiability tests; instead, HermiT uses the classification algorithm by Baader et al. (1994). All classification tests are straightforward (each test takes less than 50 ms); however, the resulting taxonomy is rather shallow, so HermiT makes an almost quadratic number of tests. Both Pellet and FaCT++, however, use more optimized versions of the classification algorithm that reduce the number of tests that need to be performed.

To summarize, although HermiT is not better than Pellet and FaCT++ on all ontologies, our results clearly demonstrate the practical potential of both the reduced nondeterminism and anywhere blocking. In fact, anywhere blocking can mean the difference between success and failure on complex ontologies, which suggests that and-branching is a more significant source of inefficiency in practice than or-branching. Anywhere blocking is applicable to tableau calculi as well (as mentioned earlier, FaCT++ already contains a preliminary version of it), so we believe that our results can be used to improve the performance of tableau reasoners as well without the need for a major redesign. Conversely, most of the optimizations used in tableau reasoners can be used with the hypertableau algorithm, and incorporating them into HermiT would make HermiT competitive with Pellet and FaCT++ in those cases where HermiT is currently slower.

## 8. Conclusion

In this paper we presented a novel reasoning algorithm for DLs. The algorithm is based on hyperresolution with anywhere blocking, which reduces the nondeterminism due to GCIs and the sizes of generated models. Furthermore, the algorithm uses a novel refinement of the *NI*-rule which can reduce the amount of nondeterminism introduced in order to handle the interaction between nominals, inverse roles, and number restrictions (Horrocks & Sattler, 2007). This refined version of the *NI*-rule is equally applicable to tableau algorithms.

We have implemented our calculus and have conducted an extensive performance comparison. Our results show that the combination of the new calculus and novel optimizations significantly increases the performance of DL reasoning in practice: our reasoner is currently the only one that can classify several complex ontologies.

Despite this advance in performance, there are still some ontologies, such as the full version of GALEN,[24] that defeat HermiT (and state-of-the-art tableau reasoners). This is because the large number of cyclic axioms in these ontologies cause HermiT to construct extremely large ABoxes and eventually exhaust all available memory. To alleviate this problem, we have developed a reasoning technique in which the $\geq$-rule is modified to non-deterministically reuse individuals from the ABox generated thus far. Initial experiments with this technique have shown very promising results (Motik & Horrocks, 2008).

Finally, we plan to extend our technique to the DL $\mathcal{SROIQ}$, which extends $\mathcal{SHOIQ}$ with more expressive role inclusion axioms that allow us to express, for example, that the brother of a person's father is also that person's uncle. This logic is of considerable interest as it underpins OWL 2—the extension of OWL currently being standardized by the W3C.

### Acknowledgments

This is an extended version of a paper published at CADE 2007 (Motik et al., 2007). We thank the anonymous reviewer for numerous comments that have contributed to the quality of this paper.

---

24. http://www.co-ode.org/galen/